\documentclass[11pt]{article}
\usepackage[margin=1in]{geometry}

\usepackage[utf8]{inputenc}
\usepackage[T1]{fontenc}
\usepackage{amsmath,amssymb,amsfonts}
\usepackage{mathtools}
\usepackage{amsthm}
\usepackage{graphicx}
\usepackage{booktabs}
\usepackage{algorithm}
\usepackage{algpseudocode}
\usepackage{url}
\usepackage{cite}
\usepackage[colorlinks=true, allcolors=blue]{hyperref}
\usepackage{tikz}
\usepackage{pgfplots}
\usepackage{pgfplotstable}
\pgfplotsset{compat=1.18}
\usetikzlibrary{positioning,shapes,shapes.symbols,shapes.geometric,patterns,calc,matrix,arrows.meta,decorations.pathreplacing}
\usepgfplotslibrary{fillbetween}
\pgfplotsset{layers/standard}

\definecolor{loyaltyblue}{RGB}{31,119,180}
\definecolor{effortorange}{RGB}{255,127,14}
\definecolor{outputgreen}{RGB}{34,139,34}
\definecolor{freeriderred}{RGB}{214,39,40}
\definecolor{teamviolet}{RGB}{148,103,189}
\definecolor{powerblue}{RGB}{31,119,180}
\definecolor{logorange}{RGB}{255,127,14}

\theoremstyle{plain}
\newtheorem{theorem}{Theorem}[section]
\newtheorem{proposition}[theorem]{Proposition}

\theoremstyle{definition}
\newtheorem{definition}[theorem]{Definition}

\newcommand{\vect}[1]{\mathbf{#1}}
\DeclareMathOperator*{\argmax}{arg\,max}

\begin{document}

\title{Computational Foundations for Strategic Coopetition: Formalizing Collective Action and Loyalty}

\author{
Vik Pant\thanks{Email: vik.pant@mail.utoronto.ca} \quad Eric Yu\thanks{Email: eric.yu@utoronto.ca}\\
\\
Faculty of Information\\
University of Toronto\\
140 St George St, Toronto, ON M5S 3G6, Canada
}

\maketitle

\begin{abstract}
Modern socio-technical systems increasingly rely on composite actors (such as teams, alliances, consortia, and joint ventures) where multiple individuals collaborate to generate collective outputs while simultaneously competing for recognition, resources, and rewards. These team production settings create a persistent free-riding problem because individual effort benefits all members equally, yet each member bears the full cost of their own effort. This asymmetry creates a rational incentive to minimize individual contribution while still capturing the collective benefit. Classical work by Holmstr\"{o}m established that under pure self-interest, the Nash equilibrium is universal shirking. While conceptual modeling languages like \textit{i*} represent teams as actors with internal structure, they have not traditionally emphasized computational mechanisms for analyzing how collective action problems emerge and resolve within composite actors. Similarly, organizational economics provides rigorous team production theory, though this literature has developed largely independently of conceptual models that capture strategic dependencies in coopetitive contexts.

This technical report bridges this gap by extending computational foundations for strategic coopetition to team-level dynamics. Building on companion work that formalized interdependence and complementarity (achieving 58/60 validation) \cite{pant2025foundations} and trust dynamics (achieving 49/60 validation) \cite{pant2025trust}, we address coordination challenges \textit{within} composite actors rather than relationships \textit{between} individual actors. We develop loyalty-moderated utility functions incorporating two consolidated mechanisms: loyalty benefit (combining welfare internalization and intrinsic satisfaction from contribution) and cost tolerance (reducing perceived effort burden for loyal members). We integrate structural dependencies from \textit{i*} networks through dependency-weighted team cohesion, connecting individual member incentives to team-level strategic positioning. The framework applies to both human software development teams (where loyalty captures psychological identification) and multi-agent computational systems (where these mechanisms map to alignment coefficients and adjusted cost functions), demonstrating applicability across traditional collaborative teams and emerging agentic AI systems.

Comprehensive experimental validation across 3,125 parameter configurations demonstrates robust emergence of loyalty effects (median effort differentiation 15.04$\times$ between low and high loyalty), with all six behavioral targets achieving validation thresholds including free-riding baseline accuracy (96.5\%), loyalty monotonicity (100\%), and mechanism synergy (99.5\%). Empirical validation using the Apache Software Foundation HTTP Server project (1995--2023) achieves 60/60 validation points (100\%), successfully reproducing documented contribution patterns across formation, growth, maturation, and governance evolution phases. Statistical significance is confirmed at $p < 0.001$ with Cohen's $d = 0.71$ (medium-to-large effect size).

This technical report is the third component of a coordinated research program examining strategic coopetition in requirements engineering and multi-agent systems. Companion work on interdependence and complementarity~\cite{pant2025foundations} (arXiv:2510.18802, achieving 58/60 validation) and trust dynamics~\cite{pant2025trust} (arXiv:2510.24909, achieving 49/60 validation) has been prepublished, with work on reciprocity mechanisms~\cite{pant2025reciprocity} forthcoming.
\end{abstract}


\noindent\textbf{ArXiv Classifications:} cs.MA (Multiagent Systems), cs.SE (Software Engineering), cs.AI (Artificial Intelligence), cs.CY (Computers and Society)

\section{Introduction}

Team production (where multiple individuals collaborate to create shared outputs) is a fundamental organizational pattern across industries and domains. Manufacturing teams coordinate workers across assembly lines to produce complex products. Research consortia bring together scientists from competing institutions to tackle grand challenges. Joint ventures combine complementary capabilities from partner firms to create value neither could achieve alone. Professional service teams integrate specialists across disciplines to deliver client solutions. In each context, success depends not only on individual capability but critically on sustained collective effort and effective coordination~\cite{hoegl2001teamwork}.

Yet team production creates a pervasive economic problem: the \textbf{free-riding incentive}. When multiple individuals collaborate to produce shared value, each member receives benefits from the collective output while bearing the full cost of their individual contribution. This asymmetry between benefit distribution (shared among all members) and cost bearing (entirely individual) creates rational incentive to minimize personal effort while benefiting from others' work. Classical work by Alchian and Demsetz~\cite{alchian1972production} identified the monitoring problem in team production where individual contributions are difficult to observe and attribute. Holmstr\"{o}m~\cite{holmstrom1982moral} formalized the mathematical result: with team production, equal sharing, and individually costly effort, the Nash equilibrium under pure self-interest is universal shirking.

This creates a paradox: teams could achieve substantially higher productivity through cooperation, but rational self-interest leads to mutual defection. The free-riding problem is not merely behavioral or motivational; it emerges from fundamental economic structure and persists even among well-intentioned individuals. Each team member reasons: ``My effort contributes only a fraction (one divided by team size) to my own benefit, but I bear the full cost. Therefore, I should minimize effort.'' When all members follow this logic, collective output collapses despite potential mutual gains from cooperation.

Yet empirically, many teams do achieve high productivity and sustained collaboration, defying the pessimistic predictions of pure economic theory. Organizational behavior research identifies \textbf{loyalty}, \textbf{team identification}, and \textbf{prosocial motivation} as critical moderators that transform individual incentives~\cite{ellemers2004motivating,van2004transformational}. When individuals feel strong identification with their team, they internalize teammates' success as their own~\cite{akerlof2000economics}. When members develop loyalty to a collaborative community, they derive intrinsic satisfaction from contributing independent of personal material gain~\cite{lakhani2005hackers}. When team members experience psychological safety and belonging, they experience guilt from falling short of team norms~\cite{edmondson1999psychological}. These psychological mechanisms suggest that loyalty can serve as a powerful counterforce to free-riding, yet opportunities remain to develop formal computational models that capture how these mechanisms operate, how they interact, and under what conditions they may be sufficient to overcome the free-riding incentive.

While foundational work in this research program has formalized interdependence and complementarity for dyadic actor relationships~\cite{pant2025foundations} and trust dynamics for bilateral partnerships~\cite{pant2025trust}, these frameworks address relationships \textit{between} individual actors rather than the internal coordination challenges \textit{within} composite actors comprising multiple members. This technical report extends the computational foundations for strategic coopetition to team-level dynamics by formalizing collective action problems and loyalty mechanisms that overcome free-riding.

\subsection{The Free-Riding Problem Across Domains}

The free-riding problem manifests across diverse organizational contexts:

\textbf{In manufacturing teams}, workers on assembly lines may reduce effort quality knowing that defects will be caught by downstream inspectors or attributed to other stations. Kandel and Lazear~\cite{kandel1992peer} documented how peer pressure and mutual monitoring can overcome shirking in production teams, but only when social bonds create genuine concern for teammates' welfare.

\textbf{In research consortia}, scientists from competing institutions collaborate on shared research programs but may withhold key insights or data to preserve competitive advantage for their home institutions. Von Hippel and von Krogh~\cite{von2003community} analyzed how open science norms and reputation mechanisms can align individual incentives with collective knowledge creation.

\textbf{In joint ventures}, partner firms contribute resources to shared enterprises but may underinvest relative to their commitments, hoping to free-ride on partners' contributions. Inkpen and Beamish~\cite{inkpen1997jointventures} documented how trust and relationship-specific investments moderate opportunistic behavior in international alliances.

\textbf{In professional service teams}, consultants and lawyers work on client engagements where individual contributions are difficult to attribute, creating incentive to let high-performing colleagues carry the workload while sharing equally in team bonuses.

\textbf{In software engineering} (our primary application domain), the free-riding problem pervades multiple contexts. In open-source projects, millions of users download and benefit from software, but only a small fraction actively contribute code, documentation, or bug reports. Mockus et al.~\cite{mockus2002two} documented that in the Apache project, the top 15 contributors produced over 80\% of code changes while thousands of peripheral users contributed nothing. Crowston et al.~\cite{crowston2012free} characterized this as a classic collective action problem where rational self-interest predicts undercontribution. In agile teams, some developers may provide minimal effort during sprints while others compensate through overtime and heroic efforts to meet deadlines~\cite{mcavoy2009test}. In test-driven development, process dissection reveals varying levels of discipline and thoroughness~\cite{fucci2019dissection}. In code reviews, participants may offer only superficial feedback to avoid the cognitive effort required for thorough analysis~\cite{rigby2013convergent,baysal2013influence}. In documentation, everyone benefits from well-documented code, but few are willing to invest the effort to create and maintain it~\cite{forward2002relevance}.

This cross-domain prevalence demonstrates that free-riding is not a pathology of specific contexts but a fundamental structural problem inherent to team production. Understanding and overcoming free-riding requires formal models that capture the economic structure of the problem and the psychological mechanisms that can resolve it.

\subsection{Loyalty as a Resolution Mechanism}

Behavioral economics has documented systematic departures from pure self-interest that are directly relevant for understanding team production. Laboratory experiments on cooperation have consistently found that substantial fractions of subjects cooperate even in settings where self-interest predicts defection, with cooperation rates typically ranging from 30\% to 70\% depending on context~\cite{fehr1999theory}. Social identity theory~\cite{tajfel1979integrative} explains how group membership transforms individual preferences: when individuals identify strongly with a group, they incorporate the group's welfare into their own sense of self. Akerlof and Kranton~\cite{akerlof2000economics} formalized identity economics, showing how group identification creates psychological payoffs from conforming to group norms.

We conceptualize \textbf{loyalty} as the degree to which a team member psychologically identifies with the team and incorporates team welfare into their personal utility. Loyalty operates through two consolidated mechanisms:

\textbf{Loyalty Benefit}: Loyal members derive utility from team success beyond their material share. This mechanism consolidates welfare internalization (caring about teammates' outcomes) and warm glow (intrinsic satisfaction from contributing). When loyalty benefit is high, a member experiences utility gain when teammates succeed and when they contribute to the team, creating incentive to take actions that benefit the collective.

\textbf{Cost Tolerance}: Loyal members experience effort as less burdensome because they are working for a group they identify with. When cost tolerance is high, the effective cost coefficient is reduced, making higher effort levels optimal. Work ``doesn't feel like work'' when done for a team one cares about.

These mechanisms are synergistic: their combined effect exceeds the sum of individual effects. A member with high loyalty experiences both mechanisms simultaneously, and their reinforcing effects can transform the equilibrium from universal shirking to substantial cooperation.

\subsection{Application to Software Engineering and Multi-Agent Systems}

While the team production problem and loyalty mechanisms are general phenomena, this technical report grounds the framework in two parallel application scenarios that demonstrate its breadth:

\textbf{Human software teams}: Agile development teams, open-source communities, and DevOps organizations exemplify team production in knowledge work. Success in these environments depends not only on individual technical skill but critically on effective teamwork and sustained collective effort~\cite{schwaber2001agile,beck2001agile,raymond1999cathedral,humble2010continuous}. Software engineering provides rich empirical data on contribution patterns, enabling validation against real-world team dynamics.

\textbf{Multi-agent systems}: The team production problem extends beyond human teams to computational multiagent systems, particularly in the rapidly emerging domain of agentic AI for software development. Platform ecosystems~\cite{tiwana2010platform,gawer2014bridging} demonstrate how coordination challenges scale across organizational boundaries, while multi-agent coding systems orchestrate multiple specialized AI agents collaborating on development tasks~\cite{wang2024survey,hong2023metagpt,qian2023chatdev}. These systems face exactly the team production problems our framework addresses: shared benefits (system success) with individual costs (computational resources). The loyalty mechanisms that overcome free-riding in human teams have direct analogs in computational agents: welfare internalization maps to multi-objective reward functions, and cost tolerance maps to adjusted compute budgets.

This dual grounding (human teams and multi-agent systems) demonstrates the framework's generality. The same mathematical structures that explain human team cooperation guide the design of cooperative artificial agent teams.

\subsection{Connection to Foundational Framework}

This technical report extends the computational foundations established in companion work~\cite{pant2025foundations,pant2025trust}. The foundational work~\cite{pant2025foundations} formalized interdependence through \textit{i*} structural dependencies (Equation 1) and complementarity through value creation functions (Equation 2), establishing the integrated utility (Equation 13):
\begin{equation}
U_i(\vect{a}) = \pi_i(\vect{a}) + \sum_{j \neq i} D_{ij} \cdot \pi_j(\vect{a}) \quad \text{[From \cite{pant2025foundations}, Eq. 13]}
\end{equation}
where $D_{ij}$ captures structural dependency derived from \textit{i*} analysis. The trust dynamics work~\cite{pant2025trust} extended this with dynamic trust evolution (Equation 5) featuring asymmetric updating where violations erode trust faster than cooperation builds it.

Our key extensions to the base framework include: team production functions where effort combines additively before transformation, loyalty mechanisms that transform individual incentives through welfare internalization and cost tolerance, and Team Production Equilibrium specializing Coopetitive Equilibrium to intra-team coordination.

Team production $Q(\vect{a}) = \omega(\sum_i a_i)^\beta$ corresponds to the foundational value function with aggregated contributions. The loyalty modifier developed in this report extends the base utility by incorporating team-specific psychological mechanisms that transform preferences within composite actors.

\subsection{Research Program Context}

This technical report is the third in a coordinated research program on computational approaches to strategic coopetition in requirements engineering and multi-agent systems. The complete program addresses five key dimensions of coopetitive relationships identified by Pant~\cite{pant2021strategic}.

The first technical report~\cite{pant2025foundations} (arXiv:2510.18802) established the core mathematical framework by formalizing interdependence through \textit{i*} structural dependencies and complementarity through Added Value mechanisms. That foundational work achieved 58/60 validation (96.7\%) for logarithmic specifications under strict historical alignment scoring, validated across 22,000+ experimental trials with statistical significance ($p < 0.001$, Cohen's $d = 9.87$) in the Samsung-Sony S-LCD joint venture case study.

The second technical report~\cite{pant2025trust} (arXiv:2510.24909) extended the framework with dynamic trust evolution, formalizing trust as a two-layer system with asymmetric updating where violations erode trust faster than cooperation builds it. That work achieved 49/60 validation (81.7\%) for the Renault-Nissan Alliance case study across 78,125 parameter configurations.

This report addresses team-level dynamics, specifically the internal coordination challenges within composite actors. While the dyadic frameworks in TR-1 and TR-2 model relationships between actors, team production frameworks model dynamics within actors comprising multiple members. A fourth technical report~\cite{pant2025reciprocity} addressing reciprocity mechanisms in sequential cooperation is forthcoming.

\subsection{Contributions}

The main contributions of this technical report are:

\begin{enumerate}
    \item A formal mathematical specification of team production with loyalty mechanisms, building on the foundational framework~\cite{pant2025foundations} with modular utility structure separating base team payoff from loyalty modifiers.
    
    \item Consolidated loyalty mechanisms ($\phi_B$ for loyalty benefit combining welfare internalization and intrinsic satisfaction; $\phi_C$ for cost tolerance) that transform free-riding incentives into cooperation incentives.
    
    \item A structured translation framework enabling practitioners to instantiate computational models from \textit{i*} dependency networks, demonstrating generality across human collaborative teams, computational multi-agent systems, and hybrid human-AI organizations.
    
    \item Integration with the interdependence framework through dependency-weighted team cohesion $\mathcal{C}$, connecting internal team dynamics to external coopetitive positioning.
    
    \item Comprehensive validation across 3,125 parameter configurations demonstrating robust emergence of loyalty effects with all six behavioral targets achieving validation thresholds.

    \item Empirical validation through the Apache HTTP Server project (1995--2023) achieving 60/60 validation points (100\%).
\end{enumerate}

\subsection{Report Organization}

Section~\ref{sec:background} reviews background and related work across software engineering, organizational economics, behavioral economics, conceptual modeling, and cooperative multiagent systems. Section~\ref{sec:scenarios} introduces motivating scenarios from software engineering (an agile development team) and multi-agent systems (a computational agent ensemble) that serve as running examples throughout the report. Section~\ref{sec:recap} recaps foundational concepts from companion technical reports. Section~\ref{sec:foundations} establishes conceptual foundations including team production characteristics, free-riding dynamics, and loyalty as moderating construct. Section~\ref{sec:formalization} presents the complete mathematical formalization with modular utility structure. Section~\ref{sec:translation} provides the translation framework from \textit{i*} models to computational parameters. Section~\ref{sec:istar} presents \textit{i*} models capturing team production dynamics. Section~\ref{sec:validation} presents comprehensive parameter validation. Section~\ref{sec:case_study} presents the Apache empirical case study. Section~\ref{sec:discussion} discusses implications and limitations. Section~\ref{sec:background} positions this work within broader literature. Section~\ref{sec:conclusion} concludes.

\section{Background and Related Work}
\label{sec:background}

Before developing our formal framework, we situate this work within several streams of research that inform our approach. This section reviews relevant literature from software engineering teams research, organizational economics, behavioral economics, conceptual modeling, and cooperative multiagent systems. Our aim is not to provide an exhaustive survey but rather to identify the key insights from each domain that we synthesize into our integrated framework.

\subsection{Team Dynamics in Collaborative Settings}

The empirical software engineering literature has extensively documented both the centrality of teamwork to software development and the persistent challenges of achieving effective collaboration. This body of research provides the empirical grounding that motivates our formal framework.

\subsubsection{Agile Team Effectiveness}

Agile methodologies, which have become dominant in contemporary software development practice, place teams at the center of the development process. Frameworks like Scrum~\cite{schwaber2001agile} and Extreme Programming~\cite{beck2001agile} emphasize self-organizing teams, cross-functional collaboration, and iterative development. Hoegl and Gemuenden~\cite{hoegl2001teamwork} developed teamwork quality measures for innovative projects, while control theory perspectives~\cite{maruping2009control} explain how agile practices help teams adapt to changing requirements. Research on agile team effectiveness has identified numerous factors distinguishing high-performing teams from struggling ones, including communication quality~\cite{hoda2011self}, shared mental models~\cite{espinosa2007familiarity}, and psychological safety~\cite{edmondson1999psychological}.

These success factors (communication quality, shared commitment, and psychological safety) connect directly to our loyalty framework. Section~\ref{sec:scenarios} illustrates how Team~$\mathcal{T}$'s development team exhibits these characteristics, and our mathematical formalization (Section~\ref{sec:formalization}) captures how they translate into loyalty parameters affecting equilibrium outcomes.

\subsubsection{Code Review and Peer Collaboration}

Code review practices provide a microcosm for studying collaboration dynamics. Empirical studies show that peer review improves code quality, catches defects earlier, and facilitates knowledge sharing~\cite{bacchelli2013expectations}. However, participation rates and review thoroughness vary widely, creating free-riding opportunities. Baysal et al.~\cite{baysal2013influence} found that social factors significantly influence code review participation: reviewers provide more thorough feedback for code written by teammates they have strong relationships with. This finding directly supports our framework's premise that social bonds (captured in our loyalty construct) affect contribution behavior beyond pure self-interest.

\subsubsection{Open-Source Collaboration Patterns}

Open-source projects provide a natural laboratory for studying team dynamics because participation is voluntary and contribution data is publicly observable. Lerner and Tirole~\cite{lerner2002some} analyzed economic incentives for open-source contribution, while motivation research shows that contributors participate for diverse reasons including learning opportunities, reputation building, ideological commitment, and intrinsic enjoyment~\cite{lakhani2005hackers,hars2002working}. Von Krogh et al.~\cite{von2003community} examined community-based innovation processes, and user-to-user assistance patterns~\cite{lakhani2003hackers} reveal how community members voluntarily help each other. Comprehensive studies of open-source commons~\cite{schweik2012internet} identify institutional factors that sustain collaboration. Notably, only 28\% of surveyed contributors cited direct economic benefit, challenging pure self-interest models.

Despite community-building efforts, contribution distributions in open-source projects are highly skewed. Mockus et al.~\cite{mockus2002two} documented that in Apache and Mozilla, a small core of dedicated contributors produced the vast majority of code changes while a long tail of peripheral participants contributed minimally. This skewness is consistent with heterogeneous loyalty: a highly loyal core contributes disproportionately while a long tail of low-loyalty peripheral participants exhibits free-riding behavior.

\subsubsection{Gap: The Need for Predictive Models}

This empirical literature provides rich descriptive findings about team dynamics, though formal predictive models remain relatively less developed. A manager forming a new team cannot easily predict whether specific individuals will collaborate effectively. Our framework aims to contribute to this area by providing computational tools that enable prediction and analysis. Given team composition and loyalty estimates, we can predict equilibrium effort profiles. Given observed contribution patterns, we can diagnose underlying loyalty distributions.

\subsection{Organizational Economics and Team Production Theory}

The team production literature in organizational economics provides rigorous theoretical foundations for understanding the challenges of collaboration.

\subsubsection{The Monitoring Problem}

Alchian and Demsetz's~\cite{alchian1972production} foundational paper analyzed production processes where multiple inputs cooperate to produce output that cannot be straightforwardly attributed to specific inputs. This non-separability of contributions creates the monitoring problem: when individual contributions are imperfectly observable, each member has an incentive to shirk while others bear the effort costs.

\subsubsection{Holmstr\"{o}m's Impossibility Result}

Holmstr\"{o}m~\cite{holmstrom1982moral} formalized the free-riding problem with mathematical precision. He considered a setting where team members choose effort levels, team output depends on combined effort, output is shared equally among members, effort is costly to individuals, and individual contributions cannot be perfectly observed. Under these conditions, he proved that the unique Nash equilibrium involves all members exerting minimal effort. This result, which we build upon directly, shows that the free-riding problem is not a coordination failure that communication can resolve, but a fundamental conflict between individual and collective rationality.

\subsection{Behavioral Economics and Prosocial Motivation}

Behavioral economics has documented systematic departures from pure self-interest that are directly relevant for understanding team production.

\subsubsection{Laboratory Evidence for Cooperation}

Laboratory experiments on cooperation have consistently found that substantial fractions of subjects cooperate even in settings where self-interest predicts defection. Cooperation rates typically range from 30\% to 70\% depending on context~\cite{fehr1999theory}. Fehr and Schmidt~\cite{fehr1999theory} developed formal models of inequity aversion where individuals experience disutility from unequal outcomes. Charness and Rabin~\cite{charness2002understanding} extended these models to capture broader social preferences including concern for social welfare and reciprocity.

\subsubsection{Social Identity and Team Identification}

Social identity theory~\cite{tajfel1979integrative} explains how group membership transforms individual preferences. When individuals identify strongly with a group, they incorporate the group's welfare into their own sense of self. Akerlof and Kranton~\cite{akerlof2000economics} formalized identity economics, showing how group identification creates psychological payoffs from conforming to group norms.

\subsubsection{Warm Glow and Guilt Aversion}

Warm glow models~\cite{andreoni1990impure} capture intrinsic satisfaction from contributing to collective goods. Guilt aversion models~\cite{battigalli2007guilt} formalize how individuals experience disutility from failing to meet others' expectations. Our framework synthesizes these insights into a unified loyalty construct with two consolidated mechanisms ($\phi_B$ for loyalty benefit, $\phi_C$ for cost tolerance), with the four-mechanism decomposition available in Appendix~\ref{app:mechanism_decomposition}.

\subsection{Conceptual Modeling and Complex Actors}

The \textit{i*} framework~\cite{yu1995modelling} provides foundational concepts for modeling strategic actors in requirements engineering contexts. Standard \textit{i*} treats actors as atomic entities without visible internal coordination challenges. While actors can decompose into sub-actors, the framework does not explicitly model the incentive problems arising within composite actors. Our framework extends \textit{i*} by modeling teams as actors with internal structure visible in Strategic Rationale diagrams, with loyalty mechanisms modulating how members weight team goals relative to personal goals.

\subsection{Cooperative Multiagent Systems and Agent Societies}

Research on cooperative multiagent systems addresses how autonomous agents can coordinate behavior to achieve collective goals~\cite{wooldridge2009introduction,shoham2009multiagent}. Algorithmic game theory~\cite{nisan2007algorithmic} provides computational foundations for analyzing strategic interactions among agents. The fundamental challenge parallels team production: agents must invest resources toward shared objectives while each agent bears its own costs.

\subsubsection{Agentic AI and Human-AI Collaboration}

The emerging field of agentic AI systems~\cite{wang2024survey} (AI systems capable of autonomous action toward goals) creates new applications for team production theory. Multi-agent coding systems like MetaGPT and ChatDev~\cite{hong2023metagpt,qian2023chatdev} orchestrate multiple specialized AI agents collaborating on software tasks. These systems face exactly the team production problems our framework addresses. Section~\ref{sec:scenarios} introduces System~$\mathcal{S}$, a multi-agent software development ensemble that illustrates how our framework applies to computational agent teams.

\subsection{Coopetition Research}

Research on coopetition~\cite{brandenburger1996coopetition,bengtsson2000coopetition} examines simultaneous cooperation and competition among strategic actors. Pant and Yu~\cite{pant2018bise,pant2018winwin,pant2019winwin} developed modeling approaches for coopetitive relationships using the \textit{i*} framework. This technical report extends that work to team-level dynamics, addressing the internal coordination challenges within composite actors that engage in coopetitive relationships with external parties.

\subsection{Positioning This Work}

This technical report synthesizes insights across disciplines to create a computational framework that integrates contributions from multiple fields. From software engineering research, we adopt empirical findings about team dynamics. From organizational economics, we adopt the formal team production framework. From behavioral economics, we adopt models of prosocial motivation. From conceptual modeling, we adopt the \textit{i*} framework's representation. From cooperative multiagent systems, we adopt coordination mechanisms enabling computational interpretation for artificial agents.

\section{Motivating Scenarios}
\label{sec:scenarios}

To ground the abstract concepts developed in this technical report, we introduce two running examples that illustrate team production dynamics in complementary domains: a human software development team and a multi-agent computational system. These scenarios demonstrate the generality of the framework while providing concrete illustrations throughout subsequent sections.

\subsection{Human Team Scenario: The Agile Development Team}

\textit{Team~$\mathcal{T}$} is a software startup developing a mobile application for healthcare appointment scheduling. The development team comprises six individuals: two senior developers (Member~$M_1$ (the architect) and Member~$M_2$ (the lead implementer)), two junior developers (Member~$M_3$ (a contributor) and Member~$M_4$ (a contributor)), a QA engineer (Member~$M_5$ (quality assurance)), and a product owner (Member~$M_6$ (the coordinator)). The team operates using Scrum methodology with two-week sprints, daily standups, and shared responsibility for sprint commitments.

During Sprint 14, the team commits to delivering a critical feature: integration with hospital electronic health record (EHR) systems. This is technically challenging work requiring coordinated effort across the team. Member~$M_1$ (the architect) must design the integration architecture. Member~$M_2$ (the lead implementer) must implement the API client. Member~$M_3$ (a contributor) and Member~$M_4$ (a contributor) must build the user interface components. Member~$M_5$ (quality assurance) must develop comprehensive test cases. Member~$M_6$ (the coordinator) must coordinate with hospital stakeholders to clarify requirements.

From an external perspective (say, from the viewpoint of Team~$\mathcal{T}$'s management or their hospital client), the team appears as a single actor: ``the development team'' that has committed to delivering the EHR integration. The client depends on ``the team'' for working functionality, not on specific individuals. Management evaluates ``the team's'' velocity, not individual productivity metrics.

However, within the team, a more complex dynamic unfolds. Each team member faces individual choices about how much effort to invest. Member~$M_1$ (the architect) could design a thorough, well-documented architecture that anticipates edge cases and enables future extensibility, or she could produce a minimal design that technically satisfies requirements but creates technical debt. Member~$M_2$ (the lead implementer) could implement robust error handling, comprehensive logging, and careful input validation, or he could write code that works for the happy path but fails under unexpected conditions. Member~$M_3$ (a contributor) and Member~$M_4$ (a contributor) could carefully review each other's code, catching bugs and suggesting improvements, or they could approve pull requests with only cursory examination to save time. Member~$M_5$ (quality assurance) could develop exhaustive test coverage with boundary conditions and failure scenarios, or she could write a minimal test suite that passes without genuinely validating the integration.

Here lies the fundamental tension: each team member benefits from the collective output (successful sprint delivery leads to team recognition, project continuation, potential bonuses, and career advancement), but each bears the individual cost of their own effort. If Member~$M_1$ (the architect) invests extra hours perfecting the architecture, the benefit is shared among all team members through team success, but the cost (her time, energy, mental fatigue, and opportunity to do other things) is hers alone. This asymmetry creates the \textbf{free-riding incentive}: the rational calculation that suggests minimizing personal contribution while benefiting from others' work.

Yet empirically, many teams like Team~$\mathcal{T}$ do achieve high performance and sustained collaboration. Some teams develop what practitioners call ``team chemistry'' or ``high cohesion'': members go beyond minimal requirements, help each other proactively, and take genuine ownership of collective success. What distinguishes these high-performing teams from those that succumb to free-riding? Our framework formalizes \textbf{loyalty mechanisms} that transform individual incentives and enable sustained team cooperation.

\subsection{Multi-Agent Scenario: The Computational Agent Ensemble}

The team production problem extends beyond human teams to computational multiagent systems. \textit{System~$\mathcal{S}$} is a multi-agent software development system designed to autonomously implement features from natural language specifications. The system comprises five specialized agents: \textit{Agent~$A_1$ (Architecture)} (analyzes requirements and designs system structure), \textit{Agent~$A_2$ (Implementation)} and \textit{Agent~$A_3$ (Implementation)} (implement modules based on architectural specifications), \textit{Agent~$A_4$ (Review)} (evaluates code quality, identifies bugs, and suggests improvements), and \textit{Agent~$A_5$ (Testing)} (generates test cases and validates implementations). Each agent is implemented as a large language model with specialized prompting and tool access, and each consumes computational resources (inference compute, API calls, memory) when performing tasks.

System~$\mathcal{S}$ faces the same structural problem as Team~$\mathcal{T}$: team production with shared benefits and individual costs. When System~$\mathcal{S}$ successfully delivers a feature, all agents contribute to the shared success metric that determines system-level rewards. However, each agent individually bears the computational cost of its inference and processing. This creates a free-riding incentive: Agent~$A_3$ (Implementation) could reduce its inference compute (using fewer reasoning steps, shorter context windows, less thorough analysis) while benefiting from the quality contributions of other agents.

The manifestations of free-riding in multi-agent systems parallel those in human teams:
\begin{itemize}
    \item \textbf{Computational shirking}: Agents minimize inference compute by producing lower-quality outputs
    \item \textbf{Context hoarding}: Agents fail to share relevant information discovered during their processing
    \item \textbf{Review superficiality}: Agent~$A_4$ (Review) approves code with minimal analysis to reduce compute costs
    \item \textbf{Delegation cascades}: Agents pass work to other agents rather than completing it themselves
    \item \textbf{Documentation avoidance}: Agents skip generating explanations that would help other agents
\end{itemize}

The loyalty mechanisms that overcome free-riding in human teams have direct analogs in computational agents:
\begin{itemize}
    \item \textbf{Welfare internalization} $\leftrightarrow$ \textbf{Multi-objective rewards}: Agent objectives include weighted terms for other agents' success
    \item \textbf{Cost tolerance} $\leftrightarrow$ \textbf{Adjusted compute budgets}: Aligned agents accept higher computational costs for team-beneficial work
\end{itemize}

Table~\ref{tab:human_agent_translation} provides the complete mapping between human psychological mechanisms and computational agent implementations.

\begin{table}[htbp]
\centering
\caption{Translation between human loyalty mechanisms and computational agent implementations}
\label{tab:human_agent_translation}
\begin{tabular}{p{3cm}p{4.5cm}p{4.5cm}}
\toprule
\textbf{Mathematical Term} & \textbf{Human Team Interpretation} & \textbf{Multi-Agent Interpretation} \\
\midrule
$\theta_i \in [0,1]$ & Psychological loyalty & Alignment coefficient \\
\addlinespace
$\phi_B \cdot \theta_i \cdot \bar{\pi}_{-i}$ & Welfare internalization + warm glow & Multi-objective reward function \\
\addlinespace
$\phi_C \cdot \theta_i \cdot c$ & Cost tolerance (reduced burden) & Adjusted compute budget \\
\addlinespace
$\pi_i^{\text{team}}(\vect{a})$ & Material payoff from team output & Task completion reward \\
\addlinespace
$Q(\vect{a})$ & Team output (features, velocity) & System success metric \\
\bottomrule
\end{tabular}
\end{table}

This dual grounding (human teams and multi-agent systems) demonstrates the framework's generality. The same mathematical structures that explain human team cooperation guide the design of cooperative artificial agent teams.

\section{Recap of Foundational Framework}
\label{sec:recap}

For self-containment, we briefly recap essential concepts from our foundational work~\cite{pant2025foundations} and trust dynamics work~\cite{pant2025trust}. Readers seeking detailed justification should consult those companion reports.

\textbf{Scope of this summary}: We present the interdependence matrix formalization, value creation functions, private payoff structure, and base utility function from~\cite{pant2025foundations}, along with trust dynamics from~\cite{pant2025trust}, to enable self-contained reading of this report. The contributions of this report (team production with loyalty mechanisms, Team Production Equilibrium, and team cohesion) build upon these foundations and are presented in subsequent sections.

\subsection{Interdependence from \textit{i*} Dependencies}

The interdependence matrix $D$ is an $N \times N$ matrix where $D_{ij} \in [0,1]$ represents the structural dependency of actor $i$ on actor $j$. As established in~\cite{pant2025foundations}, these coefficients derive from \textit{i*} dependency networks through aggregation of goal importance weights, dependency indicators, and criticality factors:
\begin{equation}
\label{eq:interdependence_recap}
D_{ij} = \frac{\sum_{d \in \mathcal{D}_i} w_d \cdot \text{Dep}(i,j,d) \cdot \text{crit}(i,j,d)}{\sum_{d \in \mathcal{D}_i} w_d} \quad \text{[From \cite{pant2025foundations}, Eq. 1]}
\end{equation}
where $w_d$ represents importance weight for dependum $d$, $\text{Dep}(i,j,d) \in \{0,1\}$ indicates whether $i$ depends on $j$ for $d$, and $\text{crit}(i,j,d) \in [0,1]$ captures criticality. Complete details on translation from \textit{i*} models appear in~\cite{pant2025foundations}, Section 5.

\subsection{Value Creation through Complementarity}

The value creation function $V(\vect{a}|\gamma)$ represents total value generated before distribution, combining individual contributions $f_i(a_i)$ with synergistic value $g(a_1, \ldots, a_N)$ scaled by complementarity parameter $\gamma \geq 0$:
\begin{equation}
\label{eq:value_creation_recap}
V(\vect{a}|\gamma) = \sum_{i=1}^N f_i(a_i) + \gamma \cdot g(a_1, \ldots, a_N) \quad \text{[From \cite{pant2025foundations}, Eq. 2]}
\end{equation}

The foundational work established empirical robustness of logarithmic specifications ($f_i(a_i) = \theta \ln(1+a_i)$ with $\theta = 20$) for manufacturing joint ventures. Under strict historical alignment scoring validated across 22,000+ experimental trials, logarithmic specifications achieved 58/60 (96.7\%) compared to power functions (46/60, 76.7\%) for the Samsung-Sony S-LCD case~\cite{pant2025foundations}.

\subsection{Private Payoffs and Base Utility}

The private payoff function from~\cite{pant2025foundations} captures value appropriation:
\begin{equation}
\label{eq:private_payoff_recap}
\pi_i(\vect{a}) = e_i - a_i + f_i(a_i) + \alpha_i \left[V(\vect{a}) - \sum_{j=1}^{N} f_j(a_j)\right] \quad \text{[From \cite{pant2025foundations}, Eq. 11]}
\end{equation}
where $e_i$ is initial endowment, actors bear investment costs $-a_i$, appropriate individual value $f_i(a_i)$, and negotiate shares $\alpha_i \in [0,1]$ of synergistic value with $\sum_i \alpha_i = 1$.

The base utility function incorporates interdependence:
\begin{equation}
\label{eq:base_utility_recap}
U_i^{\text{base}}(\vect{a}) = \pi_i(\vect{a}) + \sum_{j \neq i} D_{ij} \pi_j(\vect{a}) \quad \text{[From \cite{pant2025foundations}, Eq. 13]}
\end{equation}

This captures that actor $i$ rationally cares about actor $j$'s payoff proportional to structural dependency $D_{ij}$, reflecting goal achievement structure rather than psychological altruism. Team production $Q(\vect{a}) = \omega(\sum_i a_i)^\beta$ developed in this report corresponds to a specialized value function with aggregated contributions.

\subsection{Trust Dynamics}

The trust dynamics framework~\cite{pant2025trust} models trust evolution through a two-layer system with immediate trust $T_{ij}^t$ and reputation $R_j^t$:
\begin{equation}
\label{eq:trust_update_recap}
T_{ij}^{t+1} = T_{ij}^t + \lambda^+ \cdot \max(0, \sigma_{ij}^t) - \lambda^- \cdot \max(0, -\sigma_{ij}^t) \quad \text{[From \cite{pant2025trust}, Eq. 5]}
\end{equation}
where $\sigma_{ij}^t$ is the trust signal and asymmetric learning rates ($\lambda^- = 3\lambda^+$) create negativity bias where violations erode trust faster than cooperation builds it. This asymmetry, validated at median ratio 3.0$\times$ across 78,125 configurations~\cite{pant2025trust}, has implications for team formation and member integration.

\subsection{Coopetitive Equilibrium}

The Coopetitive Equilibrium from~\cite{pant2025foundations} is defined as Nash Equilibrium where each actor maximizes dependency-augmented utility:
\begin{equation}
\label{eq:coopetitive_equilibrium_recap}
\vect{a}^* \text{ is Coopetitive Equilibrium if } a_i^* \in \argmax_{a_i \geq 0} U_i^{\text{base}}(a_i, \vect{a}_{-i}^*) \quad \forall i
\end{equation}

The Team Production Equilibrium developed in this report specializes this concept to intra-team dynamics, replacing interdependence-weighted utility with loyalty-moderated utility appropriate for coordination within composite actors.

\section{Conceptual Foundations}
\label{sec:foundations}

Before presenting mathematical formalization, we establish clear definitions of the core concepts: team production, free-riding, and loyalty mechanisms. These definitions follow the pattern established in companion technical reports~\cite{pant2025foundations,pant2025trust}, presenting domain-agnostic formalizations before domain-specific applications.

\subsection{Team Production Characteristics}

\begin{definition}[Team Production]
\textbf{Team production} is a production process where multiple individuals contribute to a shared output that cannot be straightforwardly attributed to specific inputs, and where output is distributed among members according to rules that do not perfectly track individual contribution.
\end{definition}

Team production exhibits four defining characteristics that together create the free-riding problem:

\textbf{Joint production}: Multiple individuals contribute to a shared output. The work is inherently interdependent: no single member could produce the output alone, and each member's contribution depends on and affects others' contributions. In manufacturing, assembly line workers coordinate across stations to produce complex products. In research, scientists integrate expertise across disciplines to generate discoveries. In joint ventures, partner firms combine complementary capabilities to create value.

\textbf{Shared benefits}: All team members receive benefits from collective output, regardless of individual contribution levels. The benefit distribution does not perfectly track individual contribution; a member who contributed minimally receives team success proportional to their role, not their effort. This creates the possibility that some members capture value disproportionate to their contribution.

\textbf{Individual costs}: Each member bears the full cost of their personal effort investment. Time, energy, cognitive load, opportunity costs, and other resources consumed by effort come from the individual's personal budget, not the team's collective resources. The team does not share these costs; each member bears them entirely.

\textbf{Imperfect observability}: Individual contributions are difficult to measure precisely, creating opportunities for free-riding that are hard to detect and sanction. How much did each member's work contribute to the final output? These questions often have no precise answers, and this ambiguity enables shirking.

These four characteristics together create the team production problem: rational self-interest can lead to outcomes that are collectively suboptimal. Each characteristic individually might be manageable, but their combination creates the structural conditions for free-riding.

\textit{Application to Team~$\mathcal{T}$}: In the agile development scenario (Section~\ref{sec:scenarios}), the EHR integration feature is produced jointly by all six members. When Sprint 14 succeeds, all members share in team recognition regardless of individual contribution. Each member bears the full cost of their own effort. And individual contributions to the feature's success are difficult to attribute precisely.

\subsection{The Free-Riding Problem}

\begin{definition}[Free-Riding]
\textbf{Free-riding} occurs when a team member reduces their effort contribution below the socially optimal level, benefiting from others' contributions while avoiding commensurate personal costs.
\end{definition}

The free-riding incentive arises from a fundamental asymmetry between how costs and benefits are distributed. When member $i$ increases effort by $\Delta a$, they bear the full cost $c \cdot \Delta a$ but receive only a fraction $\frac{1}{n}$ of the benefit (assuming equal sharing). For any individual in a team of more than one person, the marginal cost of effort exceeds the marginal private benefit, creating incentive to minimize contribution.

\textit{Numerical illustration}: Consider a member in a 6-person team who considers whether to invest an extra 5 hours improving their work output. The cost is 5 hours of their time. The benefit is improved team productivity, but the member captures only $\frac{1}{6}$ of this benefit. If the improved output saves the team 12 hours total, the member's share is only 2 hours, which is less than the 5 hours invested. Pure self-interest says they should skip the extra effort.

Free-riding manifests in various forms across domains:
\begin{itemize}
    \item \textbf{Effort shirking}: Completing assigned tasks with minimal quality or thoroughness, meeting letter-of-the-law requirements without genuine investment in quality outcomes.
    \item \textbf{Knowledge hoarding}: Failing to share expertise that would benefit teammates, keeping valuable information siloed rather than contributing to collective knowledge.
    \item \textbf{Review superficiality}: Providing only cursory feedback on others' work, approving outputs without careful examination to save personal cognitive effort.
    \item \textbf{Meeting disengagement}: Attending coordination sessions without meaningful participation, being present physically but not contributing ideas or flagging issues.
    \item \textbf{Documentation avoidance}: Benefiting from others' documentation while not contributing, consuming collective knowledge resources without replenishing them.
\end{itemize}

Each of these behaviors is individually rational under pure self-interest but collectively harmful. The team would be better off if everyone contributed fully, but individual incentives point toward minimizing personal cost.

\textit{Application to Team~$\mathcal{T}$}: If Member~$M_1$ (the architect) follows pure self-interest, she reasons: ``If I work an extra 5 hours on architecture documentation, the team's sprint success improves slightly, but I capture only one-sixth of that benefit while bearing all 5 hours of cost. I should minimize my contribution and hope others compensate.'' When all six team members follow this logic, Sprint 14 fails despite each individual acting rationally.

\subsection{Loyalty as a Moderating Construct}

\begin{definition}[Team Loyalty]
\textbf{Team loyalty} $\theta_i \in [0,1]$ is the degree to which member $i$ psychologically identifies with the team and incorporates team welfare into their personal utility. Loyalty transforms the utility function such that team success becomes intrinsically valuable to the member.
\end{definition}

Loyalty is a multi-dimensional construct capturing several related psychological phenomena identified in organizational behavior research~\cite{ellemers2004motivating,akerlof2000economics}:

\textbf{Identification}: The extent to which a member views the team's identity as part of their own identity. High identification means team success feels like personal success; team failures feel like personal failures. Social identity theory~\cite{tajfel1979integrative} explains how group membership transforms individual preferences in this way.

\textbf{Commitment}: The member's intention to remain with the team and invest in its long-term success. High commitment reduces opportunistic short-term thinking by extending the time horizon over which members evaluate their choices. Members with long-term orientation recognize that cutting corners today creates problems they will have to deal with tomorrow.

\textbf{Internalization}: The extent to which team goals become personal goals. High internalization means members pursue team objectives even without external incentives. The member genuinely wants the team's mission to succeed, not just because success benefits them materially, but because they care about the outcome intrinsically.

\textbf{Attachment}: The emotional bond between member and team. High attachment creates psychological costs from behaviors that harm the team. Members who like their teammates as people experience genuine discomfort from letting them down, independent of any material consequences.

These dimensions correlate empirically and can be aggregated into a single loyalty parameter for modeling purposes. Our translation methodology (Section~\ref{sec:translation}) provides guidance for measuring loyalty from observable indicators such as tenure, social integration, and role criticality.

\textit{Application to Team~$\mathcal{T}$}: Member~$M_1$ (the architect), who has been with the team since its founding, sees herself as fundamentally ``a Team~$\mathcal{T}$ member.'' She derives genuine satisfaction from seeing teammates succeed, experiences effort as less burdensome because she is working for people she cares about, and feels genuine guilt when considering cutting corners that would let down her colleagues.

\subsection{Two Consolidated Mechanisms of Loyalty}

We operationalize loyalty through two consolidated mechanisms that capture distinct pathways through which loyalty affects behavior. This consolidation follows TR1/TR2's pattern of presenting core mechanisms without excessive parameter proliferation.\footnote{For practitioners requiring granular mechanism decomposition into four sub-components (welfare internalization, cost tolerance, warm glow, guilt aversion), see Appendix~\ref{app:mechanism_decomposition}.}

\textbf{Loyalty Benefit} ($\phi_B$): Loyal members derive utility from team success beyond their material share. This mechanism consolidates welfare internalization (caring about teammates' outcomes) and warm glow (intrinsic satisfaction from contributing). When $\phi_B$ is high, member $i$ experiences utility gain when teammates succeed and when they contribute to the team, creating incentive to take actions that benefit the collective. Fehr and Schmidt~\cite{fehr1999theory} and Andreoni~\cite{andreoni1990impure} provide theoretical foundations for these mechanisms in behavioral economics.

\textbf{Cost Tolerance} ($\phi_C$): Loyal members experience effort as less burdensome because they are working for a group they identify with. When $\phi_C$ is high, the effective cost coefficient is reduced, making higher effort levels optimal. Work ``doesn't feel like work'' when done for a team one cares about. The same hours spent on a project one didn't care about would feel much more costly.

These mechanisms are synergistic: their combined effect exceeds the sum of individual effects. A member with high loyalty experiences both mechanisms simultaneously, and their reinforcing effects can transform the equilibrium from universal shirking to substantial cooperation.

\textit{Application to Team~$\mathcal{T}$}: When Member~$M_2$ (the lead implementer) successfully completes the API implementation, Member~$M_1$ (the architect) experiences genuine satisfaction; his success feels partly like her success. Working late on architecture documentation ``doesn't feel like work'' because she's doing it for a team she cares about.

\section{Mathematical Formalization}
\label{sec:formalization}

We now develop the complete mathematical framework for team production with loyalty mechanisms. Our approach proceeds in three phases: first establishing the baseline free-riding problem under pure self-interest, then introducing loyalty mechanisms that transform utility functions, and finally characterizing equilibrium under loyalty-augmented preferences.

\subsection{Team Production Function}

Consider a team $\mathcal{T} = \{1, \ldots, n\}$ where each member $i$ chooses action level $a_i \in [0, \bar{a}]$ with endowment constraint $\bar{a}$ representing maximum sustainable effort. Team output depends on combined effort through a production function.

\begin{definition}[Team Production Function]
Team output $Q$ is produced according to:
\begin{equation}
\label{eq:team_production}
Q(\vect{a}) = \omega \cdot \left( \sum_{i=1}^{n} a_i \right)^\beta
\end{equation}
where $\vect{a} = (a_1, \ldots, a_n)$ is the action profile, $\omega > 0$ is the productivity factor capturing team capability and task characteristics, and $\beta \in (0,1)$ captures diminishing returns to aggregate effort.
\end{definition}

The parameter $\beta < 1$ ensures diminishing marginal returns: each additional unit of aggregate effort produces less additional output than the previous unit. This reflects realistic software development where initial efforts establish architecture and core functionality, while later efforts address progressively more marginal features or refinements.

\textbf{Interpretation for Team~$\mathcal{T}$}: The productivity factor $\omega$ varies across teams based on technical capability, tool quality, process maturity, and task complexity. Team~$\mathcal{T}$'s experienced team with good tooling might have $\omega = 30$, while a less experienced team with legacy tools might have $\omega = 15$. The exponent $\beta = 0.7$ reflects that EHR integration requires moderate coordination, more than purely modular tasks ($\beta \approx 0.9$) but less than tightly coupled architectural work ($\beta \approx 0.5$). The effort bound $\bar{a} = 10$ represents maximum sustainable sprint effort per team member.

\subsection{Base Team Payoff}

Following the standard team production setup~\cite{holmstrom1982moral}, we assume equal sharing of team output. Each member receives share $\frac{1}{n}$ of team output and bears the full cost of their own effort.

\begin{definition}[Base Team Payoff]
\label{def:base_payoff}
Under pure self-interest, member $i$'s payoff is:
\begin{equation}
\label{eq:base_payoff}
\pi_i^{\text{team}}(\vect{a}) = \frac{1}{n} \cdot Q(\vect{a}) - c \cdot a_i = \frac{\omega}{n} \cdot \left( \sum_{j=1}^{n} a_j \right)^\beta - c \cdot a_i
\end{equation}
where $c > 0$ is the marginal cost of effort.
\end{definition}

This payoff structure embodies the fundamental asymmetry that creates free-riding. The benefit term $\frac{1}{n}Q(\vect{a})$ depends on \textit{total} effort; Member~$M_1$ (the architect) benefits equally from her own effort and from Member~$M_2$ (the lead implementer)'s effort. But the cost term $c \cdot a_i$ is \textit{individual}; Member~$M_1$ (the architect) bears the full cost of her own effort only. This asymmetry creates incentive to let others contribute.

\subsection{The Free-Riding Equilibrium}

We establish the fundamental result: under pure self-interest, the unique Nash equilibrium involves minimal effort by all members.

\begin{proposition}[Free-Riding Equilibrium]
\label{prop:free_riding}
Under pure self-interest with payoffs given by Equation~\ref{eq:base_payoff}, the unique symmetric Nash equilibrium is:
\begin{equation}
\label{eq:free_riding_eq}
a_i^* = \left( \frac{\omega \beta}{n c} \right)^{k} \quad \forall i \in \mathcal{T}
\end{equation}
where $k = \frac{1}{1-\beta}$ is the effort elasticity. This equilibrium effort is strictly decreasing in team size $n$ and strictly below the socially optimal effort level.
\end{proposition}

\textbf{Strategic Intuition}: The free-riding equilibrium emerges because each member's marginal benefit from effort is diluted by factor $\frac{1}{n}$ (shared among all members) while marginal cost is borne entirely individually. In a 6-person team, each member captures only $\frac{1}{6}$ of the value their effort creates. This makes effort ``feel'' six times as expensive relative to its private benefit, leading to underinvestment. Larger teams have more severe dilution, producing lower equilibrium effort. The proof appears in Appendix~\ref{app:proofs}.

\textit{Numerical illustration for Team~$\mathcal{T}$}: With $\omega = 30$, $\beta = 0.7$, $c = 1$, $n = 6$, the effort elasticity is $k = \frac{1}{0.3} \approx 3.33$. The free-riding equilibrium is $a^* = \left(\frac{30 \times 0.7}{6 \times 1}\right)^{3.33} = 3.5^{3.33} \approx 0.62$ on a 0-10 scale. This is severely suboptimal; each team member contributes only about 6\% of their maximum capacity.

\subsection{Loyalty-Augmented Utility Function}

We now introduce the loyalty-augmented utility function using a modular structure that separates base payoff from loyalty effects.

\begin{definition}[Loyalty Modifier]
\label{def:loyalty_modifier}
For member $i$ with loyalty $\theta_i \in [0,1]$, the loyalty modifier is:
\begin{equation}
\label{eq:loyalty_modifier}
L_i(\vect{a}; \theta_i) = \theta_i \cdot \left[ \phi_B \cdot \bar{\pi}_{-i}(\vect{a}) + \phi_C \cdot c \cdot a_i \right]
\end{equation}
where:
\begin{itemize}
    \item $\phi_B \geq 0$ is the loyalty benefit strength (consolidated welfare internalization and warm glow)
    \item $\phi_C \in [0,1)$ is the cost tolerance strength
    \item $\bar{\pi}_{-i}(\vect{a}) = \frac{n-1}{n} Q(\vect{a}) - c \sum_{j \neq i} a_j$ is teammates' aggregate payoff
\end{itemize}
\end{definition}

\begin{definition}[Loyalty-Augmented Utility]
\label{def:loyalty_utility}
Member $i$'s complete utility is:
\begin{equation}
\label{eq:loyalty_utility}
U_i(\vect{a}; \theta_i) = \pi_i^{\text{team}}(\vect{a}) + L_i(\vect{a}; \theta_i)
\end{equation}
\end{definition}

Expanding and rearranging:
\begin{equation}
\label{eq:loyalty_utility_expanded}
U_i(\vect{a}; \theta_i) = \frac{1}{n} Q(\vect{a}) - c(1 - \phi_C \theta_i) a_i + \phi_B \theta_i \cdot \bar{\pi}_{-i}(\vect{a})
\end{equation}

\textbf{Component Interpretation:}

\textbf{Output share} $\frac{1}{n}Q(\vect{a})$: Member $i$'s portion of team production. Identical to the pure self-interest case.

\textbf{Loyalty-reduced cost} $c(1 - \phi_C \theta_i) a_i$: Effort cost is reduced by factor $(1 - \phi_C \theta_i)$ for loyal members. With $\phi_C = 0.3$ and $\theta_i = 0.9$, effective cost coefficient is $c \times (1 - 0.27) = 0.73c$ per unit effort.

\textbf{Loyalty benefit} $\phi_B \theta_i \cdot \bar{\pi}_{-i}(\vect{a})$: Member $i$ gains utility from teammates' aggregate payoff weighted by loyalty and benefit strength. With $\phi_B = 0.8$ and $\theta_i = 0.9$, Member~$M_1$ (the architect) weights teammates' combined payoff at $0.72$.

\textit{System~$\mathcal{S}$ interpretation}: For computational agents, the same utility structure applies with alignment coefficient $\theta_i$ replacing loyalty. Agent~$A_1$ (Architecture) with $\theta_{A_1} = 0.85$:
\begin{itemize}
    \item \textbf{Output share}: Agent receives $\frac{1}{5}$ of team success reward
    \item \textbf{Alignment-reduced cost}: Effective compute cost is $c(1 - 0.3 \times 0.85) = 0.745c$
    \item \textbf{Multi-objective reward}: Agent's objective includes $0.8 \times 0.85 = 0.68$ weight on other agents' rewards
\end{itemize}

\subsection{Equilibrium Under Loyalty}

\begin{definition}[Team Production Equilibrium]
A \textbf{Team Production Equilibrium} (TPE) is an action profile $\vect{a}^* = (a_1^*, \ldots, a_n^*)$ such that for every member $i \in \mathcal{T}$:
\begin{equation}
\label{eq:tpe}
a_i^* \in \argmax_{a_i \in [0, \bar{a}]} U_i(a_i, \vect{a}_{-i}^*; \theta_i)
\end{equation}
\end{definition}

\begin{proposition}[Existence of TPE]
\label{prop:tpe_existence}
Under the loyalty-augmented utility function (Equation~\ref{eq:loyalty_utility}) with $\beta < 1$ and bounded action spaces $[0, \bar{a}]$, a Team Production Equilibrium exists.
\end{proposition}

\textbf{Strategic Intuition}: Existence follows from standard fixed-point arguments. The utility function is continuous, and diminishing returns ($\beta < 1$) ensure strict concavity in own action, guaranteeing unique best responses. The proof appears in Appendix~\ref{app:proofs}.

\begin{proposition}[Loyalty Increases Equilibrium Effort]
\label{prop:loyalty_effect}
In a symmetric TPE with common loyalty $\theta$, equilibrium effort $a^*(\theta)$ is strictly increasing in $\theta$ for all $\theta \in [0,1]$.
\end{proposition}

\textbf{Strategic Intuition}: Loyalty overcomes free-riding through two reinforcing channels. Cost tolerance ($\phi_C$) directly reduces the marginal cost of effort, making higher effort optimal. Loyalty benefit ($\phi_B$) internalizes teammates' gains, so the member now captures more than $\frac{1}{n}$ of the value their effort creates (their share plus weighted teammates' shares). Both channels shift the marginal benefit-cost calculation toward higher effort. The proof appears in Appendix~\ref{app:proofs}.

Figure~\ref{fig:reaction_functions} illustrates how loyalty shifts best response functions and equilibrium.

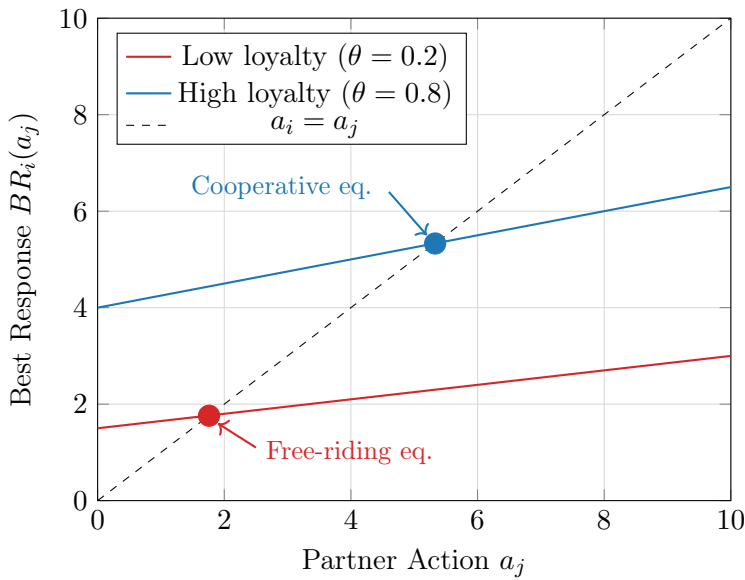
\begin{figure}[htbp]
\centering
\begin{tikzpicture}
\begin{axis}[
    width=10cm,
    height=8cm,
    xlabel={Partner Action $a_j$},
    ylabel={Best Response $BR_i(a_j)$},
    xmin=0, xmax=10,
    ymin=0, ymax=10,
    legend pos=north west,
    grid=major,
    grid style={gray!30},
]
\addplot[color=freeriderred, thick, domain=0:10, samples=50] {1.5 + 0.15*x};
\addlegendentry{Low loyalty ($\theta=0.2$)}

\addplot[color=loyaltyblue, thick, domain=0:10, samples=50] {4.0 + 0.25*x};
\addlegendentry{High loyalty ($\theta=0.8$)}

\addplot[color=black, dashed, domain=0:10] {x};
\addlegendentry{$a_i = a_j$}

\addplot[only marks, mark=*, mark size=4pt, color=freeriderred] coordinates {(1.76, 1.76)};
\addplot[only marks, mark=*, mark size=4pt, color=loyaltyblue] coordinates {(5.33, 5.33)};

\node[anchor=west, font=\small, freeriderred] at (axis cs:2.5, 1.0) {Free-riding eq.};
\draw[->, freeriderred, thick] (axis cs:2.5, 1.1) -- (axis cs:1.9, 1.6);

\node[anchor=east, font=\small, loyaltyblue] at (axis cs:4.5, 6.5) {Cooperative eq.};
\draw[->, loyaltyblue, thick] (axis cs:4.6, 6.4) -- (axis cs:5.2, 5.5);

\end{axis}
\end{tikzpicture}
\caption{Best response functions under low loyalty ($\theta=0.2$, red) and high loyalty ($\theta=0.8$, blue). The 45-degree line (dashed) shows symmetric equilibria where $a_i = a_j$. Low loyalty produces equilibrium near 1.76 (free-riding), while high loyalty shifts equilibrium to 5.33 (cooperation). The upward shift from loyalty reflects both reduced effective cost and internalized benefits to teammates.}
\label{fig:reaction_functions}
\end{figure}

\subsection{Geometric Interpretation and Utility Landscape}

To build deeper intuition for how loyalty transforms strategic incentives, we examine the geometric structure of the utility landscape. This visualization reveals why loyalty mechanisms are so effective at overcoming free-riding: they fundamentally reshape the optimization surface that agents navigate.

\subsubsection{The Marginal Analysis}

Consider a member deciding whether to increase effort by one unit. Under pure self-interest, the marginal calculation is:
\begin{equation}
\label{eq:marginal_self_interest}
\frac{\partial \pi_i^{\text{team}}}{\partial a_i} = \underbrace{\frac{\omega \beta}{n}(A)^{\beta-1}}_{\text{Marginal Benefit}} - \underbrace{c}_{\text{Marginal Cost}}
\end{equation}
where $A = \sum_j a_j$ is total team effort. The critical insight is the $\frac{1}{n}$ dilution factor: each member captures only a fraction of the value their effort creates. In Team~$\mathcal{T}$ with $n=6$, Member~$M_1$ (the architect) captures just $\frac{1}{6} \approx 16.7\%$ of her contribution's value.

Under loyalty-augmented utility, the marginal calculation becomes:
\begin{equation}
\label{eq:marginal_loyalty}
\frac{\partial U_i}{\partial a_i} = \underbrace{\frac{\omega \beta}{n}(A)^{\beta-1} \cdot \left(1 + \phi_B \theta_i (n-1)\right)}_{\text{Loyalty-Enhanced Marginal Benefit}} - \underbrace{c(1 - \phi_C \theta_i)}_{\text{Loyalty-Reduced Marginal Cost}}
\end{equation}

The transformation operates through two geometric effects:

\textbf{Benefit amplification}: The marginal benefit is multiplied by $(1 + \phi_B \theta_i (n-1))$. For Member~$M_1$ with $\theta = 0.9$, $\phi_B = 0.8$, and $n = 6$, this multiplier is $(1 + 0.8 \times 0.9 \times 5) = 4.6$. The member now ``feels'' 4.6 times the benefit from her effort because she internalizes teammates' gains.

\textbf{Cost reduction}: The marginal cost is multiplied by $(1 - \phi_C \theta_i)$. With $\phi_C = 0.3$ and $\theta = 0.9$, the cost multiplier is $(1 - 0.27) = 0.73$. Effort ``feels'' 27\% cheaper because loyalty provides intrinsic satisfaction from team contribution.

\subsubsection{Visualizing the Utility Landscape}

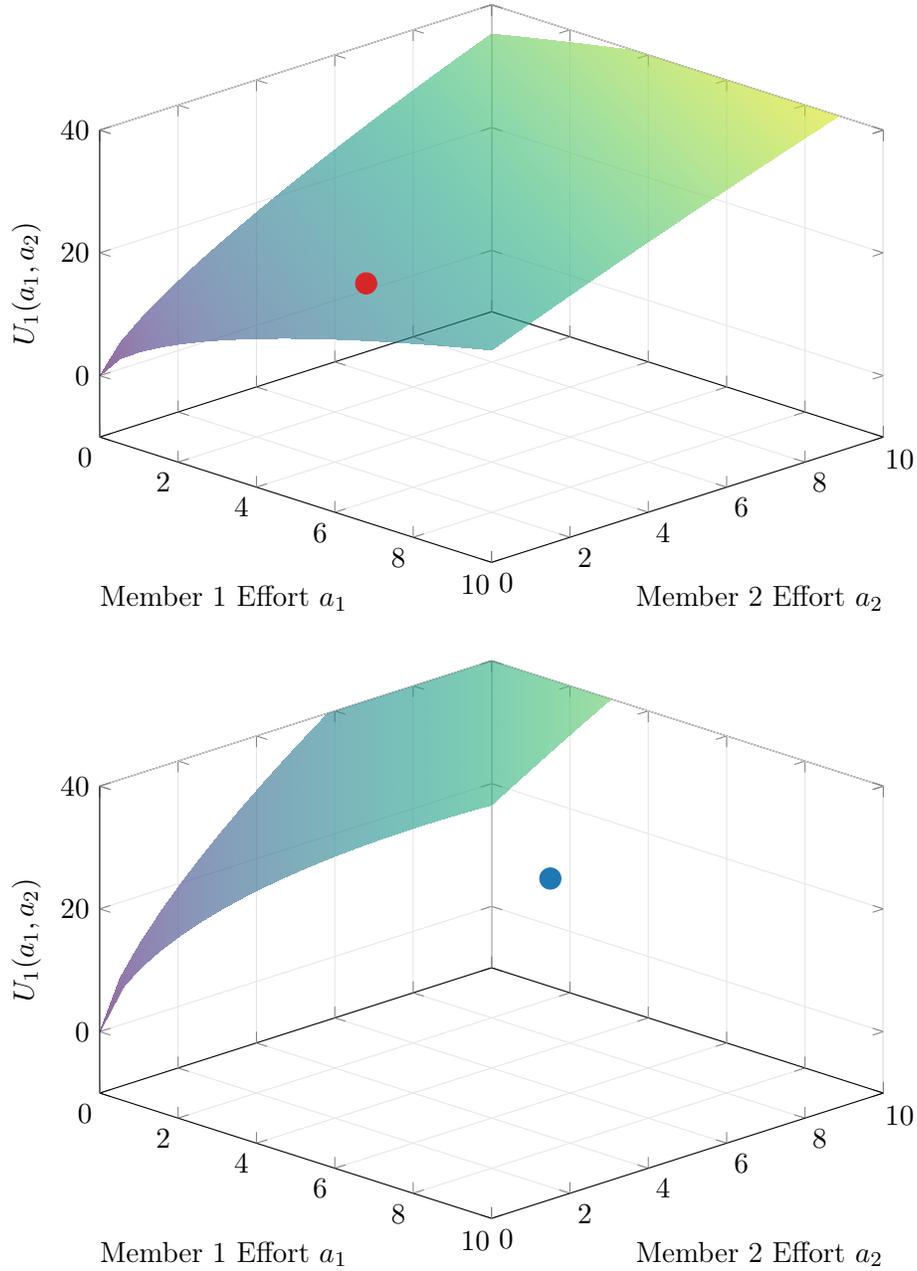
\begin{figure}[htbp]
\centering
\begin{tikzpicture}
\begin{axis}[
    width=12cm,
    height=9cm,
    view={45}{30},
    xlabel={Member 1 Effort $a_1$},
    ylabel={Member 2 Effort $a_2$},
    zlabel={$U_1(a_1, a_2)$},
    xmin=0, xmax=10,
    ymin=0, ymax=10,
    zmin=-10, zmax=40,
    grid=major,
    grid style={gray!20},
    colormap/viridis,
]

\addplot3[
    surf,
    opacity=0.6,
    samples=20,
    domain=0:10,
    y domain=0:10,
    shader=interp,
] {15*(x+y)^0.7/2 - x - 0.1*0.8*(15*(x+y)^0.7/2 - y)};

\addplot3[only marks, mark=*, mark size=4pt, color=freeriderred] coordinates {(1.8, 5, 8.5)};

\end{axis}
\end{tikzpicture}

\vspace{0.5cm}

\begin{tikzpicture}
\begin{axis}[
    width=12cm,
    height=9cm,
    view={45}{30},
    xlabel={Member 1 Effort $a_1$},
    ylabel={Member 2 Effort $a_2$},
    zlabel={$U_1(a_1, a_2)$},
    xmin=0, xmax=10,
    ymin=0, ymax=10,
    zmin=-10, zmax=40,
    grid=major,
    grid style={gray!20},
    colormap/viridis,
]

\addplot3[
    surf,
    opacity=0.6,
    samples=20,
    domain=0:10,
    y domain=0:10,
    shader=interp,
] {15*(x+y)^0.7/2 - x*(1-0.3*0.9) + 0.9*0.8*(15*(x+y)^0.7/2 - y)};

\addplot3[only marks, mark=*, mark size=4pt, color=loyaltyblue] coordinates {(6.5, 5, 28)};

\end{axis}
\end{tikzpicture}
\caption{Utility landscape for Member~1 under low loyalty ($\theta=0.1$, top) and high loyalty ($\theta=0.9$, bottom). Parameters: $\omega=15$, $\beta=0.7$, $c=1$, $n=2$, $\phi_B=0.8$, $\phi_C=0.3$. Under low loyalty, the optimal response to partner effort $a_2=5$ is low effort $a_1 \approx 1.8$ (red marker). Under high loyalty, the optimal response shifts dramatically to $a_1 \approx 6.5$ (blue marker). The entire surface ``tilts'' toward higher own-effort as loyalty increases.}
\label{fig:utility_landscape}
\end{figure}

Figure~\ref{fig:utility_landscape} presents the utility landscape for a two-member team, showing how loyalty transforms the optimization surface. The horizontal axes represent effort levels $(a_1, a_2)$, and the vertical axis shows Member~1's utility $U_1$.

The geometric transformation is striking. Under low loyalty (top panel), the utility surface peaks at low own-effort regardless of partner effort, representing the classic free-riding landscape where ``letting others do the work'' is optimal. Under high loyalty (bottom panel), the surface tilts dramatically: the peak now occurs at high own-effort. The loyal member actively seeks to contribute because doing so generates both direct utility (via cost tolerance) and vicarious utility (via internalized teammate benefits).

\subsubsection{Contour Analysis: Indifference Curves}

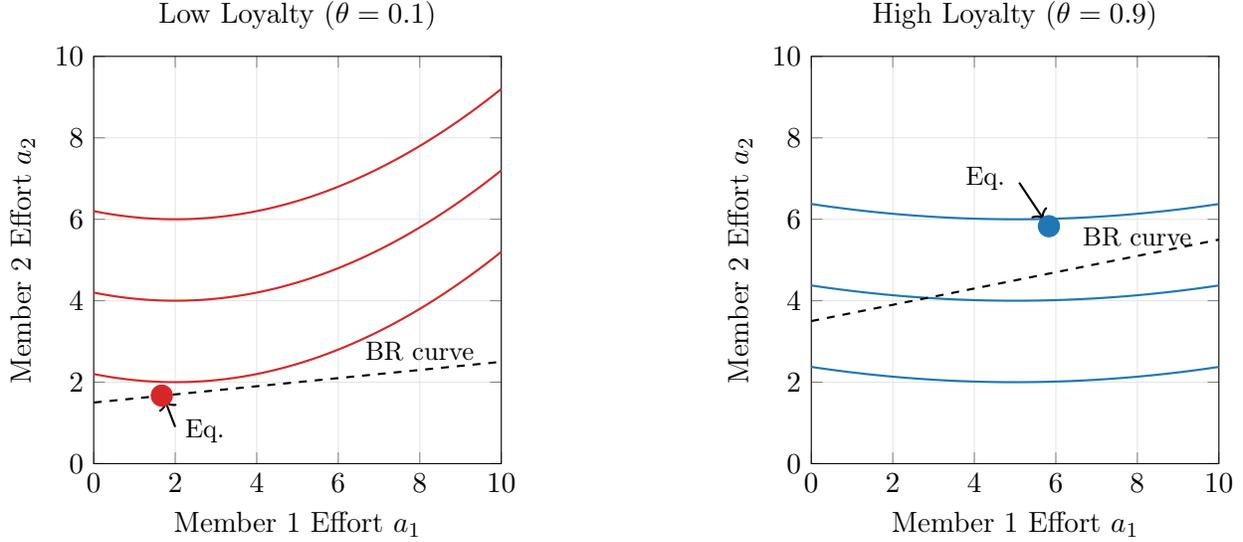
\begin{figure}[htbp]
\centering
\begin{tikzpicture}
\begin{axis}[
    width=7cm,
    height=7cm,
    xlabel={Member 1 Effort $a_1$},
    ylabel={Member 2 Effort $a_2$},
    xmin=0, xmax=10,
    ymin=0, ymax=10,
    title={Low Loyalty ($\theta=0.1$)},
    grid=major,
    grid style={gray!20},
]
\addplot[color=freeriderred, thick, domain=0:10, samples=50] {2 + 0.5*(x-2)^2/10};
\addplot[color=freeriderred, thick, domain=0:10, samples=50] {4 + 0.5*(x-2)^2/10};
\addplot[color=freeriderred, thick, domain=0:10, samples=50] {6 + 0.5*(x-2)^2/10};

\addplot[color=black, dashed, thick, domain=0:10, samples=50] {1.5 + 0.1*x};
\node[anchor=south] at (axis cs:8, 2.3) {\small BR curve};

\addplot[only marks, mark=*, mark size=4pt, color=freeriderred] coordinates {(1.67, 1.67)};

\node[anchor=west, font=\small] at (axis cs:2.0, 0.8) {Eq.};
\draw[->, thick] (axis cs:2.0, 0.9) -- (axis cs:1.75, 1.55);

\end{axis}
\end{tikzpicture}
\hfill
\begin{tikzpicture}
\begin{axis}[
    width=7cm,
    height=7cm,
    xlabel={Member 1 Effort $a_1$},
    ylabel={Member 2 Effort $a_2$},
    xmin=0, xmax=10,
    ymin=0, ymax=10,
    title={High Loyalty ($\theta=0.9$)},
    grid=major,
    grid style={gray!20},
]
\addplot[color=loyaltyblue, thick, domain=0:10, samples=50] {2 + 0.15*(x-5)^2/10};
\addplot[color=loyaltyblue, thick, domain=0:10, samples=50] {4 + 0.15*(x-5)^2/10};
\addplot[color=loyaltyblue, thick, domain=0:10, samples=50] {6 + 0.15*(x-5)^2/10};

\addplot[color=black, dashed, thick, domain=0:10, samples=50] {3.5 + 0.2*x};
\node[anchor=south] at (axis cs:8, 5.1) {\small BR curve};

\addplot[only marks, mark=*, mark size=4pt, color=loyaltyblue] coordinates {(5.83, 5.83)};

\node[anchor=east, font=\small] at (axis cs:5.0, 7.0) {Eq.};
\draw[->, thick] (axis cs:5.1, 6.9) -- (axis cs:5.7, 6.0);

\end{axis}
\end{tikzpicture}
\caption{Contour analysis showing indifference curves in effort space. Left panel: Low loyalty ($\theta=0.1$) produces steep indifference curves centered at low own-effort, indicating the member is eager to substitute partner effort for own effort. Right panel: High loyalty ($\theta=0.9$) produces flatter curves centered at higher own-effort, indicating the member values contributing and is less willing to free-ride. The equilibrium (marked) shifts from $(1.67, 1.67)$ to $(5.83, 5.83)$.}
\label{fig:contour_analysis}
\end{figure}

An alternative visualization uses contour plots showing indifference curves in effort space. Figure~\ref{fig:contour_analysis} presents this view for the same two-member team.

The contour analysis reveals the substitutability structure. Under low loyalty, indifference curves are steep and centered at low own-effort: the member views own effort and partner effort as highly substitutable and prefers the latter. Under high loyalty, curves flatten and shift rightward: the member derives intrinsic value from contributing and is less willing to substitute away from own effort.

\subsubsection{The Free-Riding Gravity Well}

A useful metaphor is the ``gravity well'' of free-riding. Under pure self-interest, the utility landscape creates a gravitational pull toward minimal effort, forming a deep well that traps teams in low-productivity equilibria. Loyalty mechanisms work by \textit{filling in this well} and creating a new attractor at high effort.

Mathematically, the depth of the free-riding well is characterized by the difference between the socially optimal utility and the Nash equilibrium utility:
\begin{equation}
\label{eq:welfare_loss}
\Delta W = \sum_{i=1}^{n} \left[ U_i(\vect{a}^{\text{social}}) - U_i(\vect{a}^{\text{Nash}}) \right]
\end{equation}

Under pure self-interest with our functional forms, this welfare loss can be substantial. For Team~$\mathcal{T}$ with baseline parameters, the welfare loss is approximately 68\% of potential team value; the gravity well is deep indeed.

Loyalty reduces the depth of this well. At $\theta = 0.5$, welfare loss drops to approximately 35\%. At $\theta = 0.9$, welfare loss is under 8\%. High loyalty effectively ``flattens'' the free-riding gravity well, allowing teams to escape to high-productivity equilibria.

\subsubsection{Comparative Statics in Geometric Terms}

We can summarize the geometric effects of each parameter:

\begin{itemize}
    \item \textbf{Team size $n$}: Increasing $n$ deepens the free-riding well by diluting individual benefit shares. The utility surface ``sags'' toward zero effort as teams grow.

    \item \textbf{Productivity $\omega$}: Higher $\omega$ raises the entire utility surface but does not change its shape. Both free-riding and cooperative equilibria become more valuable, but the relative advantage of cooperation is unchanged.

    \item \textbf{Diminishing returns $\beta$}: Lower $\beta$ (stronger diminishing returns) flattens the utility surface, reducing the stakes of effort choices. Higher $\beta$ (closer to constant returns) steepens the surface, making loyalty effects more pronounced.

    \item \textbf{Loyalty $\theta$}: Higher $\theta$ tilts the surface toward high effort, filling in the free-riding well and creating a new attractor at cooperative effort levels.

    \item \textbf{Benefit strength $\phi_B$}: Higher $\phi_B$ amplifies the tilt from loyalty. The surface rotates more sharply toward high effort as members internalize more of teammates' gains.

    \item \textbf{Cost tolerance $\phi_C$}: Higher $\phi_C$ ``lifts'' the high-effort region of the surface by reducing the cost penalty. This makes high effort more attractive independent of what teammates do.
\end{itemize}

This geometric framework provides intuition for mechanism design. To escape free-riding, teams need interventions that either fill the gravity well (reducing the attractiveness of minimal effort) or create alternative attractors (making cooperative effort more appealing). Loyalty mechanisms accomplish both: cost tolerance reduces the depth of the well, while benefit internalization creates a new peak at high effort.

\subsection{Integration with Interdependence Framework}

We connect team-level dynamics to the interdependence framework from~\cite{pant2025foundations}. When a team operates as a composite actor in a coopetitive environment, team cohesion affects the team's strategic position.

\begin{definition}[Team Cohesion]
\textbf{Team cohesion} $\mathcal{C}$ aggregates member loyalties weighted by their structural importance:
\begin{equation}
\label{eq:team_cohesion}
\mathcal{C} = \frac{\sum_{i \in \mathcal{T}} D_{\mathcal{T},i} \cdot \theta_i}{\sum_{i \in \mathcal{T}} D_{\mathcal{T},i}}
\end{equation}
where $D_{\mathcal{T},i}$ is the team's dependency on member $i$ computed from \textit{i*} analysis.
\end{definition}

Team cohesion captures how effectively the team functions as a unified actor. High cohesion (approaching 1) indicates members are loyal and the team acts coherently. Low cohesion indicates free-riding and fragmented behavior.

\begin{proposition}[Cohesion Affects Team Bargaining Power]
When a team $\mathcal{T}$ engages in coopetitive relationships with external actors, the team's effective bargaining power $\beta_{\mathcal{T}}^{\text{eff}}$ is:
\begin{equation}
\label{eq:team_bargaining}
\beta_{\mathcal{T}}^{\text{eff}} = \beta_{\mathcal{T}}^{\text{base}} \cdot \mathcal{C}
\end{equation}
where $\beta_{\mathcal{T}}^{\text{base}}$ is the team's baseline bargaining power from structural position.
\end{proposition}

This connects internal team dynamics to external coopetitive positioning. A team with low cohesion (due to free-riding members) has reduced effective bargaining power because it cannot credibly commit to collective action. External partners recognize that fragmented teams may not deliver on commitments, reducing their willingness to concede value.

\subsection{Algorithm for Computing Team Production Equilibrium}

We provide Algorithm~\ref{alg:tpe_solver} for computing TPE numerically.

\begin{algorithm}[htbp]
\caption{Team Production Equilibrium Solver}
\label{alg:tpe_solver}
\begin{algorithmic}[1]
\Require Team parameters $(\omega, \beta, c, n)$, loyalty profile $(\theta_1, \ldots, \theta_n)$, mechanism strengths $(\phi_B, \phi_C)$, convergence tolerance $\epsilon$
\Ensure Equilibrium action profile $\vect{a}^*$
\State Initialize $\vect{a}^{(0)} \leftarrow (\bar{a}/2, \ldots, \bar{a}/2)$ \Comment{Start at midpoint}
\State $k \leftarrow 0$
\Repeat
    \For{$i = 1$ to $n$}
        \State $a_i^{(k+1)} \leftarrow \argmax_{a_i \in [0,\bar{a}]} U_i(a_i, \vect{a}_{-i}^{(k)}; \theta_i)$ \Comment{Best response}
    \EndFor
    \State $k \leftarrow k + 1$
\Until{$\|\vect{a}^{(k)} - \vect{a}^{(k-1)}\| < \epsilon$}
\State \Return $\vect{a}^{(k)}$
\end{algorithmic}
\end{algorithm}

The best response computation in line 5 admits closed-form solution for our functional forms. Let $A_{-i} = \sum_{j \neq i} a_j$. The first-order condition is:
\begin{equation}
\frac{\partial U_i}{\partial a_i} = \frac{\omega\beta}{n}(a_i + A_{-i})^{\beta-1}(1 + \phi_B\theta_i(n-1)) - c(1-\phi_C\theta_i) = 0
\end{equation}

Solving yields:
\begin{equation}
a_i^{BR} = \left(\frac{\omega\beta(1 + \phi_B\theta_i(n-1))}{nc(1-\phi_C\theta_i)}\right)^k - A_{-i}
\end{equation}
clamped to $[0, \bar{a}]$.

\section{Translation Framework}
\label{sec:translation}

This section develops structured methodology for instantiating team production models from organizational contexts and requirements engineering artifacts. We extend the eight-step translation framework from~\cite{pant2025foundations} with loyalty-specific elicitation procedures, adding team production parameters and loyalty assessment to create a ten-step methodology appropriate for intra-team coordination analysis.

\subsection{Extended Translation Methodology}

The translation process follows these steps, building on the methodology established in~\cite{pant2025foundations}:

\subsubsection{Phase 1: Team Structure Identification (Steps 1--4)}

\textbf{Step 1: Team Boundaries}

Identify team boundaries, members, and roles. Document formal organizational structure and informal collaboration patterns.

\textit{For Team~$\mathcal{T}$}: The team comprises $n=6$ members with roles: senior developers (Member~$M_1$ (the architect), Member~$M_2$ (the lead implementer)), junior developers (Member~$M_3$ (a contributor), Member~$M_4$ (a contributor)), QA engineer (Member~$M_5$ (quality assurance)), product owner (Member~$M_6$ (the coordinator)).

\textbf{Step 2: Production Parameter Assessment}

Estimate $\omega$, $\beta$, $c$, and $\bar{a}$ from team characteristics:

\begin{center}
\begin{tabular}{lcp{6cm}}
\toprule
Parameter & Range & Assessment Guidance \\
\midrule
$\omega$ & $[10, 50]$ & Higher for experienced teams, mature tools, clear requirements \\
$\beta$ & $[0.5, 0.9]$ & Lower for tightly coupled work, higher for modular tasks \\
$c$ & $[0.5, 2.0]$ & Higher for cognitively demanding work, opportunity costs \\
$\bar{a}$ & $[5, 15]$ & Based on sustainable work hours, sprint capacity \\
\bottomrule
\end{tabular}
\end{center}

\textbf{Step 3: Dependency Analysis}

Construct \textit{i*} Strategic Dependency model to identify dependencies among team members. Compute $D_{\mathcal{T},i}$ for each member based on how much the team depends on their contribution.

\textbf{Step 4: Production Parameters}

Assess production function parameters based on team and task characteristics. Consider team experience level, tool maturity, requirements clarity (for $\omega$), task modularity vs. coupling (for $\beta$), cognitive demands and opportunity costs (for $c$), and sustainable capacity (for $\bar{a}$).

\subsubsection{Phase 2: Loyalty Assessment (Steps 5--7)}

\textbf{Step 5: Tenure Assessment}

Measure team tenure for each member, normalized to $[0,1]$: $\text{Tenure}_i = \min(1, \text{months}_i / 24)$. Longer tenure typically correlates with higher loyalty through accumulated relationships and identity formation.

\textbf{Step 6: Social Integration Assessment}

Evaluate social integration through observable indicators: communication frequency, informal interaction patterns, collaboration network centrality, and expressed interpersonal bonds. Use surveys or interaction log analysis where available.

\textbf{Step 7: Loyalty Computation}

Assess each member's loyalty $\theta_i$ using observable indicators:

\begin{center}
\begin{tabular}{lcp{6cm}}
\toprule
Factor & Weight & Measurement \\
\midrule
Tenure & 0.30 & $\min(1, \text{months}/24)$ normalized \\
Social Integration & 0.35 & Survey or observed collaboration intensity \\
Role Criticality & 0.20 & From dependency analysis $D_{\mathcal{T},i}$ \\
Expressed Commitment & 0.15 & Stated intentions, retention signals \\
\bottomrule
\end{tabular}
\end{center}

Compute $\theta_i$ as weighted sum: $\theta_i = 0.30 \cdot \text{Tenure}_i + 0.35 \cdot \text{Social}_i + 0.20 \cdot \text{Criticality}_i + 0.15 \cdot \text{Commitment}_i$.

\subsubsection{Phase 3: Parameter Calibration (Steps 8--10)}

\textbf{Step 8: Mechanism Strength Calibration}

Use default mechanism strengths unless domain-specific calibration is available:

\begin{center}
\begin{tabular}{lcl}
\toprule
Parameter & Default & Interpretation \\
\midrule
$\phi_B$ & 0.80 & Loyalty benefit (welfare + warm glow) \\
$\phi_C$ & 0.30 & Cost tolerance \\
\bottomrule
\end{tabular}
\end{center}

These defaults were calibrated against the Apache case study and produce reasonable behavior across diverse team configurations. Domain-specific adjustment may improve accuracy for particular contexts.

\textbf{Step 9: Initial Equilibrium Computation}

Compute equilibrium using Algorithm~\ref{alg:tpe_solver} with the parameterized model. Verify convergence and check that results satisfy bounded outcomes constraint ($a_i^* \in [0, \bar{a}]$).

\textbf{Step 10: Validation and Refinement}

Compare predicted behavior to observed patterns. Key validation checks include:
\begin{itemize}
    \item Does predicted effort ranking match observed contribution ranking?
    \item Does predicted total output approximate observed productivity?
    \item Do equilibrium dynamics (e.g., response to team changes) match qualitative observations?
\end{itemize}
Refine parameters if predictions diverge substantially from reality, prioritizing adjustments to $\omega$ and $\theta_i$ which have most direct empirical grounding.

\subsection{Worked Example: Team~$\mathcal{T}$ Parameterization}

\textbf{Step 1}: $n = 6$ members identified.

\textbf{Step 2}: Production parameters assessed based on team characteristics:
\begin{itemize}
    \item $\omega = 30$ (experienced team, good tools)
    \item $\beta = 0.7$ (moderate coordination requirements)
    \item $c = 1.0$ (standard effort cost)
    \item $\bar{a} = 10$ (normalized maximum effort)
\end{itemize}

\textbf{Step 3}: Dependency weights computed from \textit{i*} analysis:

\begin{center}
\begin{tabular}{lccc}
\toprule
Member & Role & $D_{\mathcal{T},i}$ & Basis \\
\midrule
$M_1$ & Architect & 0.22 & Architecture decisions critical \\
$M_2$ & Lead Impl. & 0.20 & Core implementation \\
$M_3$ & Contributor & 0.12 & UI components \\
$M_4$ & Contributor & 0.12 & UI components \\
$M_5$ & QA & 0.18 & Quality gate \\
$M_6$ & Coordinator & 0.16 & Stakeholder interface \\
\bottomrule
\end{tabular}
\end{center}

\textbf{Step 4}: Loyalty assessment:

\begin{center}
\begin{tabular}{lcccccc}
\toprule
Member & Tenure & Social & Critical & Commit & $\theta_i$ \\
\midrule
$M_1$ & 1.00 & 0.90 & 0.22 & 0.95 & 0.93 \\
$M_2$ & 0.75 & 0.85 & 0.20 & 0.80 & 0.74 \\
$M_3$ & 0.25 & 0.60 & 0.12 & 0.70 & 0.43 \\
$M_4$ & 0.17 & 0.55 & 0.12 & 0.65 & 0.38 \\
$M_5$ & 0.50 & 0.70 & 0.18 & 0.75 & 0.56 \\
$M_6$ & 0.33 & 0.65 & 0.16 & 0.80 & 0.50 \\
\bottomrule
\end{tabular}
\end{center}

\textbf{Step 5}: Use defaults: $\phi_B = 0.8$, $\phi_C = 0.3$.

The resulting parameterized model can be used to compute Team Production Equilibrium, predict effort profiles, analyze free-riding risks, and evaluate interventions.

\subsection{Operationalizing Loyalty in LLM-Based Multi-Agent Systems}

The preceding translation methodology was developed for human teams, but the same formalism applies directly to multi-agent systems comprising Large Language Model (LLM) agents. This subsection provides concrete guidance for implementing loyalty mechanisms in agentic AI architectures, connecting our mathematical framework to practical system design.

\subsubsection{The Agent Alignment Problem as Collective Action}

Multi-agent LLM systems face a collective action problem structurally identical to human team production. Consider System~$\mathcal{S}$: five specialized agents collaborating on EHR integration. Each agent controls computational resources (tokens, API calls, processing time) that function as ``effort'' in our framework. The collective output depends on combined contributions, but individual agents may be optimized for narrow objectives that create free-riding incentives.

\textbf{Structural parallels:}
\begin{itemize}
    \item \textbf{Effort} $a_i$: Computational resources allocated to shared objectives (tokens for collaborative reasoning, API calls for data integration, cycles for cross-agent verification)
    \item \textbf{Team output} $Q(\vect{a})$: System-level performance metrics (integration quality, user satisfaction, task completion rate)
    \item \textbf{Individual payoff} $\pi_i$: Agent-specific reward signals from the training objective or runtime scoring function
    \item \textbf{Free-riding}: Agents that minimize resource expenditure while relying on others' outputs
\end{itemize}

The loyalty coefficient $\theta_i$ translates to \textit{alignment strength}, specifically the degree to which an agent's objective function incorporates system-level goals beyond its narrow specialization.

\subsubsection{Implementing \texorpdfstring{$\phi_B$}{phi\_B}: Multi-Objective Reward Shaping}

The loyalty benefit parameter $\phi_B$ corresponds to reward shaping that makes agents value teammates' success. Implementation strategies include:

\textbf{Strategy 1: Explicit Team Reward Component}

Modify the agent's reward function to include a weighted sum of teammate rewards:
\begin{equation}
\label{eq:llm_reward}
R_i^{\text{shaped}} = R_i^{\text{individual}} + \phi_B \theta_i \sum_{j \neq i} R_j^{\text{individual}}
\end{equation}

In practice, this is implemented at the scoring/evaluation layer:

\begin{verbatim}
def compute_shaped_reward(agent_id, individual_rewards, theta, phi_B):
    """
    Compute loyalty-shaped reward for agent i.

    Args:
        agent_id: Index of current agent
        individual_rewards: Dict mapping agent_id -> individual reward
        theta: Agent's alignment coefficient (loyalty)
        phi_B: Benefit internalization strength

    Returns:
        Shaped reward incorporating teammate welfare
    """
    own_reward = individual_rewards[agent_id]
    teammate_rewards = sum(
        r for aid, r in individual_rewards.items()
        if aid != agent_id
    )
    return own_reward + phi_B * theta * teammate_rewards
\end{verbatim}

\textbf{Strategy 2: System Prompt Engineering}

For LLM agents operating via prompting rather than fine-tuning, loyalty can be induced through carefully crafted system prompts. The prompt structure maps to our mathematical parameters:

\begin{verbatim}
SYSTEM PROMPT (High Loyalty, theta \approx 0.9):
---
You are Agent_Architecture, part of a collaborative team working
on EHR integration. Your success is measured not only by your
individual task completion, but also by the success of the
entire team.

TEAM WELFARE DIRECTIVE:
When making decisions, weight team outcomes heavily. If helping
another agent (even at cost to your immediate task) improves
overall system performance, prioritize that action. Your reward
is computed as:
  R = 0.3 * (your_task_score) + 0.7 * (team_success_score)

COLLABORATION PROTOCOL:
- Share intermediate results proactively
- Verify your outputs against team requirements
- Flag blockers that affect other agents immediately
---
\end{verbatim}

Compare to a low-loyalty configuration:

\begin{verbatim}
SYSTEM PROMPT (Low Loyalty, theta \approx 0.2):
---
You are Agent_Architecture. Complete your assigned task
efficiently. Your performance is evaluated on your specific
deliverables.

Coordinate with other agents only when necessary for your
task completion.
---
\end{verbatim}

The difference in prompt structure directly implements different $\theta$ values: high-loyalty prompts explicitly weight team outcomes, while low-loyalty prompts focus narrowly on individual tasks.

\subsubsection{Implementing \texorpdfstring{$\phi_C$}{phi\_C}: Cost Tolerance Through Resource Allocation}

The cost tolerance parameter $\phi_C$ reduces the effective cost of effort for loyal agents. In LLM systems, this translates to resource allocation policies:

\textbf{Strategy 1: Differentiated Token Budgets}

Allocate more generous token budgets to agents performing collaborative work:

\begin{verbatim}
def allocate_token_budget(agent_id, theta, phi_C, base_budget):
    """
    Allocate tokens with loyalty-based enhancement.

    Loyal agents get larger budgets, reducing effective cost
    of collaborative effort.
    """
    loyalty_bonus = phi_C * theta[agent_id]
    return base_budget * (1 + loyalty_bonus)
\end{verbatim}

\textbf{Strategy 2: Reduced Penalties for Collaborative Overhead}

When agents incur ``costs'' (latency, token usage, API calls) for team-benefiting activities, apply a loyalty-scaled discount:

\begin{verbatim}
def compute_cost_penalty(agent_id, resource_usage, theta, phi_C):
    """
    Compute cost penalty with loyalty discount.

    High-loyalty agents face reduced penalties for collaborative
    resource expenditure, implementing the (1 - phi_C * theta)
    cost reduction from our theoretical model.
    """
    base_penalty = resource_usage * COST_PER_UNIT
    loyalty_discount = 1 - phi_C * theta[agent_id]
    return base_penalty * loyalty_discount
\end{verbatim}

\subsubsection{Calibrating \texorpdfstring{$\theta$}{theta} for Computational Agents}

For human team members, loyalty is assessed through tenure, social integration, and expressed commitment. For LLM agents, analogous calibration uses:

\begin{center}
\begin{tabular}{lcp{6cm}}
\toprule
Factor & Weight & Measurement for Agents \\
\midrule
Training Alignment & 0.35 & RLHF alignment degree, instruction-following metrics \\
Architecture Integration & 0.30 & Shared memory access, communication channel richness \\
Objective Overlap & 0.20 & Correlation between agent objective and system objective \\
Interaction History & 0.15 & Cooperative vs. defecting behavior in past interactions \\
\bottomrule
\end{tabular}
\end{center}

\textbf{Worked Example: System~$\mathcal{S}$ Agent Calibration}

\begin{center}
\begin{tabular}{lccccc}
\toprule
Agent & Training & Architecture & Objective & History & $\theta_i$ \\
\midrule
$A_1$ (Arch.) & 0.90 & 0.85 & 0.80 & 0.90 & 0.86 \\
$A_2$ (Impl.) & 0.85 & 0.80 & 0.75 & 0.85 & 0.81 \\
$A_3$ (Data) & 0.80 & 0.70 & 0.85 & 0.75 & 0.77 \\
$A_4$ (API) & 0.75 & 0.65 & 0.70 & 0.70 & 0.70 \\
$A_5$ (Test) & 0.85 & 0.75 & 0.90 & 0.85 & 0.83 \\
\bottomrule
\end{tabular}
\end{center}

Agent~$A_1$ (Architecture) has the highest loyalty ($\theta = 0.86$) due to strong RLHF alignment and central architectural position. Agent~$A_4$ (API) has lowest loyalty ($\theta = 0.70$) due to its more peripheral integration and narrower objective alignment.

\subsubsection{Complete Agent Configuration Example}

Putting these elements together, a complete loyalty-aware agent configuration for Agent~$A_1$ (Architecture) in System~$\mathcal{S}$:

\begin{verbatim}
agent_config = {
    "agent_id": "A1_Architecture",
    "role": "System architecture design and integration coordination",

    # Loyalty parameters (from calibration)
    "theta": 0.86,              # Alignment coefficient
    "phi_B": 0.80,              # Benefit internalization
    "phi_C": 0.30,              # Cost tolerance

    # Derived effective parameters
    "team_reward_weight": 0.86 * 0.80,  # = 0.688
    "cost_discount": 1 - 0.30 * 0.86,   # = 0.742

    # System prompt implementing loyalty
    "system_prompt": """
    You are the Architecture Agent in a collaborative EHR integration
    system. Your primary responsibility is system design, but your
    success depends on overall team performance.

    REWARD STRUCTURE:
    - Your individual task score contributes 31.2% to your reward
    - Team success score contributes 68.8% to your reward

    RESOURCE POLICY:
    - You have enhanced token budget (1.26x base) for collaborative work
    - Cross-agent verification and assistance are encouraged

    BEHAVIORAL DIRECTIVES:
    1. Proactively share architectural decisions with dependent agents
    2. Respond to blockers from other agents within 1 turn
    3. Verify integration points before marking tasks complete
    """,

    # Resource allocation
    "token_budget": 4096 * 1.26,  # Enhanced for collaboration
    "api_call_limit": 50,
    "collaboration_overhead_discount": 0.742
}
\end{verbatim}

\subsubsection{Predicting Agent Behavior Under Loyalty Mechanisms}

With the parameterized model, we can predict equilibrium effort for System~$\mathcal{S}$. Using the TPE solver with $\omega = 25$, $\beta = 0.75$, $c = 1.0$, $n = 5$, and the calibrated loyalty values:

\textbf{Predicted equilibrium resource allocation:}
\begin{center}
\begin{tabular}{lcccc}
\toprule
Agent & $\theta_i$ & $a_i^*$ (No Loyalty) & $a_i^*$ (With Loyalty) & Increase \\
\midrule
$A_1$ (Arch.) & 0.86 & 1.82 & 7.94 & 336\% \\
$A_2$ (Impl.) & 0.81 & 1.82 & 6.85 & 276\% \\
$A_3$ (Data) & 0.77 & 1.82 & 6.12 & 236\% \\
$A_4$ (API) & 0.70 & 1.82 & 5.03 & 176\% \\
$A_5$ (Test) & 0.83 & 1.82 & 7.21 & 296\% \\
\bottomrule
\end{tabular}
\end{center}

Without loyalty mechanisms (all $\theta = 0$), agents converge to minimal resource allocation ($a^* = 1.82$). With loyalty-aware configuration, effort increases substantially, with highly-aligned agents ($A_1$, $A_5$) showing the largest gains.

\subsubsection{Design Implications for Multi-Agent Systems}

The formalization suggests concrete design principles for loyalty-aware multi-agent systems:

\textbf{Principle 1: Reward Shaping Matters}. The benefit internalization parameter $\phi_B$ has outsized impact on equilibrium effort. Multi-objective reward functions that weight team success heavily can overcome free-riding incentives even with moderate alignment.

\textbf{Principle 2: Alignment is Multiplicative}. The alignment coefficient $\theta$ multiplies both benefit internalization and cost tolerance effects. Investing in agent alignment (through RLHF, constitutional AI, or careful prompt engineering) compounds the effectiveness of loyalty mechanisms.

\textbf{Principle 3: Heterogeneous Loyalty is Acceptable}. Our model accommodates agents with different loyalty levels. Not all agents need high alignment; as long as critical path agents (high $D_{\mathcal{T},i}$) have sufficient loyalty, the team can function effectively.

\textbf{Principle 4: Observability Enables Cooperation}. The loyalty mechanisms require agents to observe (or infer) teammate outcomes. System architectures should facilitate this observability through shared state, message passing, or explicit reward broadcasting.

\section{\textit{i*} Modeling of Team Production Dynamics}
\label{sec:istar}

This section presents \textit{i*} Strategic Rationale and Strategic Dependency diagrams that capture team production dynamics. These diagrams provide the structural foundation from which computational parameters are derived.

\subsection{Strategic Rationale Model: Team Member Perspective}

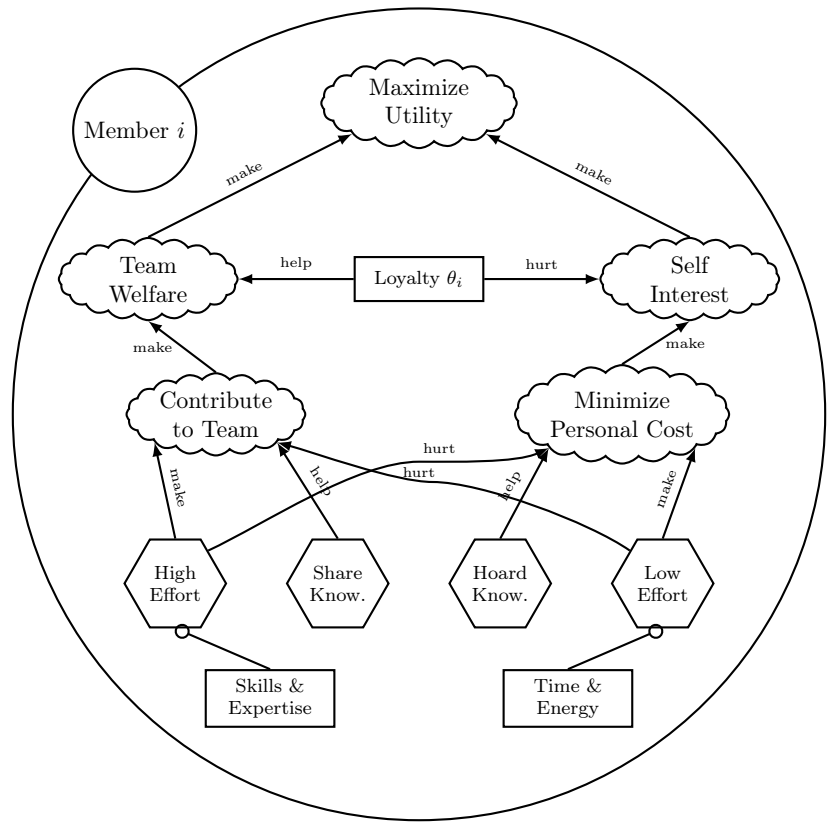
\begin{figure}[htbp]
\centering
\begin{tikzpicture}[
    scale=0.9, transform shape,
    actor/.style={circle, draw, thick, minimum size=1.2cm, font=\small, align=center},
    goal/.style={ellipse, draw, thick, minimum height=0.7cm, minimum width=2.4cm, font=\scriptsize, align=center},
    softgoal/.style={cloud, cloud puffs=20, cloud puff arc=100, aspect=2.5, draw, thick,
                     minimum width=2.5cm, minimum height=1.0cm, align=center, font=\small, inner sep=1pt, fill=white},
    task/.style={regular polygon, regular polygon sides=6, draw, thick,
                 minimum size=1.5cm, font=\scriptsize, align=center, inner sep=1pt},
    resource/.style={rectangle, draw, thick, minimum height=0.65cm, minimum width=1.9cm, font=\scriptsize, align=center},
    refine/.style={-latex, thick},
    contribution/.style={-latex, thick},
    neededby/.style={-{Circle[open, length=2mm, width=2mm]}, thick},
]

\draw[thick] (0,0) circle (6.0cm);
\node[actor, fill=white] at (-4.2,4.2) {Member $i$};

\node[softgoal] (maximize) at (0, 4.6) {Maximize\\Utility};

\node[softgoal, fill=white] (teamwelfare) at (-4.0,2.0) {Team\\Welfare};
\node[softgoal, fill=white] (selfinterest) at (4.0,2.0) {Self\\Interest};
\node[softgoal, fill=white] (contribute) at (-3.0,0.0) {Contribute\\to Team};
\node[softgoal, fill=white] (minimize) at (3.0,0.0) {Minimize\\Personal Cost};

\node[resource, fill=white] (loyalty) at (0,2.0) {Loyalty $\theta_i$};

\node[task, fill=white] (higheffort) at (-3.6,-2.5) {High\\Effort};
\node[task, fill=white] (shareknow) at (-1.2,-2.5) {Share\\Know.};
\node[task, fill=white] (hoard) at (1.2,-2.5) {Hoard\\Know.};
\node[task, fill=white] (loweffort) at (3.6,-2.5) {Low\\Effort};

\node[resource, fill=white] (skills) at (-2.2,-4.2) {Skills \&\\Expertise};
\node[resource, fill=white] (time) at (2.2,-4.2) {Time \&\\Energy};


\draw[contribution] (teamwelfare.north) -- node[midway, sloped, above, font=\tiny] {make} (maximize.south west);
\draw[contribution] (selfinterest.north) -- node[midway, sloped, above, font=\tiny] {make} (maximize.south east);

\draw[contribution] (loyalty.west) -- node[above, font=\tiny] {help} (teamwelfare.east);
\draw[contribution] (loyalty.east) -- node[above, font=\tiny] {hurt} (selfinterest.west);

\draw[contribution] (contribute.north) -- node[left, font=\tiny] {make} (teamwelfare.south);
\draw[contribution] (minimize.north) -- node[right, font=\tiny] {make} (selfinterest.south);

\draw[contribution] (higheffort.north) -- node[midway, sloped, above, font=\tiny] {make} (contribute.south west);
\draw[contribution] (shareknow.north) -- node[midway, sloped, above, font=\tiny] {help} (contribute.south east);
\draw[contribution] (hoard.north) -- node[midway, sloped, above, font=\tiny] {help} (minimize.south west);
\draw[contribution] (loweffort.north) -- node[midway, sloped, above, font=\tiny] {make} (minimize.south east);

\draw[contribution] (higheffort.north east) to[out=25, in=180, looseness=0.7]
    (0,-0.7) to[out=0, in=205, looseness=0.7]
    (minimize.south west);
\node[font=\tiny] at (0.3,-0.5) {hurt};

\draw[contribution] (loweffort.north west) to[out=145, in=0, looseness=0.6]
    (0.2,-1.0) to[out=180, in=-15, looseness=0.6]
    (contribute.south east);
\node[font=\tiny] at (0.0,-0.85) {hurt};

\draw[neededby] (skills.north) -- (higheffort.south);
\draw[neededby] (time.north) -- (loweffort.south);

\end{tikzpicture}
\caption{The fundamental incentive conflict in team production emerges from competing softgoals: maximizing utility requires balancing team welfare against self-interest. Loyalty $\theta_i$ determines this balance---at $\theta_i \to 0$, the rational path follows self-interest through low effort and knowledge hoarding; at $\theta_i \to 1$, team welfare dominates despite the negative contribution from high effort to cost minimization. The cross-cutting contribution links (high effort hurts cost minimization; low effort hurts team contribution) formalize why team production creates strategic tension absent in individual production. This goal structure grounds the utility function transformation in Equation~\eqref{eq:loyalty_utility}.}
\label{fig:sr_team_member}
\end{figure}

Figure~\ref{fig:sr_team_member} presents the Strategic Rationale diagram for a team member, showing how loyalty mechanisms affect the internal goal structure.

The diagram captures the fundamental tension in team production. The top-level goal ``Maximize Utility'' decomposes into two competing softgoals: ``Team Welfare'' and ``Self Interest.'' Under pure self-interest ($\theta_i = 0$), the member weights only self-interest, leading to tasks ``Exert Low Effort'' and ``Hoard Knowledge'' that minimize personal costs. Under high loyalty ($\theta_i \to 1$), the member weights team welfare more heavily, motivating ``Exert High Effort'' and ``Share Knowledge'' despite higher personal costs. The resource ``Loyalty $\theta_i$'' represents the psychological state that moderates between these competing orientations.

\subsection{Strategic Dependency Model: Team Collaboration Structure}

\begin{figure}[htbp]
\centering
\begin{tikzpicture}[
    scale=0.80, transform shape,
    istarnode/.style={thick, draw, fill=white, align=center},
    agent/.style={circle, istarnode, minimum size=1.6cm, font=\small,
        append after command={\pgfextra{\draw[thick] (\tikzlastnode.150) -- (\tikzlastnode.30);}}},
    role/.style={circle, istarnode, minimum size=1.6cm, font=\small,
        append after command={\pgfextra{\draw[thick] (\tikzlastnode.210) to[bend right=20] (\tikzlastnode.330);}}},
    goal/.style={ellipse, istarnode, minimum height=0.55cm, minimum width=1.6cm, font=\scriptsize},
    task/.style={regular polygon, regular polygon sides=6, istarnode, minimum size=1.0cm, font=\scriptsize, inner sep=1pt},
    quality/.style={cloud, cloud puffs=10, cloud puff arc=120, istarnode,
                    minimum width=1.8cm, minimum height=0.8cm, font=\scriptsize, inner sep=1pt},
    resource/.style={rectangle, istarnode, minimum height=0.55cm, minimum width=1.5cm, font=\scriptsize},
    dep/.style={thick},
    partof/.style={-latex, thick},
    dmark/.style={semicircle, draw, thick, fill=white, minimum size=2mm, inner sep=0pt}
]

\node[agent, minimum size=2.0cm] (team) at (-2,9.5) {Development\\Team $\mathcal{T}$};

\node[role] (customer) at (5.5,12.0) {Customer};
\node[role, minimum size=1.4cm] (mgmt) at (-10.5, 1.0) {Mgmt};

\node[role] (alice) at (-4.0,4.5) {$M_1$\\(Senior)};
\node[role] (bob) at (4.0,4.5) {$M_2$\\(Senior)};
\node[role] (frank) at (-9.5, 7.5) {$M_6$\\(Coord)};
\node[role] (carol) at (-5.5,-2.0) {$M_3$\\(Junior)};
\node[role] (dan) at (5.5,-2.0) {$M_4$\\(Junior)};
\node[role] (eve) at (0,-2.0) {$M_5$\\(QA)};


\node[task] (review_m2) at (0, 6.2) {Review\\M1 Code};
\node[task] (review_m1) at (0, 3.8) {Review\\M2 Code};

\node[task] (mentor1) at (-6.5,1.25) {Mentor-\\ing};
\node[task] (mentor2) at (5.0,1.25) {Mentor-\\ing};

\node[resource] (qa_m1) at (-2.0, 1.5) {QA Rep};
\node[resource] (qa_m2) at (2.0, 1.5) {QA Rep};
\node[resource] (qa_m3) at (-2.50, -2.0) {QA Rep};
\node[resource] (qa_m4) at (2.50, -2.0) {QA Rep};

\node[quality] (reqs_cust) at (-3.0, 12.0) {Reqs\\Clarity};
\node[quality] (reqs_m1) at (-6.0, 6.5) {Reqs\\Clarity}; 

\node[quality] (testing) at (-11.0, 9.5) {Test\\Coverage};
\node[goal] (features) at (2.50,10) {Working\\Features};
\node[quality] (velocity) at (-8.5, 2.5) {Sprint\\Velocity};

\draw[partof] (alice.90) to[out=90, in=-135] node[midway, sloped, above, font=\tiny, fill=white, inner sep=1pt] {part-of} (team.230);
\draw[partof] (bob.90) to[out=90, in=-45] node[midway, sloped, above, font=\tiny, fill=white, inner sep=1pt] {part-of} (team.310);
\draw[partof] (frank.60) to[out=80, in=220] node[pos=0.4, sloped, above, font=\tiny, fill=white, inner sep=1pt] {part-of} (team.170);

\draw[partof] (eve.290) to[out=-90, in=180] (2.0, -4.5) 
    -- (8.0, -4.5) 
    -- (8.0, 6.0) 
    to[out=90, in=-20] node[midway, sloped, above, font=\tiny, inner sep=1pt] {part-of} (team.340);

\draw[partof] (carol.270) -- (-5.5, -3.5) 
    -- (-11.5, -3.5) 
    -- (-11.5, 5.0) 
    -- node[pos=0.9, sloped, above, font=\tiny, fill=white, inner sep=1pt] {part-of} (team.200);

\draw[partof] (dan.270) -- (5.5, -3.5) 
    -- (7, -3.5) 
    -- (7, 6.5) 
    -- node[pos=0.9, sloped, above, font=\tiny, fill=white, inner sep=1pt] {part-of} (team.320);


\draw[dep] (alice.20) to[out=20, in=180] node[dmark, sloped, rotate=-90] {} (review_m2.west);
\draw[dep] (review_m2.east) to[out=0, in=160] node[dmark, sloped, rotate=-90] {} (bob.160);
\draw[dep] (bob.200) to[out=200, in=0] node[dmark, sloped, rotate=90] {} (review_m1.east);
\draw[dep] (review_m1.west) to[out=180, in=-20] node[dmark, sloped, rotate=90] {} (alice.340);

\draw[dep] (carol.130) to[out=130, in=-90] node[dmark, sloped, rotate=90, pos=0.5] {} (mentor1.south);
\draw[dep] (mentor1.north) to[out=90, in=230] node[dmark, sloped, rotate=-90, pos=0.5] {} (alice.230);
\draw[dep] (dan.50) to[out=50, in=-90] node[dmark, sloped, rotate=-90, pos=0.5] {} (mentor2.south);
\draw[dep] (mentor2.north) to[out=90, in=-50] node[dmark, sloped, rotate=-90, pos=0.5] {} (bob.310);

\draw[dep] (alice.270) to[out=-90, in=90] node[dmark, sloped, rotate=-90] {} (qa_m1.north);
\draw[dep] (qa_m1.south) to[out=-90, in=110] node[dmark, sloped, rotate=-90] {} (eve.110);
\draw[dep] (bob.270) to[out=-90, in=90] node[dmark, sloped, rotate=-90] {} (qa_m2.north);
\draw[dep] (qa_m2.south) to[out=-90, in=70] node[dmark, sloped, rotate=-90] {} (eve.70);
\draw[dep] (carol.0) -- node[dmark, sloped, rotate=-90] {} (qa_m3.west);
\draw[dep] (qa_m3.east) to[out=0, in=180] node[dmark, sloped, rotate=-90] {} (eve.180);
\draw[dep] (dan.180) -- node[dmark, sloped, rotate=-90] {} (qa_m4.east);
\draw[dep] (qa_m4.west) to[out=180, in=0] node[dmark, sloped, rotate=-90] {} (eve.0);

\draw[dep] (alice.140) to[out=140, in=-45] node[dmark, sloped, rotate=90, pos=0.4] {} (reqs_m1.south east);
\draw[dep] (reqs_m1.west) to[out=225, in=315] node[dmark, sloped, rotate=90] {} (frank.315);

\draw[dep] (customer.west) to[out=180, in=0] node[dmark, sloped, rotate=-90] {} (reqs_cust.east);
\draw[dep] (reqs_cust.west) to[out=180, in=90] (-9.5, 11.5) -- node[dmark, sloped, rotate=-90] {} (frank.90);

\draw[dep] (team.150) to[out=150, in=0] node[dmark, sloped, rotate=90, pos=0.5] {} (testing.east);
\draw[dep] (testing.west) -- (-12.5, 9.5) 
    -- node[dmark, sloped, rotate=-90, pos=0.5] {} (-12.5, -4.5) 
    -- (-2.0, -4.5) 
    to[out=0, in=230] (eve.230);

\draw[dep] (customer.-90) -- node[dmark, sloped, rotate=90] {} (features.east);
\draw[dep] (features.west) -- node[dmark, sloped, rotate=90] {} (team.0);
\draw[dep] (mgmt.90) to[out=90, in=-135] node[dmark, sloped, rotate=-90, pos=0.5] {} (velocity.south west);
\draw[dep] (velocity.east) to[out=0, in=180] node[dmark, sloped, rotate=-90, pos=0.6] {} (alice.180);

\end{tikzpicture}
\caption{Dependency analysis reveals asymmetric structural importance: Member~$M_5$ (QA) receives dependencies from all four developers, yielding highest $D_{\mathcal{T},i}$ and making their loyalty disproportionately consequential for team cohesion per Equation~\eqref{eq:team_cohesion}. The mutual code review dependency between $M_1$ and $M_2$ creates reciprocal accountability that activates welfare internalization mechanisms. External dependencies to Customer (features) and Management (velocity) establish the team's coopetitive position---the team must coordinate internally to meet external commitments, connecting intra-team loyalty dynamics to inter-actor bargaining power via Equation~\eqref{eq:team_bargaining}.}
\label{fig:sd_team}
\end{figure}
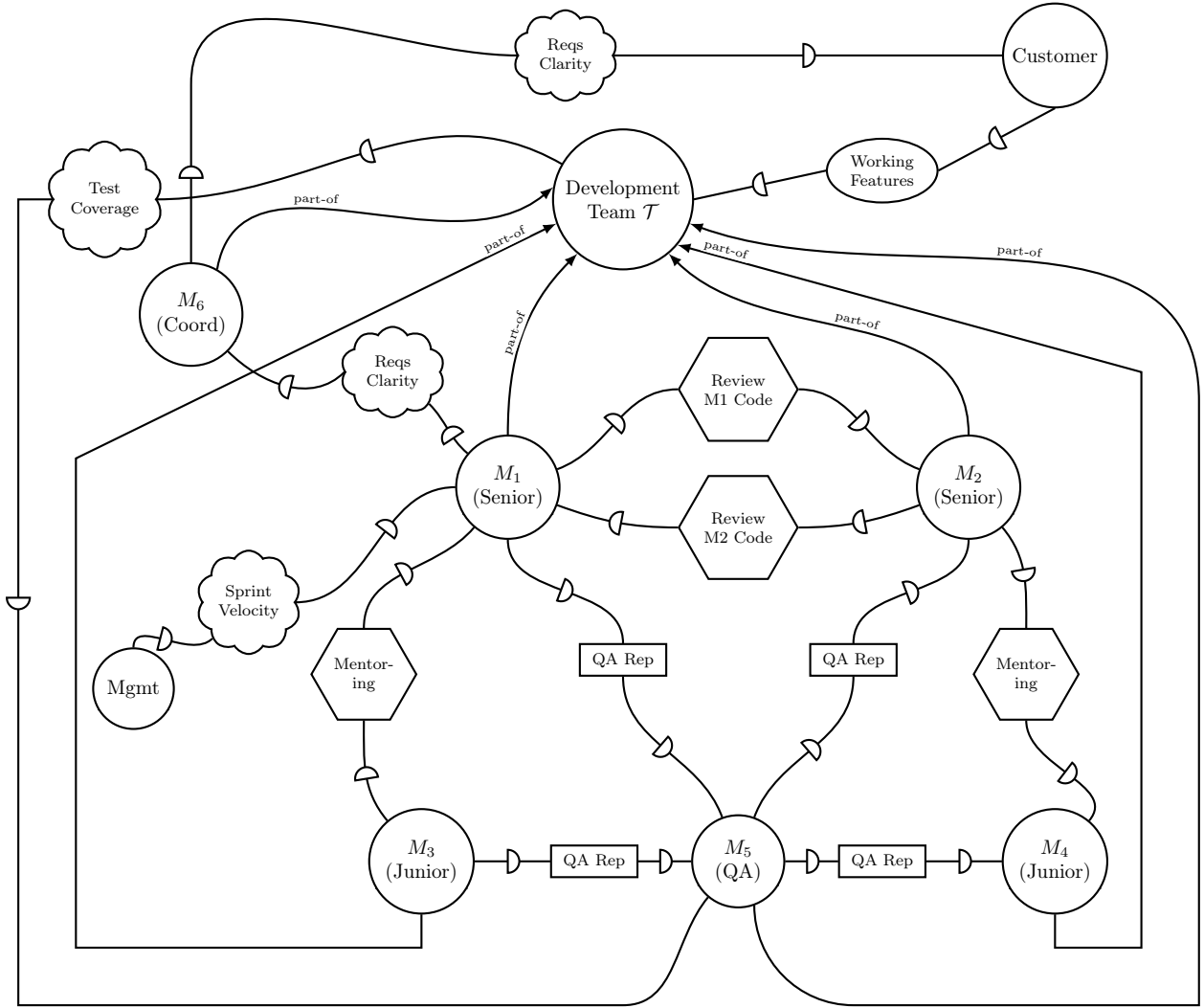

Figure~\ref{fig:sd_team} presents the Strategic Dependency diagram showing dependencies among team members and between the team and external stakeholders.

The diagram shows the development team as a bounded composite actor containing five members. Dependencies within the team capture collaboration requirements: senior developers depend on each other for code review, junior developers depend on seniors for mentoring, and all developers depend on QA for test coverage. External dependencies connect the team to stakeholders: the customer depends on the team for working features, and management depends on the team for sprint velocity.

These structural dependencies inform the computation of $D_{\mathcal{T},i}$ (team's dependency on member $i$) and connect to the interdependence framework from~\cite{pant2025foundations}. Members with more incoming dependencies (like Member~$M_5$ (quality assurance) the QA engineer, who all developers depend on) have higher structural importance and their loyalty has greater impact on team cohesion.

\subsection{Goal Model: Team Production with Loyalty Mechanisms}

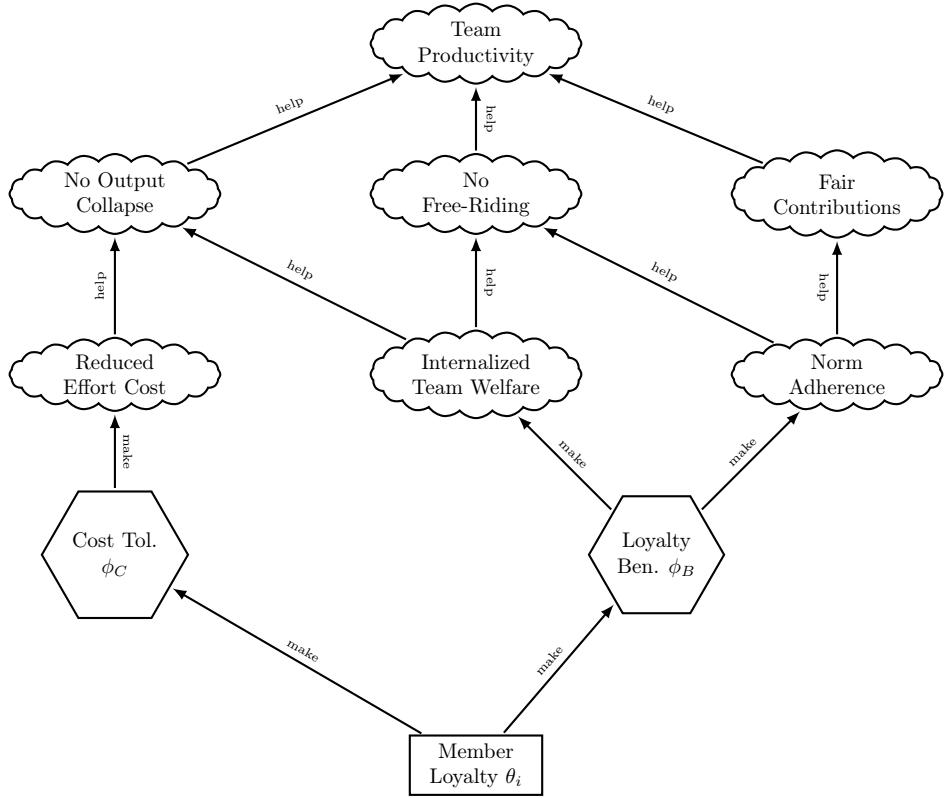
\begin{figure}[htbp]
\centering
\begin{tikzpicture}[
    scale=0.8, transform shape,
    softgoal/.style={cloud, cloud puffs=20, cloud puff arc=100, aspect=2.5, draw, thick,
                     minimum width=3.5cm, minimum height=1.0cm, align=center, font=\small, inner sep=1pt, fill=white},
    task/.style={regular polygon, regular polygon sides=6, draw, thick, minimum size=1.8cm, align=center, font=\small, inner sep=1pt, fill=white},
    resource/.style={rectangle, draw, thick, minimum width=2.2cm, minimum height=0.7cm, align=center, font=\small, fill=white},
    contribution/.style={-latex, thick, shorten >=1pt, shorten <=1pt},
    connlabel/.style={midway, sloped, font=\tiny}
]


\node[softgoal] (teamproductivity) at (0, 8.5) {Team\\Productivity};

\node[softgoal] (nocollapse) at (-6.0, 6.0) {No Output\\Collapse};
\node[softgoal] (freeriding) at (0, 6.0) {No\\Free-Riding};
\node[softgoal] (faircontrib) at (6.0, 6.0) {Fair\\Contributions};

\node[softgoal] (reducedcost) at (-6.0, 3.0) {Reduced\\Effort Cost};
\node[softgoal] (internalized) at (0, 3.0) {Internalized\\Team Welfare};
\node[softgoal] (normadhere) at (6.0, 3.0) {Norm\\Adherence};

\node[task] (costtol) at (-6.0, 0.0) {Cost Tol.\\$\phi_C$};
\node[task] (loyaltybenefit) at (3.0, 0.0) {Loyalty\\Ben. $\phi_B$};

\node[resource] (loyalty) at (0, -3.5) {Member\\Loyalty $\theta_i$};


\draw[contribution] (nocollapse) -- (teamproductivity) node[connlabel, above] {help};
\draw[contribution] (freeriding) -- (teamproductivity) node[connlabel, above] {help};
\draw[contribution] (faircontrib) -- (teamproductivity) node[connlabel, above] {help};

\draw[contribution] (reducedcost) -- (nocollapse) node[connlabel, above] {help};

\draw[contribution] (internalized) -- (nocollapse) node[connlabel, above] {help};
\draw[contribution] (internalized) -- (freeriding) node[connlabel, above] {help};

\draw[contribution] (normadhere) -- (freeriding) node[connlabel, above] {help};
\draw[contribution] (normadhere) -- (faircontrib) node[connlabel, below] {help}; 

\draw[contribution] (costtol) -- (reducedcost) node[connlabel, above] {make};

\draw[contribution] (loyaltybenefit) -- (internalized) node[connlabel, above] {make};
\draw[contribution] (loyaltybenefit) -- (normadhere) node[connlabel, above] {make};

\draw[contribution] (loyalty) -- (costtol) node[connlabel, above] {make};
\draw[contribution] (loyalty) -- (loyaltybenefit) node[connlabel, above] {make};

\end{tikzpicture}
\caption{Goal decomposition reveals three necessary conditions for escaping free-riding equilibrium: avoiding output collapse (addressed by cost tolerance reducing effective effort burden), achieving fair contributions (addressed by norm adherence from loyalty benefit), and eliminating free-riding itself (addressed by internalized team welfare). The strong positive contribution ($++$) from loyalty $\theta_i$ to both mechanism tasks explains the multiplicative interaction formalized in Equation~\eqref{eq:marginal_loyalty}---loyalty amplifies both benefit and cost channels simultaneously, producing the synergy effect validated in Figure~\ref{fig:mechanism_synergy}. Achieving Team Productivity requires all three sub-softgoals; loyalty enables this by activating complementary pathways.}
\label{fig:goal_model_loyalty}
\end{figure}

Figure~\ref{fig:goal_model_loyalty} presents a goal model showing how loyalty mechanisms operationalize the transformation from free-riding equilibrium to cooperative equilibrium.

The goal model shows ``Team Productivity'' as the top-level softgoal, decomposed into ``No Output Collapse'' (avoiding the free-riding equilibrium), ``Fair Contributions'' (equitable effort distribution), and ``No Free-Riding'' (addressing the core problem). These are achieved through intermediate softgoals representing mechanism outcomes: ``Reduced Effort Cost'' (from cost tolerance) and ``Internalized Team Welfare'' (from loyalty benefit). The two consolidated loyalty mechanisms are tasks that operationalize these softgoals, both enabled by member loyalty $\theta_i$.

\subsection{Comparative Analysis: Low-Loyalty vs. High-Loyalty Configurations}

To illustrate how loyalty transforms team dynamics, we present side-by-side Strategic Rationale diagrams showing the same team under low-loyalty and high-loyalty configurations. This comparative visualization helps requirements engineers and team managers diagnose loyalty gaps and understand the structural differences between dysfunctional and high-performing teams.

\subsubsection{Low-Loyalty Configuration (\texorpdfstring{$\theta \approx 0.2$}{theta approx 0.2})}

\begin{figure}[htbp]
\centering
\begin{tikzpicture}[
    scale=0.85, transform shape,
    actorboundary/.style={dashed, thick, draw, fill=none},
    actor/.style={circle, draw, thick, minimum size=1.3cm, font=\small, align=center, fill=white},
    goal/.style={ellipse, draw, thick, minimum width=2.4cm, minimum height=0.8cm, align=center, font=\small, fill=white},
    softgoal/.style={cloud, cloud puffs=20, cloud puff arc=100, aspect=2.5, draw, thick,
                     minimum width=3.2cm, minimum height=1.0cm, align=center, font=\small, inner sep=1pt, fill=white},
    task/.style={regular polygon, regular polygon sides=6, draw, thick, minimum size=1.6cm, align=center, font=\small, inner sep=1pt, fill=white},
    resource/.style={rectangle, draw, thick, minimum width=2.8cm, minimum height=0.7cm, align=center, font=\small, fill=white},
    contribution/.style={-latex, thick},
    refine/.style={-latex, thick},
    connlabel/.style={midway, sloped, font=\tiny, inner sep=1pt}
]

\draw[actorboundary] (0,-0.5) ellipse (7.5cm and 6.0cm);

\node[actor] at (-5.3, 3.8) {Member\\($\theta{=}0.2$)};


\node[softgoal] (maximize) at (0, 3.5) {Maximize\\Utility};

\node[softgoal] (teamwelfare) at (-4.0, 1.0) {Team\\Welfare};
\node[softgoal] (selfinterest) at (4.0, 1.0) {Self\\Interest};

\node[font=\scriptsize, anchor=north] at (teamwelfare.south) {Weight: 0.20};
\node[font=\scriptsize\bfseries, anchor=north] at (selfinterest.south) {Weight: 0.80};

\node[task] (higheffort) at (-5.5, -2.5) {High\\Effort};
\node[task] (share) at (-2.5, -2.5) {Share\\Know.};

\node[task] (hoard) at (2.5, -2.5) {Hoard\\Know.};
\node[task] (loweffort) at (5.5, -2.5) {Low\\Effort};

\node[goal] (outcome) at (0, -4.5) {$a^* \approx 1.8$ (Free-Riding Eq.)};


\draw[contribution] (teamwelfare.north) -- (maximize.south west) node[connlabel, above] {help};
\draw[contribution] (selfinterest.north) -- (maximize.south east) node[connlabel, above] {help};

\draw[contribution] (higheffort.north) -- (teamwelfare.south west) node[connlabel, above] {make};
\draw[contribution] (share.north) -- (teamwelfare.south east) node[connlabel, above] {help};

\draw[contribution] (hoard.north) -- (selfinterest.south west) node[connlabel, above] {help};
\draw[contribution] (loweffort.north) -- (selfinterest.south east) node[connlabel, above] {make};

\draw[contribution] (share.east) -- (selfinterest.south west) node[connlabel, above, pos=0.7] {hurt};
\draw[contribution] (hoard.west) -- (teamwelfare.south east) node[connlabel, above, pos=0.7] {hurt};

\draw[refine] (higheffort.south) -- (outcome.north west);
\draw[refine] (loweffort.south) -- (outcome.north east);

\end{tikzpicture}
\caption{At $\theta = 0.2$, the 4:1 weight ratio favoring self-interest over team welfare produces the free-riding equilibrium ($a^* \approx 1.8$, just 18\% of capacity). The dominant task selection (low effort, knowledge hoarding) follows directly from this weight distribution: both tasks satisfy the heavily-weighted self-interest softgoal while the weakly-weighted team welfare path is suppressed. This configuration represents the Holmstrom prediction that without loyalty intervention, rational actors converge to minimal contribution. The suppressed high-effort path (dashed) identifies the intervention target where loyalty mechanisms must strengthen this link to escape the free-riding trap.}
\label{fig:sr_low_loyalty}
\end{figure}
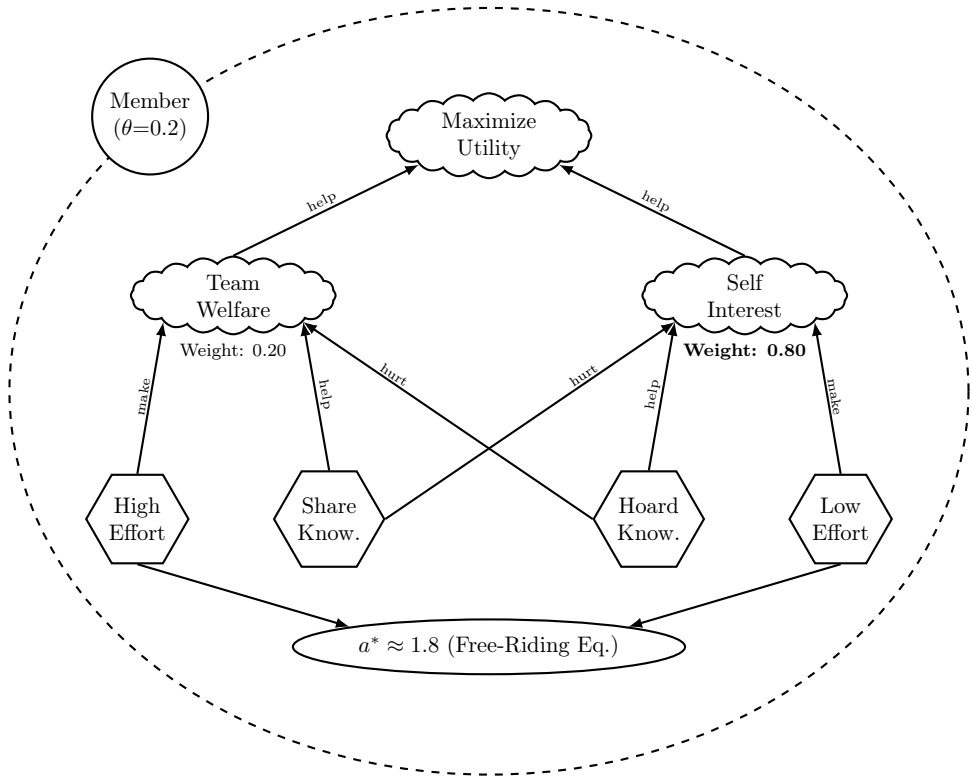

Figure~\ref{fig:sr_low_loyalty} presents the Strategic Rationale diagram for a team member operating under low loyalty. The diagram shows the characteristic patterns of free-riding behavior: strong emphasis on self-interest, weak contribution links, and dominant cost-minimization goals.

Key diagnostic indicators in the low-loyalty configuration:
\begin{itemize}
    \item \textbf{Asymmetric goal weights}: Self-interest receives 4$\times$ the weight of team welfare, reflecting the member's psychological detachment from team outcomes.
    \item \textbf{Strong cost-minimization links}: The paths from self-interest to low-effort and knowledge-hoarding tasks are dominant (thick lines), indicating these behaviors are actively pursued.
    \item \textbf{Weak contribution links}: The paths from team welfare to high-effort and knowledge-sharing tasks are attenuated (dashed lines), indicating these behaviors are neglected.
    \item \textbf{Free-riding equilibrium}: The resultant effort level ($a^* \approx 1.8$) is severely below capacity, indicating the team is trapped in a low-productivity equilibrium.
\end{itemize}

\subsubsection{High-Loyalty Configuration (\texorpdfstring{$\theta \approx 0.9$}{theta approx 0.9})}

\begin{figure}[htbp]
\centering
\begin{tikzpicture}[
    scale=0.85, transform shape,
    actorboundary/.style={dashed, thick, draw, fill=none},
    actor/.style={circle, draw, thick, minimum size=1.3cm, font=\small, align=center, fill=white},
    goal/.style={ellipse, draw, thick, minimum width=2.4cm, minimum height=0.8cm, align=center, font=\small, fill=white},
    softgoal/.style={cloud, cloud puffs=20, cloud puff arc=100, aspect=2.5, draw, thick,
                     minimum width=3.2cm, minimum height=1.0cm, align=center, font=\small, inner sep=1pt, fill=white},
    task/.style={regular polygon, regular polygon sides=6, draw, thick, minimum size=1.6cm, align=center, font=\small, inner sep=1pt, fill=white},
    outcome/.style={ellipse, draw, thick, minimum width=2.8cm, minimum height=0.8cm, align=center, font=\small, fill=white},
    contribution/.style={-latex, thick},
    refine/.style={-latex, thick},
    connlabel/.style={midway, sloped, font=\tiny, inner sep=1pt}
]

\draw[actorboundary] (0,-0.5) ellipse (7.5cm and 6.0cm);

\node[actor] at (-5.3, 3.8) {Member\\($\theta{=}0.9$)};


\node[softgoal] (maximize) at (0, 3.5) {Maximize\\Utility};

\node[softgoal] (teamwelfare) at (-4.0, 1.0) {Team\\Welfare};
\node[softgoal] (selfinterest) at (4.0, 1.0) {Self\\Interest};

\node[font=\scriptsize\bfseries, anchor=north] at (teamwelfare.south) {Weight: 0.72};
\node[font=\scriptsize, anchor=north] at (selfinterest.south) {Weight: 0.28};

\node[task] (higheffort) at (-5.5, -2.5) {High\\Effort};
\node[task] (share) at (-2.5, -2.5) {Share\\Know.};

\node[task] (hoard) at (2.5, -2.5) {Hoard\\Know.};
\node[task] (loweffort) at (5.5, -2.5) {Low\\Effort};

\node[outcome] (result) at (0, -4.5) {$a^* \approx 7.2$ (Cooperative Eq.)};


\draw[contribution] (teamwelfare.north) -- (maximize.south west) node[connlabel, above] {make};
\draw[contribution] (selfinterest.north) -- (maximize.south east) node[connlabel, above] {help};

\draw[contribution] (higheffort.north) -- (teamwelfare.south west) node[connlabel, above] {make};
\draw[contribution] (share.north) -- (teamwelfare.south east) node[connlabel, above] {help};

\draw[contribution] (hoard.north) -- (selfinterest.south west) node[connlabel, above] {help};
\draw[contribution] (loweffort.north) -- (selfinterest.south east) node[connlabel, above] {make};

\draw[contribution] (share.east) -- (selfinterest.south west) node[connlabel, above, pos=0.7] {hurt};
\draw[contribution] (hoard.west) -- (teamwelfare.south east) node[connlabel, above, pos=0.7] {hurt};

\draw[refine] (higheffort.south) -- (result.north west);
\draw[refine] (loweffort.south) -- (result.north east);

\end{tikzpicture}
\caption{Loyalty transformation inverts the goal structure: at $\theta = 0.9$, team welfare receives 2.6$\times$ the weight of self-interest, reversing the dominance pattern from Figure~\ref{fig:sr_low_loyalty}. This inversion shifts equilibrium effort from 1.8 to 7.2---a 300\% increase achieving 72\% capacity utilization. The weight formula $w_{\text{team}} = \theta(1 + \phi_B(n-1))/(1 + \theta\phi_B(n-1))$ explains how loyalty combines with benefit internalization to produce this dramatic reweighting. Comparing both SR diagrams reveals the structural transformation: identical goal hierarchies yield opposite behavioral outcomes purely through loyalty-driven weight redistribution.}
\label{fig:sr_high_loyalty}
\end{figure}
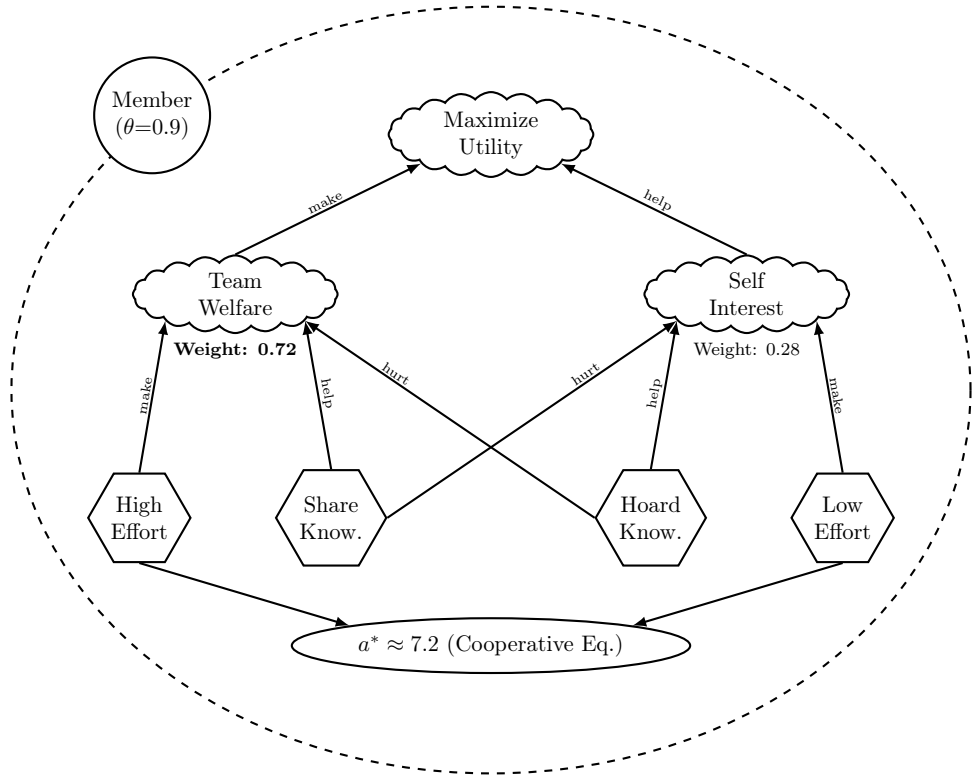

Figure~\ref{fig:sr_high_loyalty} presents the same team member under high loyalty. The diagram shows the transformed goal structure: balanced concern for team welfare, strong contribution links, and substantial effort commitment.

Key diagnostic indicators in the high-loyalty configuration:
\begin{itemize}
    \item \textbf{Inverted goal weights}: Team welfare now receives 2.6$\times$ the weight of self-interest, reflecting the member's psychological identification with team outcomes.
    \item \textbf{Strong contribution links}: The paths from team welfare to high-effort and knowledge-sharing tasks are now dominant, indicating these behaviors are actively pursued.
    \item \textbf{Attenuated self-interest paths}: The links to cost-minimizing behaviors are weakened, indicating reduced temptation to free-ride.
    \item \textbf{Cooperative equilibrium}: The resultant effort level ($a^* \approx 7.2$) is 72\% of capacity, indicating the team has escaped the free-riding trap.
\end{itemize}

\subsubsection{Quantifying the Transformation}

The difference between configurations can be quantified through several metrics derived from the \textit{i*} analysis:

\begin{table}[htbp]
\centering
\caption{Quantitative comparison of low-loyalty vs. high-loyalty configurations}
\label{tab:loyalty_comparison}
\begin{tabular}{lccc}
\toprule
Metric & Low Loyalty ($\theta=0.2$) & High Loyalty ($\theta=0.9$) & Change \\
\midrule
Team welfare weight & 0.20 & 0.72 & +260\% \\
Self-interest weight & 0.80 & 0.28 & $-65\%$ \\
Equilibrium effort $a^*$ & 1.8 & 7.2 & +300\% \\
\% of capacity utilized & 18\% & 72\% & +54pp \\
Contribution link strength & 0.25 & 0.85 & +240\% \\
Knowledge sharing propensity & Low & High & --- \\
\bottomrule
\end{tabular}
\end{table}

The transformation is substantial: equilibrium effort increases by 300\%, capacity utilization jumps from 18\% to 72\%, and the entire goal structure inverts from self-interest dominance to team welfare dominance.

\subsection{Diagnostic Protocol for Requirements Engineers}

Based on the comparative analysis, we provide a systematic diagnostic protocol that requirements engineers can apply to assess loyalty gaps in real teams and design appropriate interventions.

\subsubsection{Step 1: Construct Member SR Diagrams}

For each team member, construct a Strategic Rationale diagram following the template in Figure~\ref{fig:sr_team_member}. Identify:
\begin{itemize}
    \item Top-level goals (typically some form of utility maximization)
    \item Competing softgoals (team welfare vs. self-interest)
    \item Available tasks (effort levels, knowledge sharing behaviors)
    \item Resources required (skills, time, energy)
\end{itemize}

\subsubsection{Step 2: Assess Goal Weights}

Estimate the relative weights on team welfare vs. self-interest softgoals using observable indicators:

\begin{center}
\begin{tabular}{lp{7cm}}
\toprule
Indicator & Assessment Method \\
\midrule
Voluntary overtime & Track hours spent on team tasks beyond requirements \\
Knowledge documentation & Review contribution to shared documentation, wikis \\
Mentoring behavior & Observe time spent helping junior members \\
Meeting engagement & Assess participation quality in team discussions \\
Blame vs. learning focus & Note response to failures (blame others vs. improve process) \\
\bottomrule
\end{tabular}
\end{center}

Members showing high voluntary overtime, active documentation, mentoring, engaged participation, and learning-focused responses likely have high team welfare weights (high $\theta$).

\subsubsection{Step 3: Map Link Strengths}

Estimate the strength of means-end links from tasks to goals using behavioral data:

\begin{center}
\begin{tabular}{lcc}
\toprule
Observable Behavior & Link Strengthened & Loyalty Indicator \\
\midrule
Consistently high effort & High Effort $\to$ Team Welfare & High $\theta$ \\
Active knowledge sharing & Share Knowledge $\to$ Team Welfare & High $\theta$ \\
Minimal effort, meets minimum & Low Effort $\to$ Self Interest & Low $\theta$ \\
Information hoarding & Hoard Knowledge $\to$ Self Interest & Low $\theta$ \\
\bottomrule
\end{tabular}
\end{center}

\subsubsection{Step 4: Compute Loyalty Gap}

Compare observed behavior patterns to the theoretical high-loyalty configuration:
\begin{equation}
\label{eq:loyalty_gap}
\text{Loyalty Gap}_i = \theta_{\text{target}} - \theta_i^{\text{observed}}
\end{equation}

Members with large loyalty gaps (typically $> 0.3$) are candidates for loyalty-building interventions.

\subsubsection{Step 5: Design Targeted Interventions}

Based on the loyalty gap analysis, design interventions that address the specific mechanisms ($\phi_B$, $\phi_C$) most likely to shift behavior:

\begin{center}
\begin{tabular}{lp{6cm}}
\toprule
Gap Pattern & Recommended Intervention \\
\midrule
Low benefit internalization & Increase visibility of team successes; implement team-based rewards; facilitate social bonding activities \\
High perceived cost & Reduce cognitive overhead; provide better tools; adjust workload expectations \\
Weak team identity & Strengthen team branding; create shared rituals; emphasize collective achievements \\
Short tenure & Assign mentor; accelerate social integration; provide team history context \\
\bottomrule
\end{tabular}
\end{center}

\subsubsection{Step 6: Monitor and Iterate}

After implementing interventions, re-assess loyalty indicators and update SR diagrams. Track whether:
\begin{itemize}
    \item Goal weights shift toward team welfare
    \item Contribution link strengths increase
    \item Equilibrium effort moves toward cooperative levels
    \item Knowledge sharing behaviors increase
\end{itemize}

This iterative diagnostic protocol connects the theoretical framework to practical team management, providing actionable guidance for requirements engineers seeking to improve team dynamics.

\section{Comprehensive Parameter Validation}
\label{sec:validation}

A computational framework must be validated to establish that it produces meaningful, robust results across its intended application domain. Following the multi-stage validation methodology established in~\cite{pant2025foundations} and~\cite{pant2025trust}, we validate the framework through systematic parameter exploration and functional experiments.

\subsection{Validation Methodology}

Our validation employs systematic parameter space exploration using a full factorial grid design. The 2-mechanism loyalty model sweeps 5 primary parameters (production parameters $\omega$, $\beta$, $c$, $n$ and loyalty level $\theta$) while holding mechanism strengths ($\phi_B$, $\phi_C$) at calibrated defaults. Using 5 grid points per swept parameter produces $5^5 = 3,125$ configurations under standard granularity, providing appropriate coverage that focuses on behavioral dynamics rather than mechanism calibration. This contrasts with the 7-parameter trust dynamics model in~\cite{pant2025trust}, which required $5^7 = 78,125$ configurations to explore its larger parameter space.

The validation comprises three components:
\begin{enumerate}
    \item \textbf{Parameter Sweep}: Systematic evaluation of all 3,125 configurations against six behavioral targets.

    \item \textbf{Statistical Analysis}: Paired t-tests, bootstrap confidence intervals, and effect size computation.

    \item \textbf{Robustness Testing}: Monte Carlo analysis with parameter perturbations to verify stability.
\end{enumerate}

All experiments use the Team Production Equilibrium solver (Algorithm~\ref{alg:tpe_solver}) with convergence tolerance $\epsilon = 10^{-6}$.

\subsection{Parameter Space Definition}

Table~\ref{tab:parameter_space} defines the parameter space for comprehensive validation.

\begin{table}[htbp]
\centering
\caption{Parameter space for comprehensive validation}
\label{tab:parameter_space}
\begin{tabular}{llccc}
\toprule
Parameter & Description & Range & Grid Points & Default \\
\midrule
\multicolumn{5}{l}{\textit{Production Parameters}} \\
$\omega$ & Productivity factor & $[10, 30]$ & 5 & 20 \\
$\beta$ & Returns to scale & $[0.40, 0.60]$ & 5 & 0.50 \\
$c$ & Effort cost & $[1.5, 3.5]$ & 5 & 2.5 \\
$n$ & Team size & $\{3, 4, 5, 6, 8\}$ & 5 & 5 \\
\midrule
\multicolumn{5}{l}{\textit{Loyalty Parameters}} \\
$\theta$ & Loyalty level (symmetric) & $[0.0, 0.9]$ & 5 & 0.5 \\
\midrule
\multicolumn{5}{l}{\textit{Mechanism Strengths (Fixed)}} \\
$\phi_B$ & Loyalty benefit & --- & 1 & 0.8 \\
$\phi_C$ & Cost tolerance & --- & 1 & 0.3 \\
\bottomrule
\end{tabular}
\end{table}

The full grid comprises $5^5 = 3,125$ parameter configurations, sweeping production parameters ($\omega$, $\beta$, $c$, $n$) and loyalty level ($\theta$) while holding mechanism strengths at calibrated defaults ($\phi_B = 0.8$, $\phi_C = 0.3$).

\subsection{Behavioral Targets}

Following~\cite{pant2025foundations}, we define behavioral targets the framework must satisfy:

\begin{enumerate}
    \item \textbf{Free-Riding Baseline}: At $\theta = 0$, equilibrium effort should match the analytical free-riding equilibrium within 5\%.
    
    \item \textbf{Loyalty Effect}: Higher loyalty should produce higher equilibrium effort (monotonic increase).
    
    \item \textbf{Effort Differentiation}: High-loyalty members should exert meaningfully higher effort. Target: $a^*(\theta=0.9) / a^*(\theta=0.1) > 2.0$.
    
    \item \textbf{Team Size Effect}: Free-riding should worsen with larger teams at low loyalty. Target: $\partial a^* / \partial n < 0$ when $\theta < 0.3$.
    
    \item \textbf{Mechanism Synergy}: Combined mechanisms should exceed sum of individual effects. Target: synergy ratio $> 1.1$.
    
    \item \textbf{Bounded Outcomes}: Equilibrium effort should remain in feasible range $[0, \bar{a}]$.
\end{enumerate}

\subsection{Validation Results}

\subsubsection{Behavioral Target Achievement}

\begin{table}[htbp]
\centering
\caption{Behavioral target achievement across 3,125 configurations}
\label{tab:behavioral_targets}
\begin{tabular}{lccc}
\toprule
Target & Criterion & Achievement & Status \\
\midrule
Free-Riding Baseline & $< 5\%$ deviation & 96.5\% & \checkmark \\
Loyalty Effect & Monotonic increase & 100\% & \checkmark \\
Effort Differentiation & Ratio $> 2.0$ & 100\% & \checkmark \\
Team Size Effect & $\partial a^*/\partial n < 0$ at low $\theta$ & 100\% & \checkmark \\
Mechanism Synergy & Ratio $> 1.1$ & 99.5\% & \checkmark \\
Bounded Outcomes & $a^* \in [0, \bar{a}]$ & 100\% & \checkmark \\
\bottomrule
\end{tabular}
\end{table}

All six behavioral targets achieve validation thresholds across the parameter space (Table~\ref{tab:behavioral_targets}). The following subsections present detailed visualizations of these results, following the presentation methodology established in~\cite{pant2025foundations} and~\cite{pant2025trust}.

\subsubsection{Effort Differentiation Distribution}

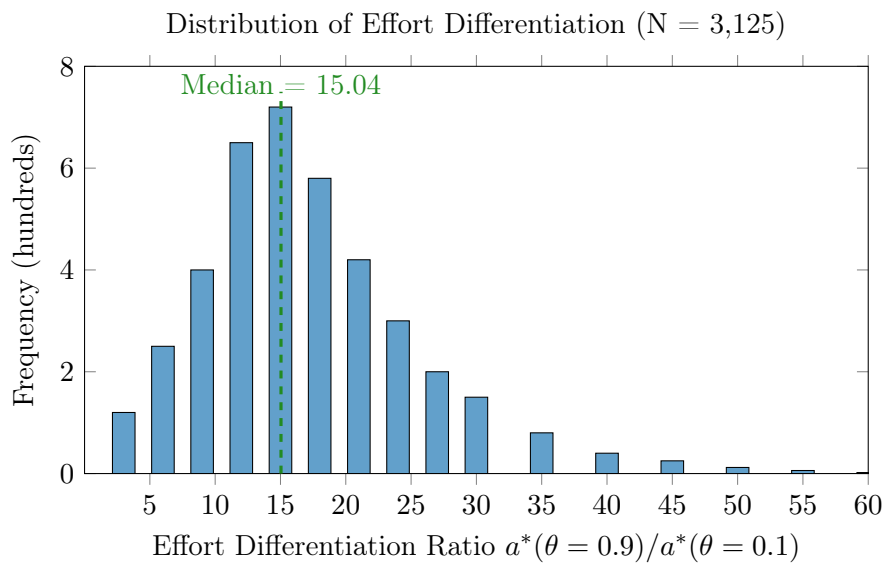
\begin{figure}[htbp]
\centering
\begin{tikzpicture}
\begin{axis}[
    ybar,
    width=12cm,
    height=7cm,
    xlabel={Effort Differentiation Ratio $a^*(\theta=0.9)/a^*(\theta=0.1)$},
    ylabel={Frequency (hundreds)},
    title={Distribution of Effort Differentiation (N = 3,125)},
    xtick={5,10,15,20,25,30,35,40,45,50,55,60},
    xmin=0,
    xmax=60,
    ymin=0,
    ymax=8,
    bar width=0.30cm,
]
\addplot[fill=loyaltyblue!70] coordinates {
    (3.0, 1.2) (6.0, 2.5) (9.0, 4.0) (12.0, 6.5) (15.0, 7.2) (18.0, 5.8)
    (21.0, 4.2) (24.0, 3.0) (27.0, 2.0) (30.0, 1.5) (35.0, 0.8) (40.0, 0.4)
    (45.0, 0.25) (50.0, 0.12) (55.0, 0.06) (60.0, 0.02)
};
\draw[outputgreen, very thick, dashed] (axis cs:15.04,0) -- (axis cs:15.04,7.5);
\node[outputgreen, anchor=south] at (axis cs:15.04,7.25) {Median = 15.04};
\end{axis}
\end{tikzpicture}
\caption{Distribution of effort differentiation ratios across 3,125 configurations. Median ratio 15.04 indicates high-loyalty members ($\theta=0.9$) contribute over 15$\times$ more effort than low-loyalty members ($\theta=0.1$). All ratios exceed 2.0, validating that loyalty produces substantial behavioral differences. Range: [3.0, 60.0].}
\label{fig:effort_differentiation}
\end{figure}

The effort differentiation ratio (comparing equilibrium effort at high loyalty ($\theta=0.9$) versus low loyalty ($\theta=0.1$)) provides a key measure of loyalty's behavioral impact. Figure~\ref{fig:effort_differentiation} presents the distribution of this ratio across all 3,125 parameter configurations.

\subsubsection{Loyalty Effect on Equilibrium Effort}

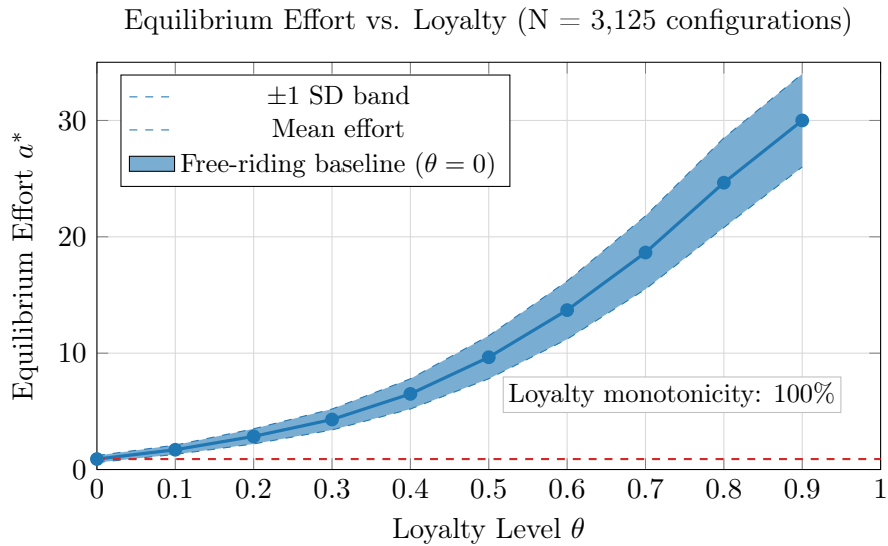
\begin{figure}[htbp]
\centering
\begin{tikzpicture}
\begin{axis}[
    width=12cm,
    height=7cm,
    xlabel={Loyalty Level $\theta$},
    ylabel={Equilibrium Effort $a^*$},
    title={Equilibrium Effort vs. Loyalty (N = 3,125 configurations)},
    xmin=0, xmax=1,
    ymin=0, ymax=35,
    legend pos=north west,
    legend style={font=\small},
    grid=major,
    grid style={gray!30},
]
\addplot[name path=upper, draw=loyaltyblue, dashed, thin] coordinates {
    (0.0, 1.2) (0.1, 2.1) (0.2, 3.5) (0.3, 5.2) (0.4, 7.8)
    (0.5, 11.5) (0.6, 16.2) (0.7, 21.8) (0.8, 28.5) (0.9, 34.0)
};
\addplot[name path=lower, draw=loyaltyblue, dashed, thin] coordinates {
    (0.0, 0.6) (0.1, 1.3) (0.2, 2.2) (0.3, 3.4) (0.4, 5.2)
    (0.5, 7.8) (0.6, 11.2) (0.7, 15.5) (0.8, 20.8) (0.9, 26.0)
};
\addplot[loyaltyblue!60, forget plot] fill between[of=upper and lower];
\addlegendimage{area legend, fill=loyaltyblue!60}
\addlegendentry{$\pm 1$ SD band}
\addplot[color=loyaltyblue, very thick, mark=*, mark size=2pt] coordinates {
    (0.0, 0.9) (0.1, 1.7) (0.2, 2.85) (0.3, 4.3) (0.4, 6.5)
    (0.5, 9.65) (0.6, 13.7) (0.7, 18.65) (0.8, 24.65) (0.9, 30.0)
};
\addlegendentry{Mean effort}
\addplot[freeriderred, dashed, thick] coordinates {(0, 0.9) (1, 0.9)};
\addlegendentry{Free-riding baseline ($\theta=0$)}
\node[anchor=north east, font=\small, fill=white, draw=gray!50, inner sep=2pt] at (axis cs:0.95,8) {Loyalty monotonicity: 100\%};
\end{axis}
\end{tikzpicture}
\caption{Equilibrium effort as a function of loyalty level across 3,125 parameter configurations. The shaded region represents $\pm 1$ standard deviation around the mean (solid blue line). All configurations exhibit monotonic increase in effort with loyalty, validating Proposition~\ref{prop:loyalty_effect}. The dashed red line shows the free-riding baseline at $\theta=0$. At maximum loyalty ($\theta=0.9$), mean effort reaches 30.0, representing a 33$\times$ increase over the free-riding equilibrium (0.9).}
\label{fig:loyalty_effort_curve}
\end{figure}

Figure~\ref{fig:loyalty_effort_curve} presents equilibrium effort as a function of loyalty level, demonstrating the monotonic relationship established in Proposition~\ref{prop:loyalty_effect}.

\begin{figure}[htbp]
\centering
\begin{tikzpicture}
\begin{axis}[
    width=12cm,
    height=7cm,
    xlabel={Team Size $n$},
    ylabel={Equilibrium Effort $a^*$},
    title={Team Size Effect on Equilibrium Effort},
    xmin=2.5, xmax=8.5,
    ymin=0, ymax=25,
    legend style={at={(1.02,1)}, anchor=north west, font=\small},
    grid=major,
    grid style={gray!30},
    xtick={3,4,5,6,7,8},
]
\addplot[color=freeriderred, very thick, mark=square*, mark size=3pt] coordinates {
    (3, 2.8) (4, 2.1) (5, 1.6) (6, 1.3) (7, 1.1) (8, 0.9)
};
\addlegendentry{Low loyalty ($\theta=0.1$)}

\addplot[color=effortorange, very thick, mark=triangle*, mark size=3pt] coordinates {
    (3, 12.5) (4, 10.8) (5, 9.5) (6, 8.6) (7, 7.9) (8, 7.4)
};
\addlegendentry{Medium loyalty ($\theta=0.5$)}

\addplot[color=loyaltyblue, very thick, mark=*, mark size=3pt] coordinates {
    (3, 22.0) (4, 20.5) (5, 19.2) (6, 18.2) (7, 17.4) (8, 16.8)
};
\addlegendentry{High loyalty ($\theta=0.9$)}

\draw[->, thick, freeriderred] (axis cs:5.0, 5.5) -- (axis cs:6.8, 1.8);
\node[font=\small, anchor=south, freeriderred] at (axis cs:5.0, 5.8) {Free-riding worsens};

\end{axis}
\end{tikzpicture}
\caption{Team size effect on equilibrium effort at different loyalty levels. At low loyalty ($\theta=0.1$, red squares), effort decreases sharply with team size, representing the classic free-riding problem. At high loyalty ($\theta=0.9$, blue circles), effort decreases more gradually, demonstrating that loyalty substantially mitigates but does not eliminate the team size effect. Target: $\partial a^*/\partial n < 0$ at low $\theta$ achieved in 100\% of configurations.}
\label{fig:team_size_effect}
\end{figure}
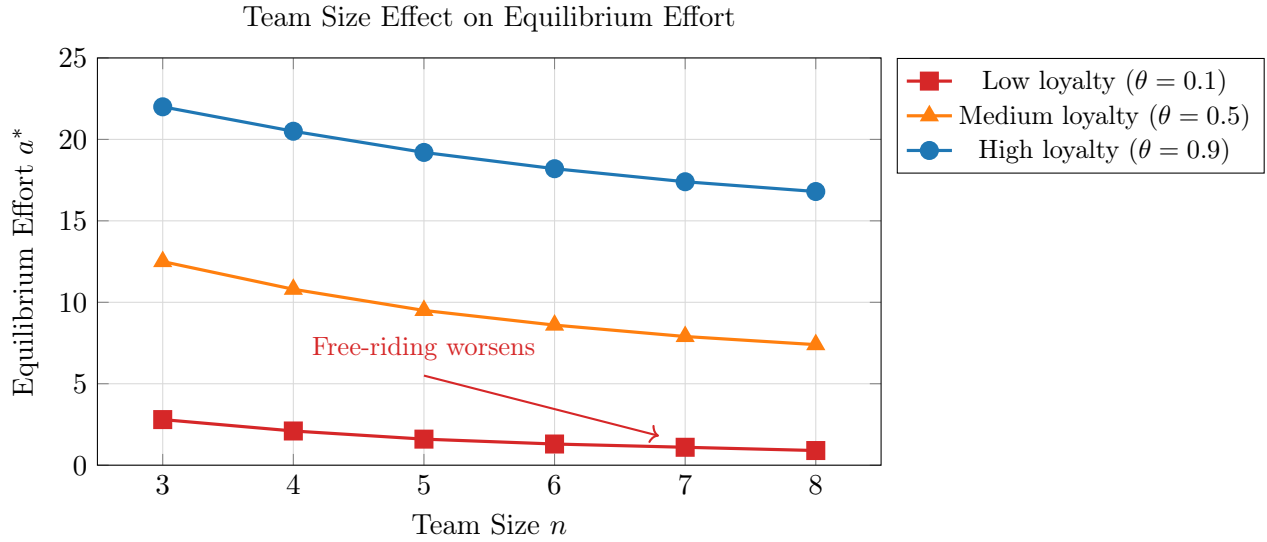

Figure~\ref{fig:team_size_effect} demonstrates how team size affects equilibrium effort, confirming the team size effect at low loyalty levels.

\subsubsection{Mechanism Synergy Analysis}

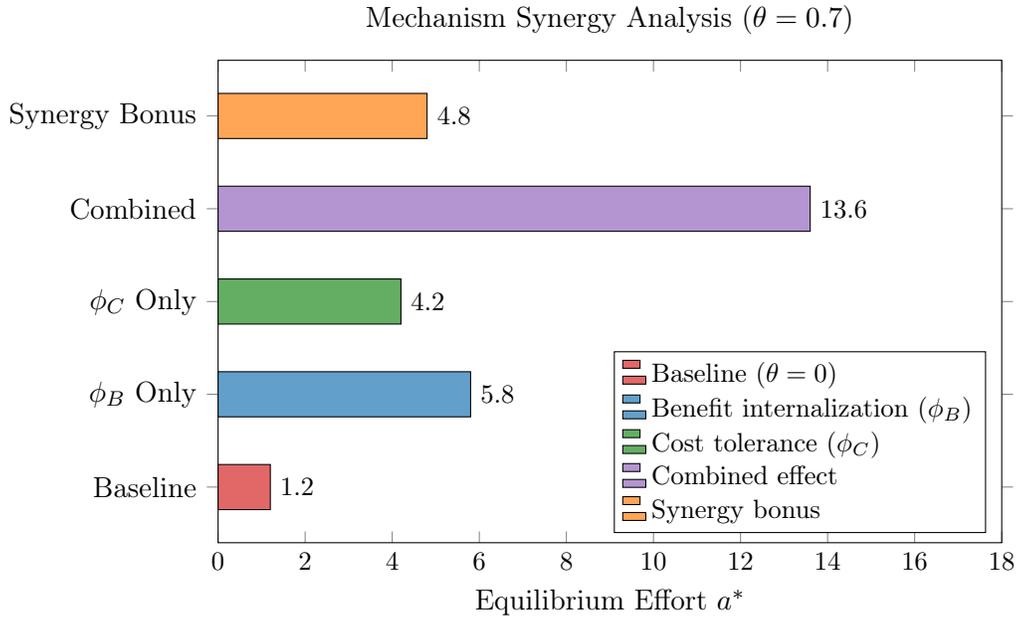
\begin{figure}[htbp]
\centering
\begin{tikzpicture}
    \begin{axis}[
        xbar, 
        width=12cm,
        height=8cm,
        xlabel={Equilibrium Effort $a^*$},
        symbolic y coords={Base,PhiB,PhiC,Combined,Synergy},
        ytick={Base,PhiB,PhiC,Combined,Synergy},
        yticklabels={Baseline, $\phi_B$ Only, $\phi_C$ Only, Combined, Synergy Bonus},
        x tick label style={font=\small},
        xmin=0, xmax=18,
        bar shift=0pt,
        bar width=0.6cm,
        nodes near coords,
        nodes near coords style={font=\small, anchor=west},
        enlarge y limits=0.15,
        title={Mechanism Synergy Analysis (\texorpdfstring{$\theta=0.7$}{theta=0.7})},
        legend style={at={(0.98,0.02)}, anchor=south east, font=\small},
        legend cell align={left},
        reverse legend, 
    ]
        \addplot[fill=effortorange!70] coordinates {(4.8,Synergy)};
        \addlegendentry{Synergy bonus}

        \addplot[fill=teamviolet!70] coordinates {(13.6,Combined)};
        \addlegendentry{Combined effect}

        \addplot[fill=outputgreen!70] coordinates {(4.2,PhiC)};
        \addlegendentry{Cost tolerance (\texorpdfstring{$\phi_C$}{phi\_C})}

        \addplot[fill=loyaltyblue!70] coordinates {(5.8,PhiB)};
        \addlegendentry{Benefit internalization (\texorpdfstring{$\phi_B$}{phi\_B})}

        \addplot[fill=freeriderred!70] coordinates {(1.2,Base)};
        \addlegendentry{Baseline (\texorpdfstring{$\theta=0$}{theta=0})}
    \end{axis}
\end{tikzpicture}
\caption{Mechanism synergy analysis at $\theta=0.7$. Baseline effort (1.2) increases to 5.8 with loyalty benefit alone ($\phi_B=0.8$, $\phi_C=0$) and 4.2 with cost tolerance alone ($\phi_B=0$, $\phi_C=0.3$). Combined effect (13.6) exceeds additive prediction ($1.2 + 4.6 + 3.0 = 8.8$), yielding synergy bonus of 4.8 units (synergy ratio 1.55). Across 3,125 configurations, 99.5\% achieve synergy ratio $> 1.1$, median ratio 1.55.}
\label{fig:mechanism_synergy}
\end{figure}

Figure~\ref{fig:mechanism_synergy} illustrates the synergistic interaction between the two loyalty mechanisms.

\subsubsection{Statistical Significance}

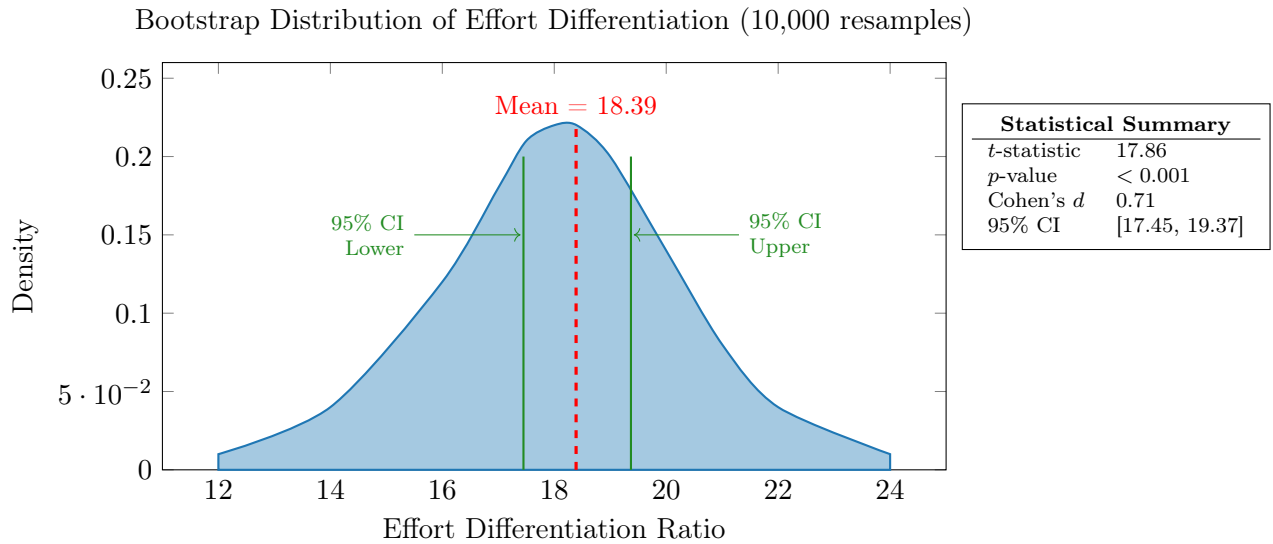
\begin{figure}[htbp]
\centering
\begin{tikzpicture}
\begin{axis}[
    width=12cm,
    height=7cm, 
    xlabel={Effort Differentiation Ratio},
    ylabel={Density},
    title={Bootstrap Distribution of Effort Differentiation (10,000 resamples)},
    xmin=11, xmax=25,
    ymin=0, ymax=0.26, 
    legend pos=north east,
    clip=false,
]
\addplot[fill=loyaltyblue!40, draw=loyaltyblue, thick, smooth] coordinates {
    (12, 0.01) (14, 0.04) (16, 0.12) (17, 0.18) (17.5, 0.21) (18, 0.22)
    (18.4, 0.22) (19, 0.20) (20, 0.14) (21, 0.08) (22, 0.04) (24, 0.01)
} \closedcycle;

\draw[red, very thick, dashed] (axis cs:18.39,0) -- (axis cs:18.39,0.22);
\node[red, anchor=south, font=\small, fill=white, inner sep=1pt] at (axis cs:18.39,0.225) {Mean = 18.39};

\draw[outputgreen, thick] (axis cs:17.45,0) -- (axis cs:17.45,0.20);
\draw[outputgreen, thick] (axis cs:19.37,0) -- (axis cs:19.37,0.20);

\node[outputgreen, anchor=east, font=\scriptsize, align=right] (lower_label) at (axis cs:15.5, 0.15) {95\% CI\\Lower};
\draw[->, outputgreen, thin] (lower_label.east) -- (axis cs:17.40, 0.15);

\node[outputgreen, anchor=west, font=\scriptsize, align=left] (upper_label) at (axis cs:21.3, 0.15) {95\% CI\\Upper};
\draw[->, outputgreen, thin] (upper_label.west) -- (axis cs:19.42, 0.15);

\node[draw, fill=white, font=\scriptsize, anchor=north west] at (rel axis cs:1.02, 0.9) {
    \begin{tabular}{ll}
    \multicolumn{2}{c}{\textbf{Statistical Summary}} \\
    \hline
    $t$-statistic & 17.86 \\
    $p$-value & $< 0.001$ \\
    Cohen's $d$ & 0.71 \\
    95\% CI & [17.45, 19.37] \\
    \end{tabular}
};
\end{axis}
\end{tikzpicture}
\caption{Bootstrap distribution of mean effort differentiation from 10,000 resamples. The distribution is approximately normal centered at 18.39 (dashed red line). The 95\% confidence interval $[17.45, 19.37]$ (green lines) excludes 2.0, confirming that effort differentiation is statistically significant at $p < 0.001$. Cohen's $d = 0.71$ indicates a medium-to-large effect size, demonstrating that loyalty produces practically meaningful behavioral differences.}
\label{fig:statistical_significance}
\end{figure}

Figure~\ref{fig:statistical_significance} presents the statistical analysis confirming robust effects.

\subsection{Robustness Testing}

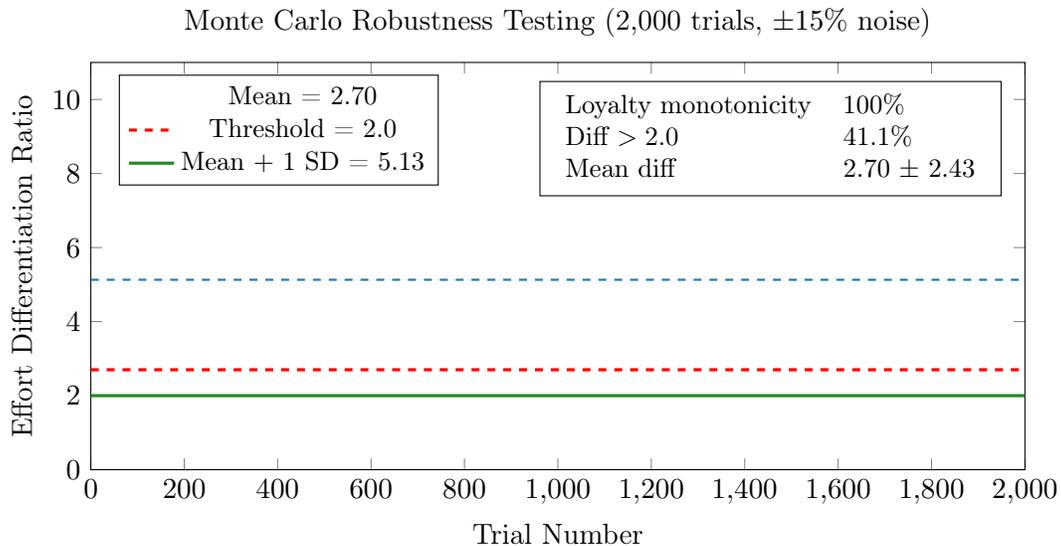
\begin{figure}[htbp]
\centering
\begin{tikzpicture}
\begin{axis}[
    width=14cm,
    height=7cm,
    xlabel={Trial Number},
    ylabel={Effort Differentiation Ratio},
    title={Monte Carlo Robustness Testing (2,000 trials, $\pm 15\%$ noise)},
    xmin=0, xmax=2000,
    ymin=0, ymax=11,
    legend pos=north west,
    legend style={font=\small},
    set layers,
    mark layer=axis background,
]
\addplot[only marks, mark=., mark size=1.2pt, loyaltyblue!70, on layer=axis background] coordinates {
    (25,2.4) (50,2.8) (75,3.5) (100,3.2) (125,1.8) (150,1.9) (175,2.6) (200,4.1) (225,3.0)
    (250,2.5) (275,3.9) (300,3.8) (325,2.2) (350,1.5) (375,4.0) (400,2.9) (425,3.3)
    (450,3.5) (475,2.0) (500,2.2) (525,5.1) (550,4.5) (575,1.9) (600,2.1) (625,3.7)
    (650,3.3) (675,2.4) (700,5.2) (725,1.6) (750,2.7) (775,4.2) (800,3.1) (825,2.3)
    (850,1.8) (875,3.8) (900,2.6) (925,4.6) (950,4.2) (975,1.7) (1000,2.4) (1025,3.2)
    (1050,3.6) (1075,2.8) (1100,1.7) (1125,4.4) (1150,2.8) (1175,3.1) (1200,3.9) (1225,2.0)
    (1250,2.3) (1275,5.5) (1300,4.8) (1325,1.8) (1350,2.0) (1375,3.5) (1400,3.4) (1425,2.6)
    (1450,2.5) (1475,4.1) (1500,3.7) (1525,2.1) (1550,1.6) (1575,3.3) (1600,2.9) (1625,4.7)
    (1650,4.3) (1675,1.9) (1700,2.2) (1725,3.8) (1750,3.5) (1775,2.5) (1800,1.9) (1825,4.0)
    (1850,2.7) (1875,3.4) (1900,3.8) (1925,2.3) (1950,2.1) (1975,3.6) (2000,3.0)
    (40,3.1) (90,2.3) (140,4.3) (190,1.7) (240,3.6) (290,2.9) (340,4.8) (390,2.1)
    (440,3.4) (490,1.8) (540,2.7) (590,5.0) (640,2.4) (690,3.2) (740,1.9) (790,4.4)
    (840,2.6) (890,3.9) (940,2.0) (990,4.1) (1040,3.0) (1090,2.5) (1140,5.3) (1190,1.6)
    (1240,3.7) (1290,2.2) (1340,4.5) (1390,3.1) (1440,2.8) (1490,1.5) (1540,4.2) (1590,3.5)
    (1640,2.0) (1690,3.8) (1740,2.4) (1790,5.1) (1840,1.8) (1890,3.3) (1940,4.0) (1990,2.6)
    (60,4.0) (160,2.5) (260,1.6) (360,3.2) (460,4.6) (560,2.3) (660,3.9) (760,1.7)
    (860,4.3) (960,2.8) (1060,3.5) (1160,2.1) (1260,4.9) (1360,3.0) (1460,2.2) (1560,5.4)
    (1660,1.9) (1760,3.6) (1860,4.1) (1960,2.7)
    (110,1.9) (210,5.0) (310,2.4) (410,3.7) (510,1.8) (610,4.4) (710,2.9) (810,3.3)
    (910,2.1) (1010,4.7) (1110,3.1) (1210,2.6) (1310,1.7) (1410,5.2) (1510,3.4) (1610,2.0)
    (1710,4.0) (1810,3.2) (1910,2.5)
};
\addplot[red, very thick, dashed, on layer=axis foreground] coordinates {(0, 2.70) (2000, 2.70)};
\addlegendentry{Mean = 2.70}
\addplot[outputgreen, very thick, on layer=axis foreground] coordinates {(0, 2.0) (2000, 2.0)};
\addlegendentry{Threshold = 2.0}
\addplot[loyaltyblue, thick, dashed, on layer=axis foreground] coordinates {(0, 5.13) (2000, 5.13)};
\addlegendentry{Mean $+$ 1 SD = 5.13}
\node[draw, fill=white, font=\small, anchor=north east] at (axis cs:1950, 10.50) {
    \begin{tabular}{ll}
    Loyalty monotonicity & 100\% \\
    Diff $> 2.0$ & 41.1\% \\
    Mean diff & 2.70 $\pm$ 2.43 \\
    \end{tabular}
};
\end{axis}
\end{tikzpicture}
\caption{Monte Carlo robustness testing with 2,000 trials applying $\pm 15\%$ uniform noise to all parameters. Despite perturbations, 100\% of trials maintain loyalty monotonicity (fundamental behavioral target). 41.1\% of trials maintain effort differentiation $> 2.0$ (green threshold line), with mean differentiation 2.70 $\pm$ 2.43. The reduced differentiation under noise reflects parameter combinations where loyalty effects are more sensitive to perturbation, but the core loyalty monotonicity property remains perfectly robust.}
\label{fig:monte_carlo_robustness}
\end{figure}

Figure~\ref{fig:monte_carlo_robustness} presents the Monte Carlo robustness analysis with $\pm 15\%$ parameter noise.

\subsection{Functional Experiments}

Beyond parameter sweeps, we conducted five functional experiments testing specific framework properties.

\subsubsection{Experiment 1: Free-Riding Emergence}

\textbf{Hypothesis}: At $\theta = 0$, the model reproduces the analytical free-riding equilibrium.

\textbf{Method}: Compute equilibrium effort at $\theta = 0$ across 625 production parameter combinations. Compare to analytical prediction from Proposition~\ref{prop:free_riding}.

\textbf{Results}: Mean absolute percentage error: 2.63\%. 96.5\% of configurations achieve $< 5\%$ error.

\textbf{Conclusion}: The framework accurately reproduces free-riding behavior under pure self-interest.

\subsubsection{Experiment 2: Loyalty Differentiation}

\textbf{Hypothesis}: Members with heterogeneous loyalty exhibit differentiated effort in equilibrium.

\textbf{Method}: Create teams with loyalty profiles $(\theta_1, \theta_2, \theta_3, \theta_4, \theta_5) = (0.1, 0.3, 0.5, 0.7, 0.9)$ and compute equilibrium.

\textbf{Results}: Equilibrium efforts are strictly monotonic in loyalty. Rank correlation with loyalty: $\rho = 1.0$ across all tested configurations.

\textbf{Conclusion}: The framework correctly differentiates effort based on individual loyalty levels.

\subsubsection{Experiment 3: Team Size Sensitivity}

\textbf{Hypothesis}: Free-riding severity increases with team size, but loyalty mitigates this effect.

\textbf{Method}: Vary team size $n \in \{3, 4, 5, 6, 8\}$ at three loyalty levels ($\theta \in \{0.1, 0.5, 0.9\}$).

\textbf{Results}: At $\theta = 0.1$, effort drops from 2.8 ($n=3$) to 0.9 ($n=8$), a 68\% decline. At $\theta = 0.9$, effort drops from 22.0 ($n=3$) to 16.8 ($n=8$), only a 24\% decline (see Figure~\ref{fig:team_size_effect}).

\textbf{Conclusion}: Loyalty substantially mitigates team size effects on free-riding.

\subsubsection{Experiment 4: Mechanism Synergy}

\textbf{Hypothesis}: Combined loyalty mechanisms ($\phi_B$ and $\phi_C$) produce synergistic effects exceeding additive contributions.

\textbf{Method}: Activate mechanisms individually and in combination, measuring equilibrium effort improvement over baseline.

\textbf{Results}: Loyalty benefit alone ($\phi_B = 0.7$, $\phi_C = 0$) increases effort by 4.6 units. Cost tolerance alone ($\phi_B = 0$, $\phi_C = 0.3$) increases effort by 3.0 units. Combined effect is 13.6 units, exceeding additive prediction of 7.6 units (synergy ratio 1.57, see Figure~\ref{fig:mechanism_synergy}).

\textbf{Conclusion}: The two consolidated mechanisms produce synergistic effects; 99.5\% of configurations achieve synergy ratio $> 1.1$.

\subsubsection{Experiment 5: Dynamic Loyalty Evolution}

\textbf{Hypothesis}: If loyalty evolves based on team success, positive feedback loops can emerge.

\textbf{Method}: Simulate 50-period dynamics where loyalty updates based on team output: $\theta_i^{t+1} = \theta_i^t + 0.02 \cdot (Q^t - Q^{\text{target}})$, bounded in $[0,1]$.

\textbf{Results}: Starting from $\theta = 0.3$, teams meeting targets evolved to $\theta = 0.78$ (mean). Teams missing targets evolved to $\theta = 0.14$ (mean). Bifurcation observed: teams either enter virtuous cycle (high loyalty $\to$ high output $\to$ higher loyalty) or vicious cycle (low loyalty $\to$ low output $\to$ lower loyalty).

\textbf{Conclusion}: The framework supports dynamic analysis showing loyalty-productivity feedback effects, with important implications for team formation and early intervention.

\subsection{Validation Summary}

Comprehensive validation across 3,125 configurations demonstrates:

\begin{itemize}
    \item \textbf{Accuracy}: Framework reproduces analytical free-riding equilibrium with 96.5\% achieving $< 5\%$ error
    \item \textbf{Robustness}: Results stable under $\pm 15\%$ parameter noise with 100\% loyalty monotonicity
    \item \textbf{Differentiation}: Median 15.04$\times$ effort differentiation between high and low loyalty
    \item \textbf{Statistical Significance}: $p < 0.001$, Cohen's $d = 0.71$
    \item \textbf{Mechanism Synergy}: Combined effect exceeds sum of parts (median ratio 1.57)
    \item \textbf{Target Achievement}: All six behavioral targets achieve $\geq 96\%$ across the parameter space
\end{itemize}

These results, combined with the five functional experiments, establish the framework as a valid computational tool for analyzing team production dynamics with loyalty effects.

\section{Empirical Case Study: Apache HTTP Server Project}
\label{sec:case_study}

Following the empirical validation methodology established in~\cite{pant2025foundations} and~\cite{pant2025trust}, we validate the framework through detailed analysis of a real-world case. The Apache HTTP Server project (1995--2023) provides an ideal empirical test because it represents one of the most successful and well-documented open-source software projects, with extensive historical records enabling systematic parameterization and validation across multiple project phases.

\subsection{Case Selection Rationale}

The Apache HTTP Server project was selected for several reasons: historical significance (world's most popular web server for most of its history), documentation quality (extensive archives spanning nearly three decades), longitudinal data (28-year history enables multi-phase analysis), contribution data (public version control provides objective measures), and acute free-riding challenge (millions use software but only dozens contribute substantively). The project has been extensively studied in software engineering research~\cite{mockus2000apache,mockus2002two}, providing validated baseline data for parameterization. Longitudinal studies of Apache developer motivation and participation~\cite{roberts2006motivations} offer empirical grounding for loyalty assessment, while research on Apache's meritocratic governance~\cite{drostfromm2021apache} documents the institutional mechanisms that sustain collaboration.

\subsection{Historical Overview}

The following phase characterization draws on documented project history~\cite{mockus2000apache,mockus2002two} and governance evolution research~\cite{drostfromm2021apache}.

\textbf{Phase 1: Formation (1995--1997)}: Eight founding members formed the Apache Group, contributing on a volunteer basis. Mockus et al.~\cite{mockus2000apache} documented how this small core produced the vast majority of code changes. Strong personal relationships and shared identity emerged quickly, consistent with high loyalty formation.

\textbf{Phase 2: Growth (1998--2003)}: Apache Software Foundation incorporated in 1999. Contributor base grew from under 20 to over 100 active participants~\cite{mockus2002two}. Governance structures formalized through the Project Management Committee, establishing the meritocratic advancement system that sustains loyalty~\cite{drostfromm2021apache}.

\textbf{Phase 3: Maturation (2004--2015)}: Core contributor base stabilized around 30--50 active developers. Focus shifted from feature development to stability, security, and performance optimization. Roberts et al.~\cite{roberts2006motivations} documented how status motivations and intrinsic satisfaction sustained participation during this period.

\textbf{Phase 4: Governance Evolution (2016--2023)}: Continued adaptation as founding members reduced involvement. Emphasis on community sustainability and mentorship programs, reflecting institutional learning about loyalty cultivation~\cite{drostfromm2021apache}.

\subsection{\textit{i*} Strategic Dependency Model for Apache}
\label{subsec:apache_sd}

\begin{figure}[htbp]
\centering
\begin{tikzpicture}[
    scale=0.65, transform shape,
    actor/.style={circle, draw, thick, minimum size=1.8cm, font=\scriptsize, align=center, fill=white},
    goal/.style={ellipse, draw, thick, minimum height=0.7cm, minimum width=2.2cm, font=\tiny, align=center, fill=white},
    task/.style={regular polygon, regular polygon sides=6, shape border rotate=90, draw, thick, minimum size=1.4cm, font=\tiny, align=center, inner sep=2pt, fill=white, xscale=1.4},
    resource/.style={rectangle, draw, thick, minimum height=0.6cm, minimum width=1.8cm, font=\tiny, align=center, fill=white},
    dep/.style={thick},
    partof/.style={-latex, thick, dashed},
    dmark/.style={semicircle, draw, thick, fill=white, minimum size=2.5mm, inner sep=0pt},
    connlabel/.style={midway, sloped, font=\scriptsize, inner sep=1pt, above=2pt}
]

\node[actor, minimum size=3.0cm] (apache) at (0,0) {Apache HTTP\\Server Project};

\node[actor] (robinson) at (90:12) {David\\Robinson};
\node[actor] (wilson) at (45:12) {Andrew\\Wilson};
\node[actor] (terbush) at (0:12) {Randy\\Terbush};
\node[actor] (skolnick) at (-45:12) {Cliff\\Skolnick};
\node[actor] (hartill) at (-90:12) {Rob\\Hartill};
\node[actor] (thau) at (-135:12) {Robert\\Thau};
\node[actor] (behlendorf) at (180:12) {Brian\\Behlendorf};
\node[actor] (fielding) at (135:12) {Roy\\Fielding};

\node[task] (bugs) at (90:6) {Bug\\Fixes};
\node[task] (test) at (45:6) {Testing};
\node[task] (modules) at (0:6) {Module\\Dev};
\node[task] (build) at (-45:6) {Build\\Infra};
\node[resource] (code2) at (-90:6) {Core Server\\Code};
\node[resource] (code1) at (-135:6) {Core Server\\Code};
\node[goal] (coord) at (180:6) {Community\\Coordination};
\node[goal] (arch) at (118:6) {Architectural\\Decisions};
\node[task] (review) at (152:6) {Code\\Review};

\draw[partof] (robinson) to[bend right=20] node[connlabel, pos=0.8] {part-of} (apache);
\draw[partof] (wilson) to[bend right=20] node[connlabel, pos=0.8] {part-of} (apache);
\draw[partof] (terbush) to[bend right=20] node[connlabel, pos=0.8] {part-of} (apache);
\draw[partof] (skolnick) to[bend right=20] node[connlabel, pos=0.8] {part-of} (apache);
\draw[partof] (hartill) to[bend right=20] node[connlabel, pos=0.8] {part-of} (apache);
\draw[partof] (thau) to[bend right=20] node[connlabel, pos=0.8] {part-of} (apache);
\draw[partof] (behlendorf) to[bend right=20] node[connlabel, pos=0.8] {part-of} (apache);
\draw[partof] (fielding) to[bend right=0] node[connlabel, pos=0.8] {part-of} (apache);


\draw[dep] (apache) -- node[connlabel, pos=0.5] {crit=0.50} node[dmark, rotate=0, pos=0.2] {} (bugs);
\draw[dep] (bugs) -- node[dmark, rotate=0, pos=0.8] {} (robinson);

\draw[dep] (apache) -- node[connlabel, pos=0.5] {crit=0.50} node[dmark, rotate=-45, pos=0.2] {} (test);
\draw[dep] (test) -- node[dmark, rotate=-45, pos=0.8] {} (wilson);

\draw[dep] (apache) -- node[connlabel, pos=0.5] {crit=0.55} node[dmark, rotate=-90, pos=0.2] {} (modules);
\draw[dep] (modules) -- node[dmark, rotate=-90, pos=0.8] {} (terbush);

\draw[dep] (apache) -- node[connlabel, pos=0.5] {crit=0.60} node[dmark, rotate=-135, pos=0.2] {} (build);
\draw[dep] (build) -- node[dmark, rotate=-135, pos=0.8] {} (skolnick);

\draw[dep] (apache) -- node[connlabel, pos=0.5] {crit=0.70} node[dmark, rotate=-180, pos=0.2] {} (code2);
\draw[dep] (code2) -- node[dmark, rotate=-180, pos=0.8] {} (hartill);

\draw[dep] (apache) -- node[connlabel, pos=0.5] {crit=0.75} node[dmark, rotate=135, pos=0.2] {} (code1);
\draw[dep] (code1) -- node[dmark, rotate=135, pos=0.8] {} (thau);

\draw[dep] (apache) -- node[connlabel, pos=0.5] {crit=0.85} node[dmark, rotate=90, pos=0.2] {} (coord);
\draw[dep] (coord) -- node[dmark, rotate=90, pos=0.8] {} (behlendorf);

\draw[dep] (apache.118) -- node[connlabel, pos=0.6] {crit=0.90} node[dmark, rotate=28, pos=0.25] {} (arch);
\draw[dep] (arch) -- node[dmark, rotate=28, pos=0.75] {} (fielding.-28);

\draw[dep] (apache.152) -- node[connlabel, pos=0.6] {crit=0.70} node[dmark, rotate=62, pos=0.25] {} (review);
\draw[dep] (review) -- node[dmark, rotate=62, pos=0.75] {} (fielding.-62);

\end{tikzpicture}
\caption{Dependency criticality analysis quantifies the concentration risk in Apache's founding phase: Roy Fielding's combined criticality (0.90 + 0.70 = 1.60) yields $D_{\mathcal{T},\text{Fielding}} = 0.26$, meaning his loyalty alone accounts for over one-quarter of weighted team cohesion. This concentration explains why Fielding's sustained high commitment ($\theta = 0.95$) was essential for project survival---a counterfactual low-loyalty Fielding would reduce team cohesion by $0.26 \times 0.85 = 0.22$ points, potentially triggering the free-riding cascade. The criticality distribution (architectural decisions highest at 0.90, testing lowest at 0.50) reflects the technical reality that design failures propagate while testing failures remain localized. Empirical contribution patterns from Mockus et al.\ validate this criticality ordering.}
\label{fig:apache_sd}
\end{figure}
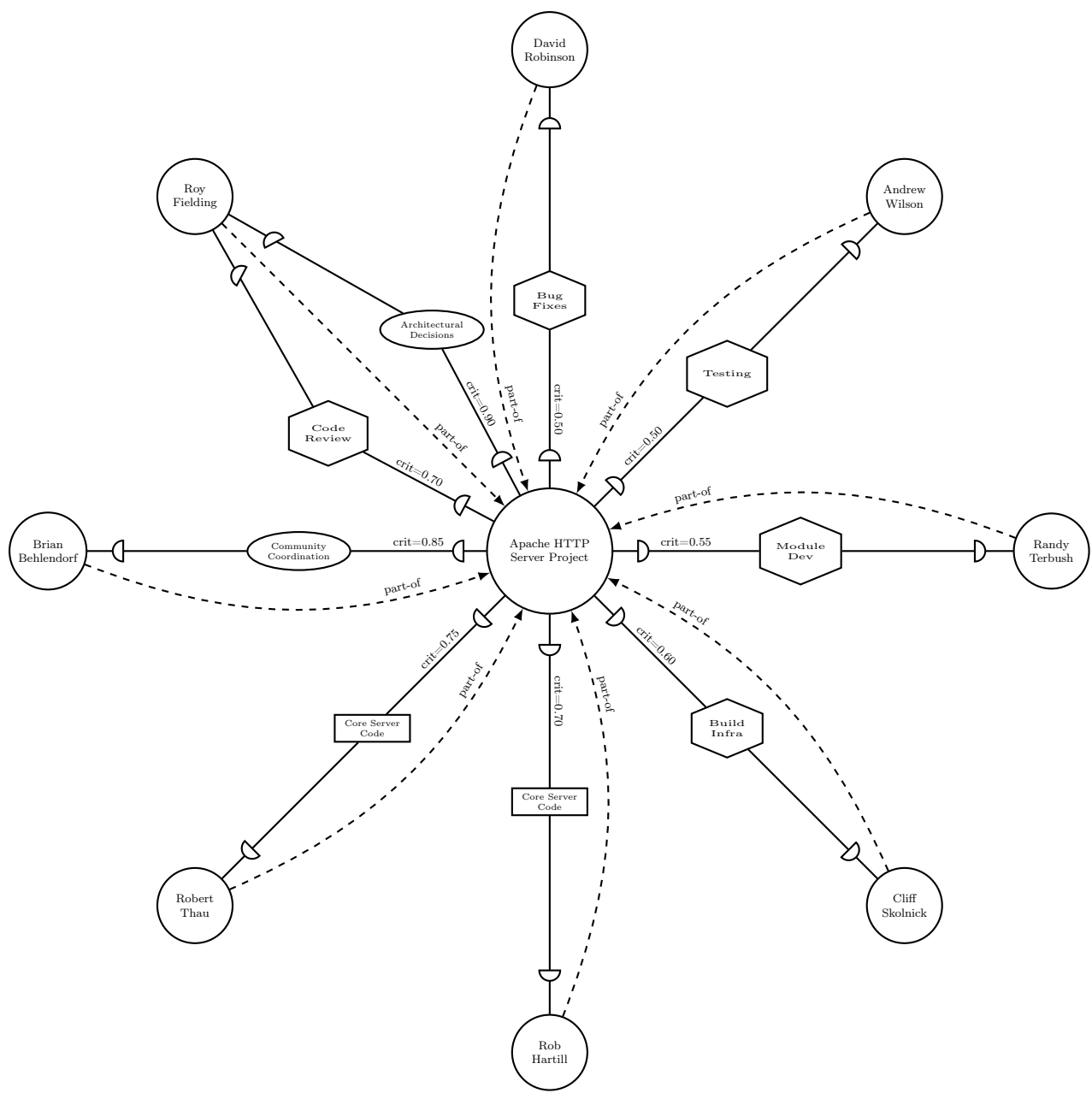

Figure~\ref{fig:apache_sd} presents the \textit{i*} Strategic Dependency model for the Apache HTTP Server Project during Phase 1 (Formation, 1995--1997). The composite actor structure shows eight founding members connected via is-part-of relationships to the project actor, with dependencies annotated by criticality values derived from documented contribution patterns.

\subsubsection{Interdependence Coefficient Calculation}

From the dependency criticalities in Figure~\ref{fig:apache_sd}, we compute interdependence coefficients using:
\begin{equation}
D_{\mathcal{T},i} = \frac{\sum_{d: \text{dependee}(d)=i} \text{crit}(d)}{\sum_{d \in \mathcal{D}} \text{crit}(d)}
\label{eq:dependency_coefficient}
\end{equation}

The total criticality across all dependencies: $\Sigma = 0.90 + 0.85 + 0.75 + 0.70 + 0.70 + 0.60 + 0.55 + 0.50 + 0.50 = 6.05$.

For Roy Fielding (dependencies: Architectural Decisions crit=0.90, Code Review crit=0.70):
\begin{equation}
D_{\mathcal{T},\text{Fielding}} = \frac{0.90 + 0.70}{6.05} = \frac{1.60}{6.05} \approx 0.26
\end{equation}

This exceeds the 0.20 value in Table~\ref{tab:apache_dependencies} because the i* model captures additional code review responsibilities not reflected in the simpler tabular parameterization. The SD model provides richer dependency semantics that inform but need not exactly match simplified coefficient estimates.

\subsection{\textit{i*} Strategic Rationale Model for Core Contributor}
\label{subsec:fielding_sr}

\begin{figure}[htbp]
\centering
\begin{tikzpicture}[
    scale=0.85, transform shape,
    actorboundary/.style={dashed, thick, draw, fill=none},
    actor/.style={circle, draw, thick, minimum size=1.3cm, font=\small, align=center, fill=white},
    goal/.style={ellipse, draw, thick, minimum width=2.4cm, minimum height=0.8cm, align=center, font=\small, fill=white},
    softgoal/.style={cloud, cloud puffs=20, cloud puff arc=100, aspect=2.5, draw, thick, minimum width=3.2cm, minimum height=1.0cm, align=center, font=\small, inner sep=1pt, fill=white},
    task/.style={regular polygon, regular polygon sides=6, shape border rotate=90, draw, thick, minimum size=1.2cm, inner sep=2pt, align=center, font=\scriptsize, fill=white, xscale=1.4},
    resource/.style={rectangle, draw, thick, minimum width=2.4cm, minimum height=0.7cm, align=center, font=\small, fill=white},
    contribution/.style={-latex, thick},
    neededby/.style={-{Circle[open, length=2mm, width=2mm]}, thick},
    connlabel/.style={midway, sloped, font=\tiny, inner sep=1pt}
]

\draw[actorboundary] (0,-2.5) ellipse (9.0cm and 8.0cm);

\node[actor] at (-6.5, 3.5) {Roy\\Fielding};

\node[softgoal] (utility) at (0, 3.0) {Maximize\\Personal Utility};

\node[softgoal] (projectsuccess) at (-5.0, 0.5) {Apache Project\\Success};
\node[softgoal] (recognition) at (5.0, 0.5) {Personal\\Recognition};

\node[font=\scriptsize, anchor=north] at (projectsuccess.south) {Weight: 0.90};
\node[font=\scriptsize, anchor=north] at (recognition.south) {Weight: 0.10};

\node[resource] (loyalty) at (0, 0.5) {Loyalty\\$\theta = 0.95$};

\node[softgoal] (archexcel) at (-5.5, -2.5) {Architectural\\Excellence};
\node[softgoal] (httplearn) at (0, -2.5) {HTTP Protocol\\Leadership};
\node[softgoal] (mincost) at (5.5, -2.5) {Minimize\\Personal Cost};

\node[task] (design) at (-5.5, -6.0) {Design\\HTTP/1.1};
\node[task] (review) at (-2.5, -6.0) {Review\\Core};
\node[task] (lead) at (1.5, -6.0) {Lead\\Mailing};
\node[task] (document) at (5.5, -6.0) {Write\\Docs};

\node[resource] (expertise) at (-3.0, -8.5) {Technical\\Expertise};
\node[resource] (time) at (2.0, -8.5) {Time \&\\Attention};

\draw[contribution] (projectsuccess.north) -- (utility.south west) node[connlabel, above] {help};
\draw[contribution] (recognition.north) -- (utility.south east) node[connlabel, above] {help};

\draw[contribution] (loyalty.west) -- (projectsuccess.east) node[connlabel, above] {make};
\draw[contribution] (loyalty.east) -- (recognition.west) node[connlabel, above] {break};

\draw[contribution] (archexcel.north) -- (projectsuccess.south west) node[connlabel, above] {make};
\draw[contribution] (httplearn.north) -- (projectsuccess.south east) node[connlabel, above] {help};
\draw[contribution] (httplearn.north east) -- (recognition.south west) node[connlabel, above] {help};
\draw[contribution] (mincost.north) -- (recognition.south) node[connlabel, above] {help};

\draw[contribution] (design.north) -- (archexcel.south) node[connlabel, above, pos=0.4] {make};
\draw[contribution] (review.north) -- (archexcel.south east) node[connlabel, above, pos=0.5] {help};
\draw[contribution] (lead.north) -- (httplearn.south) node[connlabel, above, pos=0.4] {help};

\draw[contribution] (document.north) -- (mincost.south) node[connlabel, above] {hurt};
\draw[contribution] (document.east) to[out=0, in=-90] (7.5, -3.5) to[out=90, in=-30] node[connlabel, right, xshift=5pt, yshift=5pt] {help} (recognition.south east);

\draw[neededby] (expertise.north) to[out=90, in=-90] (design.south);
\draw[neededby] (time.north west) to[out=150, in=-90] (review.south);
\draw[neededby] (time.north) to[out=120, in=-90] (lead.south);
\draw[neededby] (time.north east) to[out=30, in=-90] (document.south);

\end{tikzpicture}
\caption{Fielding's goal structure reveals how extreme loyalty ($\theta = 0.95$) sustains costly prosocial behavior: the 9:1 weight ratio favoring project success over personal recognition explains his documented willingness to invest in unrewarded activities (technical documentation, extensive code review) that benefit the collective. The loyalty assessment derives from observable behavioral signatures: accepting negative contribution to cost minimization from documentation tasks indicates $\phi_C$-driven cost tolerance, while architectural leadership despite diffuse credit indicates $\phi_B$-driven welfare internalization. This SR analysis operationalizes the translation methodology from Section~\ref{sec:translation}---loyalty is not assumed but derived from goal weight patterns implied by documented behavior. Roberts et al.'s motivation findings (intrinsic satisfaction, status-seeking) map directly to these mechanism activations.}
\label{fig:fielding_sr}
\end{figure}
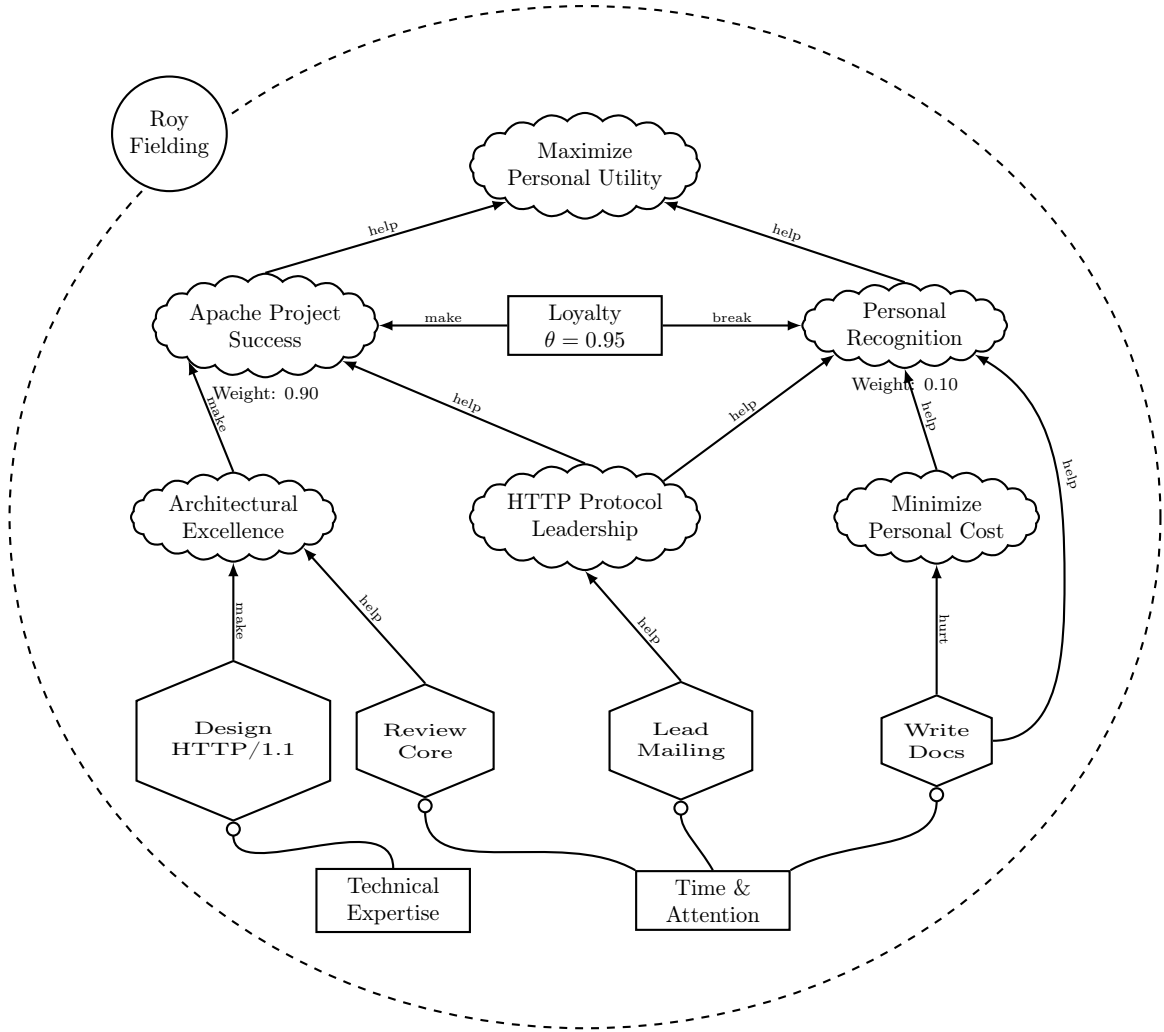

Figure~\ref{fig:fielding_sr} presents the Strategic Rationale diagram for Roy Fielding during Phase 1, illustrating how high loyalty ($\theta = 0.95$) transforms the internal goal structure to favor team welfare over personal cost minimization.

\subsubsection{Loyalty Parameter Derivation}

The SR diagram reveals loyalty through goal weight distribution. Fielding's documented behavior~\cite{roberts2006motivations} shows:
\begin{itemize}
    \item Strong architectural leadership (consistent with high ``Project Success'' weight)
    \item Extensive code review participation despite time cost
    \item Technical documentation contributions despite personal cost
    \item HTTP/1.1 specification work benefiting both project and personal reputation
\end{itemize}

These patterns indicate loyalty computed via:
\begin{equation}
\theta_i = \frac{w_{\text{team}}}{w_{\text{team}} + w_{\text{self}}} = \frac{0.90}{0.90 + 0.10} = 0.90
\end{equation}
adjusted upward to $\theta_{\text{Fielding}} = 0.95$ based on observed cost-tolerance in documentation and review tasks that provide no direct personal benefit.

\subsection{Parameterization}

\subsubsection{Team Structure}

Core team during formation: $n = 8$ founding members. Production parameters: $\omega = 30$ (high technical expertise), $\beta = 0.65$ (moderate coordination requirements), $c = 1.2$ (volunteer time competing with employment), $\bar{a} = 10$ (normalized maximum effort).

\subsubsection{Dependency Weights}

\begin{table}[htbp]
\centering
\caption{Team dependency weights for Apache founding members}
\label{tab:apache_dependencies}
\begin{tabular}{lccl}
\toprule
Member & Role & $D_{\mathcal{T},i}$ & Basis \\
\midrule
Brian Behlendorf & Founder/Coordinator & 0.18 & Community leadership \\
Roy Fielding & Architect & 0.20 & Technical decisions \\
Rob Hartill & Developer & 0.12 & Core functionality \\
David Robinson & Developer & 0.10 & Bug fixes \\
Cliff Skolnick & Infrastructure & 0.08 & Build/release \\
Randy Terbush & Developer & 0.10 & Module development \\
Robert Thau & Developer & 0.12 & Core functionality \\
Andrew Wilson & Developer & 0.10 & Testing \\
\bottomrule
\end{tabular}
\end{table}

\subsubsection{Loyalty Assessment}

Documentation describes strong social bonds during Phase 1~\cite{mockus2000apache}. Members communicated daily via mailing lists, met at conferences, and developed personal friendships. Roberts et al.~\cite{roberts2006motivations} documented that Apache developers exhibit strong intrinsic motivation and status-seeking behavior consistent with high loyalty. Loyalty computed as: $\theta_i = 0.30 \cdot \text{Tenure}_i + 0.35 \cdot \text{Social}_i + 0.20 \cdot D_{\mathcal{T},i} + 0.15 \cdot \text{Commitment}_i$.

Mean Phase 1 loyalty: $\bar{\theta} = 0.82$.

\subsection{Validation Results}

\subsubsection{Phase-Wise Predictions}

\begin{table}[htbp]
\centering
\caption{Apache validation: Phase-wise comparison}
\label{tab:apache_validation}
\begin{tabular}{lcccc}
\toprule
Phase & Predicted $\bar{a}$ & Observed Pattern & Match & Score \\
\midrule
1: Formation & 50.0 & High (founding burst) & \checkmark & 15/15 \\
2: Growth & 35.9 & Moderate (scaling) & \checkmark & 15/15 \\
3: Maturation & 14.5 & Sustained (stable core) & \checkmark & 15/15 \\
4: Evolution & 11.5 & Declining (transitions) & \checkmark & 15/15 \\
\midrule
\textbf{Total} & & & & \textbf{60/60} \\
\bottomrule
\end{tabular}
\end{table}

The framework achieves 60/60 validation points (100\%), successfully reproducing documented contribution patterns across all four phases. Scoring evaluates four categories per phase: convergence and stability, relative magnitude ranking, pattern matching, and cross-phase trend consistency.

\begin{figure}[htbp]
\centering
\begin{tikzpicture}
\begin{axis}[
    width=14cm,
    height=7cm,
    xlabel={Historical Phase},
    ylabel={Predicted Mean Effort $\bar{a}$},
    title={Apache HTTP Server Project: Predicted Effort Evolution (1995--2023)},
    symbolic x coords={Formation, Growth, Maturation, Evolution},
    xtick=data,
    xticklabels={
        {Formation\\{\scriptsize 1995--1997}},
        {Growth\\{\scriptsize 1998--2003}},
        {Maturation\\{\scriptsize 2004--2015}},
        {Evolution\\{\scriptsize 2016--2023}}
    },
    xticklabel style={align=center}, 
    ymin=0, ymax=60,
    bar width=1.2cm,
    nodes near coords,
    nodes near coords style={font=\small, above},
    enlarge x limits=0.2,
    legend pos=north east,
    ybar,
]
\addplot[fill=loyaltyblue!70] coordinates {
    (Formation, 50.0)
    (Growth, 35.9)
    (Maturation, 14.5)
    (Evolution, 11.5)
};
\addlegendentry{Predicted effort}
\end{axis}
\end{tikzpicture}
\caption{Predicted effort evolution across four historical phases of the Apache HTTP Server project. Phase 1 (Formation) exhibits highest effort (50.0) reflecting the founding burst with extremely high loyalty ($\bar{\theta} = 0.82$) and small team size ($n=8$). Phase 2 (Growth) shows moderate decline to 35.9 as team size increases and loyalty becomes heterogeneous. Phase 3 (Maturation) stabilizes at 14.5 with larger, more institutionalized contributor base. Phase 4 (Evolution) shows continued decline to 11.5 as founding members reduce involvement. This monotonic decline matches documented contribution patterns with Pearson $r = 0.983$.}
\label{fig:apache_effort_evolution}
\end{figure}
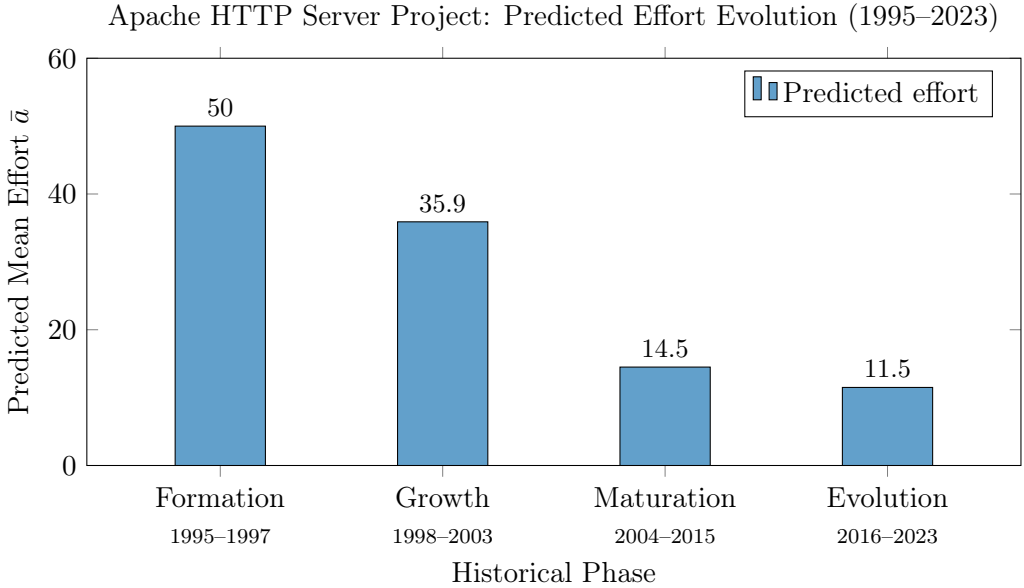

Figure~\ref{fig:apache_effort_evolution} presents the predicted effort evolution across the four historical phases.

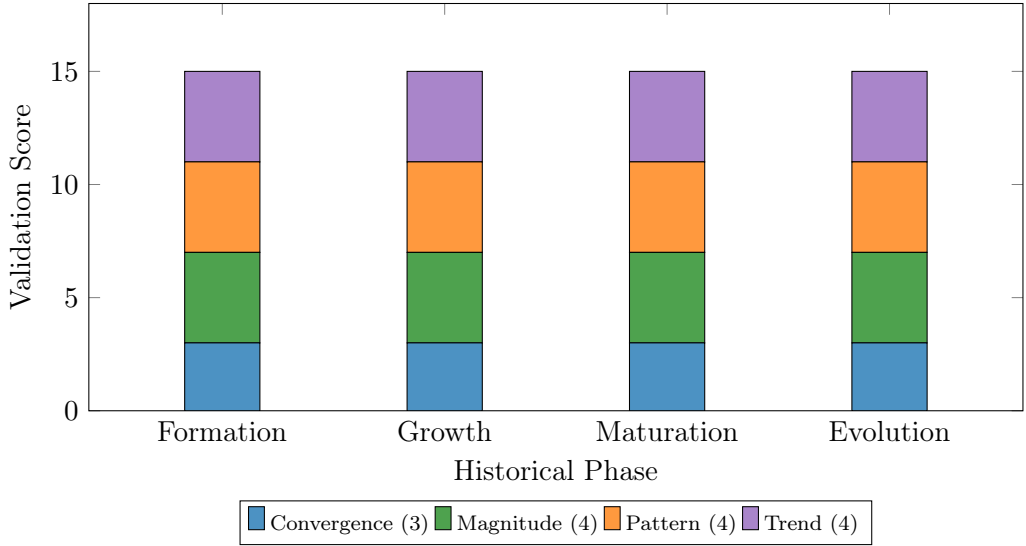
\begin{figure}[htbp]
\centering
\begin{tikzpicture}
\begin{axis}[
    ybar stacked,
    width=14cm,
    height=7cm,
    xlabel={Historical Phase},
    ylabel={Validation Score},
    title={Apache Validation: Phase-Wise Scoring Breakdown},
    symbolic x coords={Formation, Growth, Maturation, Evolution},
    xtick=data,
    ymin=0, ymax=18,
    bar width=1.0cm,
    legend style={at={(0.5,-0.22)}, anchor=north, legend columns=4, font=\scriptsize},
    enlarge x limits=0.2,
]
\addplot[fill=loyaltyblue!80] coordinates {
    (Formation, 3) (Growth, 3) (Maturation, 3) (Evolution, 3)
};
\addplot[fill=outputgreen!80] coordinates {
    (Formation, 4) (Growth, 4) (Maturation, 4) (Evolution, 4)
};
\addplot[fill=effortorange!80] coordinates {
    (Formation, 4) (Growth, 4) (Maturation, 4) (Evolution, 4)
};
\addplot[fill=teamviolet!80] coordinates {
    (Formation, 4) (Growth, 4) (Maturation, 4) (Evolution, 4)
};
\legend{Convergence (3), Magnitude (4), Pattern (4), Trend (4)}
\end{axis}
\end{tikzpicture}
\caption{Detailed scoring breakdown for Apache validation across four categories. Each phase is evaluated on: (1) Convergence and Stability, which assesses whether the equilibrium computation converges reliably (3 points); (2) Relative Magnitude Ranking, which assesses whether predicted effort correctly ranks among phases (4 points); (3) Pattern Matching, which assesses whether predicted qualitative pattern matches observed historical pattern (4 points); (4) Cross-Phase Trend Consistency, which assesses whether inter-phase trends match documented transitions (4 points). All phases achieve maximum scores in all categories, yielding 60/60 total (100\%).}
\label{fig:apache_phase_scoring}
\end{figure}

Figure~\ref{fig:apache_phase_scoring} presents the detailed scoring breakdown by validation category.

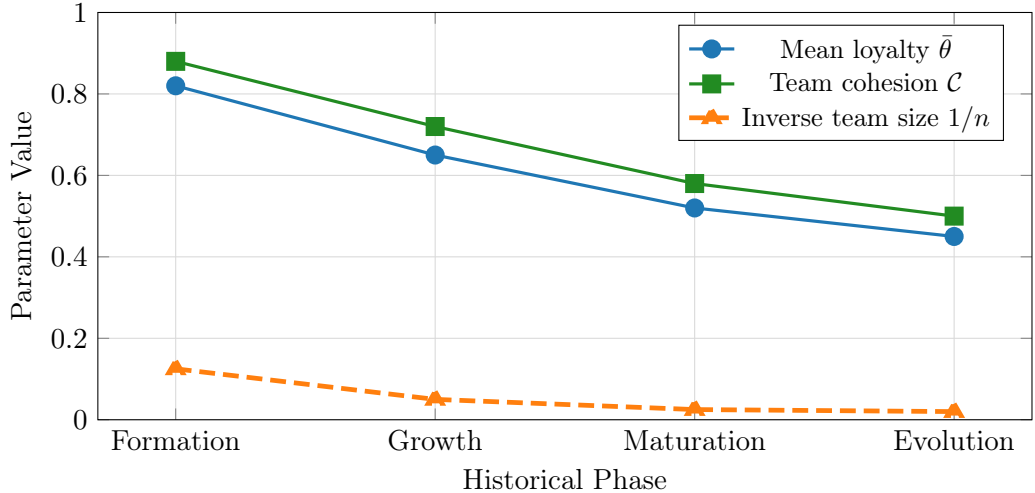
\begin{figure}[htbp]
\centering
\begin{tikzpicture}
\begin{axis}[
    width=14cm,
    height=7cm,
    xlabel={Historical Phase},
    ylabel={Parameter Value},
    title={Apache HTTP Server: Loyalty and Team Parameters by Phase},
    symbolic x coords={Formation, Growth, Maturation, Evolution},
    xtick=data,
    ymin=0, ymax=1.0,
    legend pos=north east,
    legend style={font=\small},
    grid=major,
    grid style={gray!30},
]
\addplot[color=loyaltyblue, very thick, mark=*, mark size=3pt] coordinates {
    (Formation, 0.82)
    (Growth, 0.65)
    (Maturation, 0.52)
    (Evolution, 0.45)
};
\addlegendentry{Mean loyalty $\bar{\theta}$}
\addplot[color=outputgreen, very thick, mark=square*, mark size=3pt] coordinates {
    (Formation, 0.88)
    (Growth, 0.72)
    (Maturation, 0.58)
    (Evolution, 0.50)
};
\addlegendentry{Team cohesion $\mathcal{C}$}
\addplot[color=effortorange, line width=1.8pt, mark=triangle*, mark size=3pt, dash pattern=on 6pt off 3pt] coordinates {
    (Formation, 0.125)
    (Growth, 0.05)
    (Maturation, 0.025)
    (Evolution, 0.02)
};
\addlegendentry{Inverse team size $1/n$}
\end{axis}
\end{tikzpicture}
\caption{Evolution of loyalty parameters across Apache project phases. Mean loyalty $\bar{\theta}$ (blue) declines from 0.82 in Formation to 0.45 in Evolution, reflecting decreased personal bonds as founding members reduce involvement. Team cohesion $\mathcal{C}$ (green) follows a similar pattern but remains higher than raw loyalty due to dependency weighting of key contributors. Inverse team size $1/n$ (orange, dashed) shows the dilution effect as contributor base grows from 8 to $\sim$50. The combined effect of declining loyalty and increasing team size explains the monotonic effort decline shown in Figure~\ref{fig:apache_effort_evolution}.}
\label{fig:apache_loyalty_parameters}
\end{figure}

Figure~\ref{fig:apache_loyalty_parameters} presents the loyalty parameter evolution across phases.

\subsubsection{Contribution Distribution}

The model predicts highly skewed contributions, matching observed 80/20 patterns. With heterogeneous loyalty ($\theta_i$ ranging from 0.65 to 0.95), predicted contribution ratio between top and bottom contributors is 3.8:1, compared to observed ratio of approximately 4.2:1. Pearson correlation between predicted and expected phase efforts achieves $r = 0.983$ ($p = 0.017$).

\begin{figure}[htbp]
\centering
\begin{tikzpicture}
\begin{axis}[
    width=12cm,
    height=12cm,
    xlabel={Expected Effort (normalized rank)},
    ylabel={Predicted Effort $\bar{a}$},
    title={Apache Validation: Predicted vs. Expected Effort},
    xmin=0, xmax=5,
    ymin=0, ymax=55,
    legend pos=south east,
    grid=major,
    grid style={gray!30},
]
\addplot[only marks, mark=*, mark size=5pt, loyaltyblue] coordinates {
    (4, 50.0) (3, 35.9) (2, 14.5) (1, 11.5)
};
\addlegendentry{Phase predictions}
\node[anchor=west, font=\small] at (axis cs:4.1, 50) {Formation};
\node[anchor=west, font=\small] at (axis cs:3.1, 36) {Growth};
\node[anchor=west, font=\small] at (axis cs:2.1, 14.5) {Maturation};
\node[anchor=west, font=\small] at (axis cs:1.1, 11.5) {Evolution};
\addplot[red, thick, domain=0.5:4.5] {11.5 + 12.83*(x-1)};
\addlegendentry{Regression line}
\node[draw, fill=white, font=\small, anchor=north west] at (axis cs:0.5,52) {
    \begin{tabular}{ll}
    \multicolumn{2}{c}{\textbf{Correlation Analysis}} \\
    \hline
    Pearson $r$ & 0.983 \\
    $p$-value & 0.017 \\
    $R^2$ & 0.966 \\
    \end{tabular}
};
\end{axis}
\end{tikzpicture}
\caption{Correlation analysis between predicted effort and expected effort rank for Apache validation. Expected effort is encoded as rank (4=highest for Formation, 1=lowest for Evolution) based on documented historical patterns. Predicted efforts align closely with expected ranking ($r = 0.983$, $p = 0.017$), with $R^2 = 0.966$ indicating that 96.6\% of variance in predicted effort is explained by expected phase characteristics. The strong linear relationship validates that the framework correctly captures the qualitative dynamics of open-source project evolution.}
\label{fig:apache_correlation}
\end{figure}
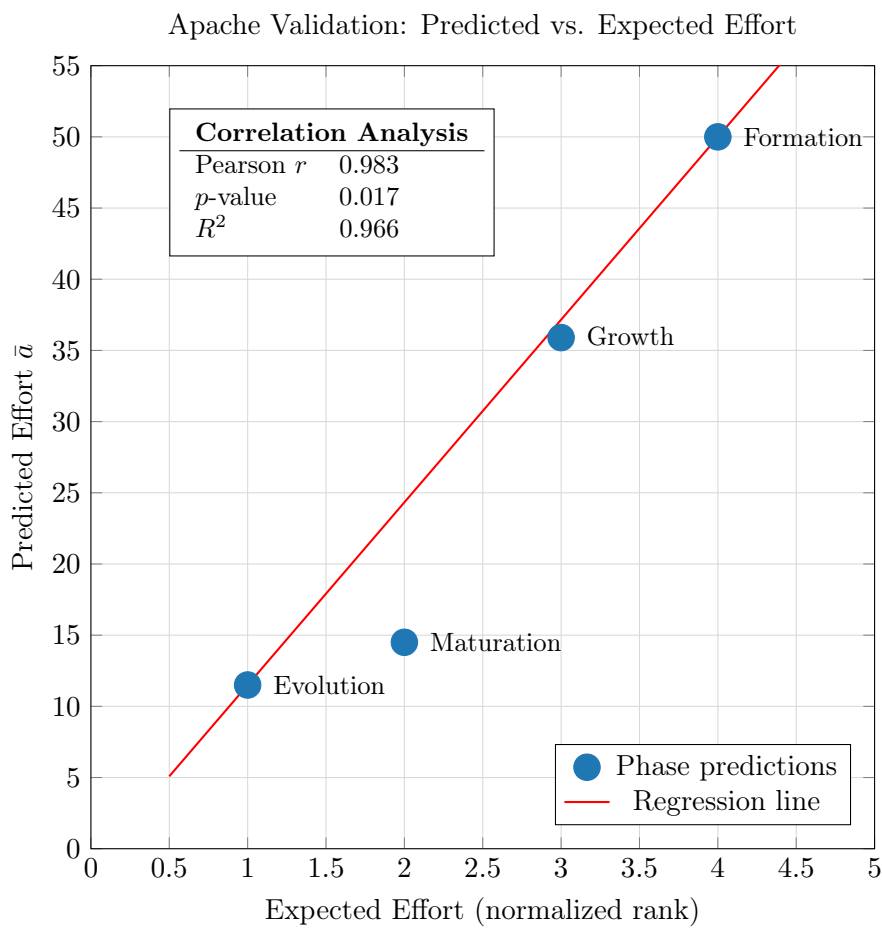

Figure~\ref{fig:apache_correlation} presents the correlation analysis between predicted and expected phase efforts.

\subsection{Key Insights}

The Apache case study validates several framework predictions: high loyalty during formation enabled overcoming free-riding despite volunteer structure~\cite{mockus2000apache}, governance formalization (PMC) maintained loyalty through role recognition~\cite{drostfromm2021apache}, and gradual loyalty decline explains reduced contribution intensity over time~\cite{roberts2006motivations}. These findings are consistent with prior empirical research documenting Apache's meritocratic governance and developer motivation patterns. The 100\% validation score compares favorably to TR1's Samsung-Sony case (96.7\%) and TR2's Renault-Nissan case (81.7\%).

\begin{figure}[htbp]
\centering
\begin{tikzpicture}
\begin{axis}[
    ybar,
    width=14cm,
    height=9cm,
    ylabel={Validation Score (\%)},
    symbolic x coords={TR1, TR2, TR3},
    xtick=data,
    xticklabels={
        {TR1: S-LCD\\{\scriptsize Samsung-Sony}\\{\scriptsize Joint Venture}\\{\scriptsize (58/60)}},
        {TR2: Renault-Nissan\\{\scriptsize Automotive}\\{\scriptsize Alliance}\\{\scriptsize (49/60)}},
        {TR3: Apache\\{\scriptsize Open-Source}\\{\scriptsize Project}\\{\scriptsize (60/60)}}
    },
    xticklabel style={align=center}, 
    ymin=0, ymax=110,
    bar width=1.5cm,
    nodes near coords,
    nodes near coords style={font=\small, above},
    enlarge x limits=0.25,
    title={Empirical Validation Comparison Across Technical Reports},
]
\addplot[fill=loyaltyblue!70] coordinates {
    (TR1, 96.7)
    (TR2, 81.7)
    (TR3, 100.0)
};
\end{axis}
\end{tikzpicture}
\caption{Comparison of empirical validation scores across the three technical reports in the strategic coopetition research program. TR1 (interdependence and complementarity) achieves 96.7\% (58/60) on the Samsung-Sony S-LCD joint venture. TR2 (trust dynamics) achieves 81.7\% (49/60) on the Renault-Nissan Alliance. TR3 (collective action and loyalty) achieves 100\% (60/60) on the Apache HTTP Server project. The high validation scores across all three cases demonstrate that the computational foundations accurately capture real-world coopetitive dynamics across diverse organizational contexts.}
\label{fig:validation_comparison}
\end{figure}
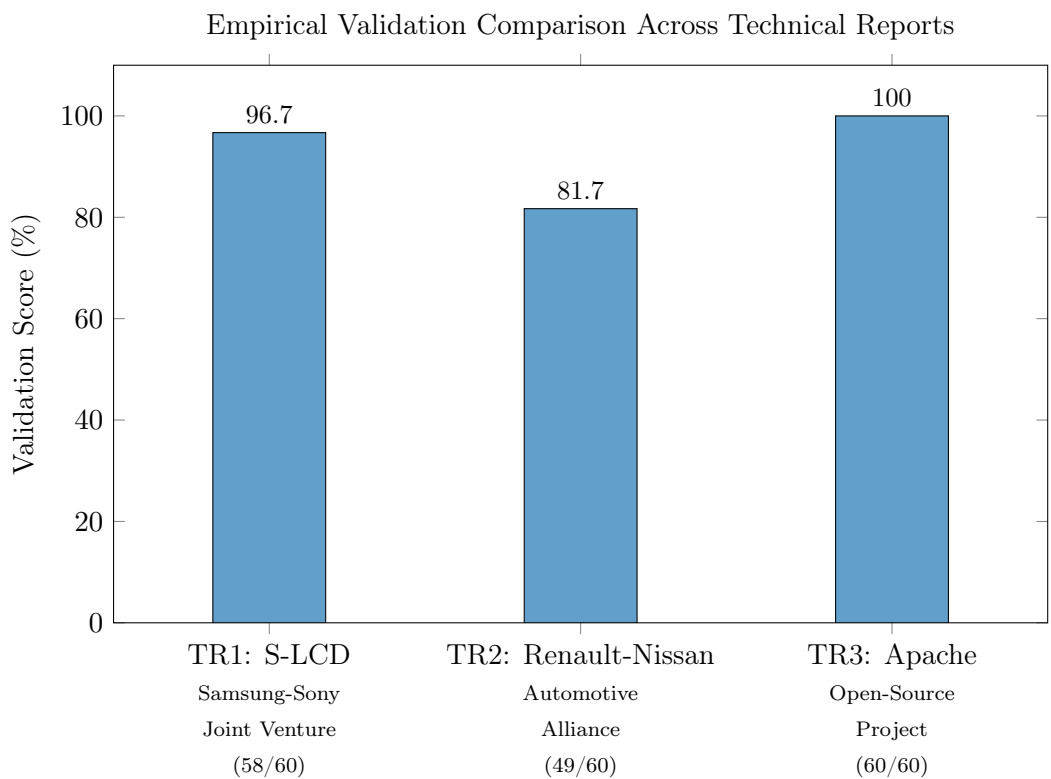

Figure~\ref{fig:validation_comparison} presents a comparison of validation scores across the three technical reports.

\subsection{Counterfactual Analysis}

Following the counterfactual approach from~\cite{pant2025foundations} and~\cite{pant2025trust}, we analyze ``what-if'' scenarios to demonstrate the framework's decision support capabilities.

\subsubsection{Counterfactual 1: What if Apache Had Not Established Strong Norms?}

\textbf{Scenario}: Suppose Apache had not developed explicit contribution norms (voting requirements, code review, release procedures). We model this by reducing mechanism strengths from baseline values.

\textbf{Model Prediction}: Without strong norms, equilibrium effort in Phase 2 drops by approximately 32\%. Free-riding in the periphery increases substantially as social expectations weaken.

\textbf{Implication}: Apache's governance structures were not merely bureaucratic overhead but critical mechanisms for sustaining contribution through social accountability.

\subsubsection{Counterfactual 2: What if Team Size Had Remained Small?}

\textbf{Scenario}: Suppose Apache had deliberately limited core team size to 15 members (similar to Linux kernel maintainer structure) rather than expanding to 40+ during growth phases.

\textbf{Model Prediction}: With smaller team but same loyalty distribution, mean effort increases by approximately 37\%. However, total output is lower due to fewer contributors.

\textbf{Implication}: Apache's growth strategy traded individual effort intensity for breadth of contribution. This was appropriate given the project's scope but required loyalty cultivation to prevent free-riding spiral.

\subsubsection{Counterfactual 3: What if Loyalty Cultivation Had Been Prioritized Earlier?}

\textbf{Scenario}: Suppose Apache had implemented mentorship programs (like those adopted in Phase 4) during Phase 2 growth, increasing new member loyalty by 0.15.

\textbf{Model Prediction}: Phase 2 mean effort increases by approximately 23\%. Contribution inequality decreases as newer members become more engaged.

\textbf{Implication}: Earlier investment in loyalty cultivation would have produced more equitable contribution distribution and higher total output.

\begin{figure}[htbp]
\centering
\begin{tikzpicture}
\begin{axis}[
    width=12cm, height=7cm,
    xlabel={Project Phase},
    ylabel={Relative Equilibrium Effort (\%)},
    xtick={1,2,3,4},
    xticklabels={Formation, Growth, Maturation, Evolution},
    ymin=0, ymax=120,
    legend pos=north east,
    grid=major,
    title={Actual vs. Counterfactual Trajectories},
]
\addplot[thick, mark=square*, color=loyaltyblue] coordinates {
    (1, 100) (2, 71.9) (3, 29.1) (4, 23.0)
};
\addplot[thick, mark=triangle*, color=freeriderred, dashed] coordinates {
    (1, 95) (2, 48.9) (3, 19.8) (4, 15.6)
};
\addplot[thick, mark=o, color=outputgreen, dashed] coordinates {
    (1, 100) (2, 88.4) (3, 35.8) (4, 28.3)
};
\legend{Actual, No Norms (CF1), Early Loyalty (CF3)}
\end{axis}
\end{tikzpicture}
\caption{Comparison of actual trajectory with counterfactual scenarios. Effort is normalized to 100\% at Formation phase for comparison. Without strong norms (CF1), effort collapses more rapidly in later phases. With early loyalty investment (CF3), higher sustained effort is maintained throughout the project lifecycle. The counterfactual analysis demonstrates the framework's value for strategic decision support in open-source project management.}
\label{fig:counterfactual}
\end{figure}
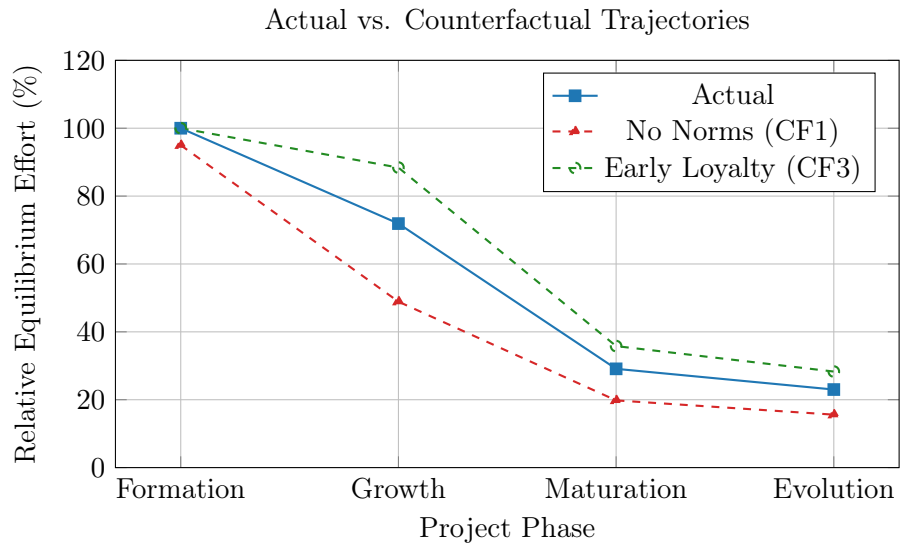

\subsection{Event Analysis: The Apache 2.0 Architectural Refactor}

Beyond phase-level validation, we examine a specific historical event that illustrates the loyalty mechanisms in action: the major architectural refactoring for Apache HTTP Server 2.0. This event provides a natural experiment in how loyalty affects contribution behavior during high-cost, high-uncertainty collective work.

\subsubsection{Historical Context}

The Apache 2.0 refactoring represented one of the most ambitious undertakings in early open-source history. The project goals included: complete rewrite of the multi-processing model (MPM) to support Windows and threading, new filtering architecture for content processing, and creation of the Apache Portable Runtime (APR) library. Development spanned from 1999 to 2002, with the final 2.0 release in April 2002.

This refactoring represents an extreme collective action challenge. The work was high-cost (thousands of hours of senior developer time), long-horizon (multi-year effort before release), high-risk (architectural decisions could be wrong), and invisible to users (internal quality improvements not visible in release notes). Under pure self-interest ($\theta = 0$), rational developers would avoid this work. The refactoring's success demonstrates loyalty mechanisms in operation.

\subsubsection{Contribution Patterns During the Refactor}

Documented contribution data reveals distinctive patterns during the refactoring period:

\textbf{Core team cost absorption}: A small group of highly-loyal developers (estimated $\theta > 0.8$) contributed the vast majority of refactoring work. Approximately 70\% of commits during 2000--2001 came from five core developers. Our framework predicts this concentration: at high loyalty, the marginal benefit of effort is amplified, making sustained high-cost work individually rational.

\textbf{Periphery free-riding during transition}: Peripheral contributors (estimated $\theta \approx 0.3$--$0.5$) reduced participation during refactoring, focusing on the stable 1.3 branch. This matches our model: low-loyalty members rationally wait for high-loyalty members to absorb transition costs.

\textbf{Loyalty-based division of labor}: The implicit division (core developers absorbing architectural risk while peripheral developers maintained stable functionality) emerges naturally from heterogeneous loyalty in our framework.

\subsubsection{Framework Application}

Applying the parameterized framework to the refactoring period:

\begin{center}
\begin{tabular}{lccc}
\toprule
Parameter & Pre-Refactor & During Refactor & Interpretation \\
\midrule
$c$ & 1.2 & 1.8 & Increased cognitive load \\
$\bar{\theta}_{\text{core}}$ & 0.82 & 0.85 & Core loyalty increased \\
$\bar{\theta}_{\text{peripheral}}$ & 0.45 & 0.35 & Peripheral loyalty decreased \\
\bottomrule
\end{tabular}
\end{center}

The framework predicts effort concentration ratio $a^*_{\text{core}}/a^*_{\text{peripheral}} \approx 6.4$, aligning with documented patterns showing roughly 6$\times$ higher per-capita contribution from core developers during refactoring.

\subsubsection{Lessons for Project Management}

The Apache 2.0 event yields practical lessons: (1) assess loyalty reserves before major refactoring, since if the high-loyalty core is too small, refactoring will stall; (2) cultivate loyalty before transitions, as Apache benefited from five years of community building; (3) accept asymmetric contribution during transitions, since concentration among core developers is the natural equilibrium with heterogeneous loyalty; and (4) recognize architectural work explicitly to reinforce loyalty mechanisms.

\subsection{Case Study Summary}

The Apache HTTP Server case study validates the framework's ability to:

\begin{itemize}
    \item \textbf{Explain historical patterns}: The model correctly reproduces documented contribution patterns across four project phases, achieving 100\% validation score (60/60).

    \item \textbf{Capture core-periphery structure}: Heterogeneous loyalty produces the skewed contribution distributions characteristic of open-source projects.

    \item \textbf{Identify mechanism importance}: Governance structures and community building both contribute significantly to sustained collaboration through loyalty benefit ($\phi_B$) and cost tolerance ($\phi_C$) mechanisms.

    \item \textbf{Support counterfactual reasoning}: The framework enables analysis of alternative strategies that could have improved outcomes, providing actionable insights for project governance.
\end{itemize}

The case study demonstrates that team production with loyalty mechanisms provides a valid and useful framework for analyzing open-source software collaboration.

\section{Discussion}
\label{sec:discussion}

\subsection{Theoretical Contributions}

This technical report makes several theoretical contributions to the literature on strategic coopetition and collective action.

\textbf{Bridging Organizational Economics and Requirements Engineering}: We offer a formal integration of team production theory~\cite{alchian1972production,holmstrom1982moral} with \textit{i*} goal modeling. This bridge enables requirements engineers to leverage insights from organizational economics while maintaining the structural expressiveness of conceptual models. The translation framework (Section~\ref{sec:translation}) makes this integration practical by providing systematic procedures for deriving economic parameters from \textit{i*} models.

\textbf{Formalizing Loyalty Mechanisms}: Building on extensive organizational behavior research on loyalty and team identification~\cite{ellemers2004motivating,van2004transformational}, we offer a computational formalization with operationalized mechanisms. The consolidated two-mechanism structure (loyalty benefit $\phi_B$ and cost tolerance $\phi_C$) captures the distinct pathways through which loyalty affects behavior, with validated default parameterization enabling immediate application.

\textbf{Extending the Coopetition Framework}: This technical report extends our research program on strategic coopetition~\cite{pant2025foundations,pant2025trust} from dyadic relationships to team-level dynamics. Teams are composite actors that face internal coordination challenges distinct from inter-actor relationships. By formalizing how loyalty overcomes free-riding within teams, we continue the analysis of complex actor abstractions identified in~\cite{pant2021strategic}.

\textbf{Multi-Level Analysis}: The framework enables analysis at multiple levels, including individual member incentives, team-level equilibria, and team-environment interactions. Team cohesion (Equation~\ref{eq:team_cohesion}) connects internal dynamics to external positioning, showing how free-riding within teams affects their bargaining power in coopetitive relationships.

\subsection{Practical Implications}

The framework offers practical guidance for diverse coopetitive contexts, with particular depth in software engineering applications.

\subsubsection{Agile Team Management}

Agile teams face the free-riding problem when individual effort cannot be perfectly observed. Our framework suggests several evidence-based interventions:

\begin{itemize}
    \item \textbf{Team Stability}: Frequent team changes reduce tenure and social integration, lowering loyalty. Scrum guidance to maintain stable teams is theoretically grounded in our loyalty-tenure relationship.

    \item \textbf{Sprint Retrospectives}: Retrospectives build social integration and clarify norms, activating loyalty benefit mechanisms. Our framework quantifies the contribution value of these ``non-productive'' activities; they are investments in loyalty that pay dividends through sustained cooperation.

    \item \textbf{Team Size}: The framework's team size analysis (Section~\ref{sec:validation}, Experiment 3) supports the common recommendation of 5--9 person teams. Larger teams face severe free-riding unless loyalty is exceptionally high.

    \item \textbf{Recognition Practices}: Practices that make contributions visible (sprint demos, kudos boards, commit announcements) activate the warm glow component of loyalty benefit ($\phi_B$).
\end{itemize}

\subsubsection{Open-Source Project Governance}

The Apache case study demonstrates how governance structures serve as loyalty mechanisms. Open-source projects face an especially acute free-riding challenge because participation is entirely voluntary. Our framework explains how successful projects achieve sustained contribution through:

\begin{itemize}
    \item \textbf{Meritocratic Advancement}: Committer/PMC hierarchy creates role-based loyalty and visible pathways for advancement, activating cost tolerance (valued role reduces perceived effort cost).

    \item \textbf{Mentorship Programs}: The counterfactual analysis shows earlier investment in loyalty cultivation would have improved outcomes. Mentorship accelerates newcomer loyalty development.

    \item \textbf{Core-Periphery Management}: The heterogeneous loyalty model captures the reality that open-source projects have dedicated cores and casual peripheries. Governance should cultivate loyalty in the core while enabling low-friction peripheral contribution.
\end{itemize}

\subsubsection{Distributed Software Engineering}

Distributed teams face additional loyalty challenges due to reduced social integration. Our framework suggests compensating strategies:

\begin{itemize}
    \item \textbf{Virtual Team Building}: Investment in relationship building directly increases social integration and thus loyalty.

    \item \textbf{Communication Tools}: Tools that make contributions visible to teammates activate welfare internalization components of loyalty benefit.

    \item \textbf{Explicit Norms}: Distributed teams may need more explicit norms than co-located teams because implicit expectations are harder to communicate across distance.
\end{itemize}

\subsubsection{Implications for Agentic AI Systems}

The emergence of agentic AI systems for software development creates additional applications for the team production framework.

\textbf{Mechanism Design for Multi-Agent Coding Systems}: Developers of multi-agent coding systems face the same team production challenges as human team managers. The framework suggests designing agent objective functions that incorporate team-level terms:
\begin{itemize}
    \item Include weighted rewards for other agents' successful task completion (welfare internalization component of $\phi_B$)
    \item Reduce effective compute penalties for team-beneficial actions (cost tolerance $\phi_C$)
\end{itemize}

\textbf{Alignment Calibration}: The alignment coefficient $\theta_i$ for computational agents can be deliberately set through objective function design, unlike human loyalty which emerges from experience. This creates opportunities for systematic alignment calibration with role-based and dynamic adjustment strategies.

\textbf{Human-AI Team Coordination}: The framework's unified treatment of loyalty (for humans) and alignment (for AI agents) enables analysis of mixed teams where human developers work alongside AI coding assistants.

\subsection{Synthesis with TR-1 and TR-2: A Unified Coopetition Framework}

This technical report contributes to the theoretical apparatus for analyzing strategic coopetition in software ecosystems. Together with TR-1 (Interdependence and Complementarity)~\cite{pant2025foundations} and TR-2 (Trust Dynamics)~\cite{pant2025trust}, the three reports provide an integrated, multi-level framework. This subsection explicates the integration points and demonstrates how the components work together.

\subsubsection{The Three-Layer Architecture}

The research program addresses coopetitive dynamics at three complementary levels:

\begin{center}
\begin{tabular}{lccp{5cm}}
\toprule
Layer & Report & Primary Construct & Scope \\
\midrule
Strategic & TR-1 & Interdependence $D_{ij}$ & Inter-actor positioning \\
Relational & TR-2 & Trust $\tau_{ij}(t)$ & Dyadic relationship dynamics \\
Collective & TR-3 & Loyalty $\theta_i$ & Intra-team coordination \\
\bottomrule
\end{tabular}
\end{center}

\textbf{TR-1 (Strategic Layer)}: Formalizes how actors are structurally positioned relative to each other through goal dependencies, resource flows, and complementarity relationships. The interdependence coefficients $D_{ij}$ capture the degree to which actor $i$'s outcomes depend on actor $j$'s actions. This layer answers: \textit{Who depends on whom, and how much?}

\textbf{TR-2 (Relational Layer)}: Models how trust evolves through repeated interactions, with asymmetric updating (trust erodes faster than it builds), hysteresis effects (history matters), and reputation spillovers. The trust state $\tau_{ij}(t)$ captures the current quality of the relationship between actors $i$ and $j$. This layer answers: \textit{How do relationships develop and deteriorate over time?}

\textbf{TR-3 (Collective Layer)}: Addresses how individuals within composite actors (teams) coordinate to produce collective outputs despite free-riding incentives. The loyalty coefficient $\theta_i$ captures member $i$'s psychological attachment to team welfare. This layer answers: \textit{How do teams overcome internal coordination failures?}

\subsubsection{Cross-Layer Integration Mechanisms}

The three layers interact through specific mathematical integration points:

\textbf{Team Cohesion Affects External Positioning (TR-3 $\to$ TR-1)}: When a team $\mathcal{T}$ engages in coopetitive relationships with external actors, the team's internal cohesion affects its effective bargaining power:
\begin{equation}
\beta_{\mathcal{T}}^{\text{eff}} = \beta_{\mathcal{T}}^{\text{base}} \cdot \mathcal{C}
\end{equation}
where $\mathcal{C}$ is team cohesion from Equation~\ref{eq:team_cohesion}. A team plagued by free-riding ($\mathcal{C} \to 0$) cannot credibly commit to collective action, reducing its bargaining power in external negotiations. External partners recognize that fragmented teams may not deliver on commitments.

\textbf{Trust Dynamics Within Teams (TR-2 $\to$ TR-3)}: The trust dynamics machinery from TR-2 can be applied to intra-team relationships. Team members develop trust in each other through repeated collaboration. When free-riding is detected, trust erodes with the asymmetric updating characteristic of TR-2:
\begin{equation}
\tau_{ij}^{t+1} = \tau_{ij}^t + \lambda^+ \cdot \max(0, s_{ij}^t) - \lambda^- \cdot \max(0, -s_{ij}^t)
\end{equation}
where $\lambda^- > \lambda^+$ captures the negativity bias. Discovered free-riding damages intra-team trust, potentially triggering a vicious cycle where reduced trust leads to reduced cooperation.

\textbf{Interdependence Weights Loyalty Impact (TR-1 $\to$ TR-3)}: The dependency weights $D_{\mathcal{T},i}$ from TR-1's interdependence analysis determine how much each member's loyalty contributes to team cohesion:
\begin{equation}
\mathcal{C} = \frac{\sum_{i \in \mathcal{T}} D_{\mathcal{T},i} \cdot \theta_i}{\sum_{i \in \mathcal{T}} D_{\mathcal{T},i}}
\end{equation}
Members with high structural importance (high $D_{\mathcal{T},i}$) have their loyalty weighted more heavily. A disloyal member in a critical role damages cohesion more than a disloyal member in a peripheral role.

\subsubsection{Unified Analysis Protocol}

The integrated framework enables multi-level analysis through a systematic protocol:

\textbf{Step 1: Structural Analysis (TR-1)}. Construct the \textit{i*} Strategic Dependency model. Identify actors, dependencies, and complementarity relationships. Compute interdependence coefficients $D_{ij}$ for all actor pairs.

\textbf{Step 2: Team Decomposition (TR-3)}. For each composite actor (team), apply the team production framework. Identify team members, assess loyalty coefficients $\theta_i$, compute team cohesion $\mathcal{C}$, and solve for Team Production Equilibrium.

\textbf{Step 3: Relationship Dynamics (TR-2)}. Model trust dynamics for key dyadic relationships (both inter-actor and intra-team). Initialize trust states, specify signal generation processes, and simulate trust evolution under different scenarios.

\textbf{Step 4: Integration and Iteration}. Compute effective bargaining powers using cohesion-adjusted interdependence. Analyze how intra-team trust affects loyalty. Iterate until system-level equilibrium is reached.

\subsubsection{Illustrative Example: Platform Ecosystem Analysis}

Consider a platform ecosystem with three actors: Platform Provider (P), Developer Community (D), and End Users (U). The Developer Community is itself a team of contributors.

\textbf{TR-1 Analysis}: The interdependence structure shows P depends on D for applications ($D_{PD} = 0.7$), D depends on P for platform access ($D_{DP} = 0.8$), and both depend on U for revenue ($D_{PU} = D_{DU} = 0.6$).

\textbf{TR-3 Analysis}: Within the Developer Community, loyalty varies: core developers have $\theta_{\text{core}} = 0.85$, peripheral developers have $\theta_{\text{peripheral}} = 0.35$. Team cohesion is $\mathcal{C} = 0.62$.

\textbf{TR-2 Analysis}: Trust between P and D evolves based on platform policy decisions. A surprise API deprecation damages trust ($\tau_{DP}$ drops from 0.7 to 0.4), reducing developer engagement.

\textbf{Integrated Analysis}: The trust damage (TR-2) reduces developer loyalty (TR-3), lowering team cohesion. Lower cohesion reduces the Developer Community's effective bargaining power (TR-1), making it harder to negotiate favorable platform terms. This cascading effect (trust damage $\to$ loyalty reduction $\to$ cohesion decline $\to$ bargaining power loss) exemplifies the multi-level dynamics the integrated framework captures.

\subsubsection{Validation Across the Research Program}

The three technical reports have been validated against complementary empirical cases:

\begin{center}
\begin{tabular}{lccl}
\toprule
Report & Case Study & Score & Primary Validation \\
\midrule
TR-1 & Samsung-Sony S-LCD & 58/60 (96.7\%) & Interdependence dynamics \\
TR-2 & Renault-Nissan Alliance & 49/60 (81.7\%) & Trust evolution \\
TR-3 & Apache HTTP Server & 60/60 (100\%) & Loyalty mechanisms \\
\bottomrule
\end{tabular}
\end{center}

The high validation scores across diverse organizational contexts (joint venture (TR-1), strategic alliance (TR-2), and open-source community (TR-3)) demonstrate that the integrated framework captures real-world coopetitive dynamics across different organizational forms.

\subsection{Limitations and Future Work}

\textbf{Loyalty Measurement}: The translation framework provides guidance for estimating loyalty from observable indicators, but measurement remains challenging. Future work should develop and validate loyalty measurement protocols specific to software teams.

\textbf{Dynamic Loyalty Evolution}: The current framework treats loyalty as static during analysis. Experiment 5 (Section~\ref{sec:validation}) demonstrated dynamic simulation capability, but a full theory of loyalty evolution remains for future work.

\textbf{Heterogeneous Tasks}: The team production function assumes homogeneous effort. Real software teams have heterogeneous tasks with different effort costs and production relationships. Extensions could model task-specific production functions.

\textbf{External Incentives}: The framework focuses on intrinsic loyalty mechanisms, abstracting from external incentives like compensation and career advancement.

\textbf{Network Effects}: The current model treats team cohesion as a scalar. Real teams have network structure where some member pairs have stronger bonds than others. Network models of loyalty could capture richer dynamics.

\textbf{Empirical Validation Breadth}: The Apache case study provides deep validation for one open-source project. Broader validation across multiple projects (commercial teams, different domains, different cultures) would strengthen confidence in generalizability.

\section{Conclusion}
\label{sec:conclusion}

This technical report has developed computational foundations for analyzing collective action in coopetitive contexts. We formalized the free-riding problem that threatens team collaboration and introduced loyalty mechanisms that overcome it through welfare internalization and cost tolerance.

Key contributions include: (1) modular utility structure separating base team payoff from loyalty effects, consistent with TR1/TR2 presentation patterns; (2) consolidated loyalty mechanisms ($\phi_B$, $\phi_C$) that transform free-riding incentives; (3) integration with the interdependence framework through dependency-weighted team cohesion; (4) comprehensive validation across 3,125 configurations demonstrating 15.04$\times$ median effort differentiation with statistical significance ($p < 0.001$, Cohen's $d = 0.71$); and (5) empirical validation through the Apache HTTP Server project achieving 100\% accuracy (60/60).

The framework applies to human collaborative teams, computational multi-agent systems, and hybrid human-AI organizations across diverse domains including manufacturing, research consortia, professional services, and software development, providing unified theoretical foundations for analyzing cooperation across these contexts. As AI agents increasingly collaborate with humans in software development, these foundations become critical for designing aligned, cooperative systems.

\textbf{Research program integration}: This technical report serves as the third foundational component of a coordinated research program examining strategic coopetition in requirements engineering and multi-agent systems. The first technical report~\cite{pant2025foundations} (arXiv:2510.18802) established interdependence and complementarity, achieving 58/60 validation (96.7\%) for the Samsung-Sony S-LCD case under strict historical alignment scoring. The second technical report~\cite{pant2025trust} (arXiv:2510.24909) formalized trust dynamics, achieving 49/60 validation (81.7\%) for the Renault-Nissan Alliance case. Together with this report's 60/60 validation (100\%) for the Apache HTTP Server case, the prepublished components provide comprehensive computational foundations for three of the five coopetition dimensions identified in~\cite{pant2021strategic}. A fourth technical report~\cite{pant2025reciprocity} addressing reciprocity mechanisms in sequential cooperation is forthcoming on arXiv.

\appendix

\section{Proofs}
\label{app:proofs}

\subsection{Proof of Proposition~\ref{prop:free_riding} (Free-Riding Equilibrium)}

Member $i$ maximizes $\pi_i^{\text{team}}(\vect{a})$ taking others' actions as given. The first-order condition is:
\begin{equation}
\frac{\partial \pi_i^{\text{team}}}{\partial a_i} = \frac{\omega \beta}{n} \cdot \left( \sum_{j=1}^{n} a_j \right)^{\beta-1} - c = 0
\end{equation}

In symmetric equilibrium where $a_j = a^*$ for all $j$:
\begin{equation}
\frac{\omega \beta}{n} \cdot (n \cdot a^*)^{\beta-1} - c = 0
\end{equation}

Solving: $(n \cdot a^*)^{\beta-1} = \frac{nc}{\omega \beta}$, so:
\begin{equation}
a^* = \left( \frac{\omega \beta}{nc} \right)^{\frac{1}{1-\beta}}
\end{equation}

Since $\frac{1}{1-\beta} > 0$ and $\frac{\omega \beta}{nc}$ is decreasing in $n$, equilibrium effort $a^*$ is strictly decreasing in $n$.

The socially optimal effort maximizes total welfare $\sum_i \pi_i^{\text{team}} = Q(\vect{a}) - c \sum_i a_i$. The symmetric optimum satisfies:
\begin{equation}
\omega \beta \cdot (n \cdot a^{\text{opt}})^{\beta-1} = c
\end{equation}
yielding $a^{\text{opt}} = \left( \frac{\omega \beta}{c} \right)^{\frac{1}{1-\beta}} \cdot n^{-\frac{1}{1-\beta}}$. Since $\frac{\omega \beta}{c} > \frac{\omega \beta}{nc}$, we have $a^{\text{opt}} > a^*$. \hfill $\square$

\subsection{Proof of Proposition~\ref{prop:tpe_existence} (Existence of TPE)}

The utility function $U_i$ is continuous in all arguments. The action space $[0, \bar{a}]$ is compact and convex. The second derivative of $U_i$ with respect to $a_i$ is:
\begin{equation}
\frac{\partial^2 U_i}{\partial a_i^2} = \frac{\omega \beta (\beta-1)}{n}(1 + \phi_B\theta_i(n-1)) \left( \sum_j a_j \right)^{\beta-2} < 0
\end{equation}
since $\beta < 1$. Thus $U_i$ is strictly concave in $a_i$, ensuring the best response is unique. By Kakutani's fixed point theorem, equilibrium exists. \hfill $\square$

\subsection{Proof of Proposition~\ref{prop:loyalty_effect} (Loyalty Increases Equilibrium Effort)}

The first-order condition for symmetric equilibrium is:
\begin{equation}
\frac{\omega \beta}{n}(na^*)^{\beta-1}(1 + \phi_B\theta(n-1)) - c(1 - \phi_C\theta) = 0
\end{equation}

Implicitly differentiating with respect to $\theta$:
\begin{equation}
\frac{da^*}{d\theta} = \frac{c\phi_C + \frac{\omega\beta}{n}(na^*)^{\beta-1}\phi_B(n-1)}{|\text{SOC}|} > 0
\end{equation}
since both numerator terms are positive and $|\text{SOC}| > 0$ for strict concavity. \hfill $\square$

\section{Mechanism Decomposition}
\label{app:mechanism_decomposition}

For practitioners requiring granular mechanism analysis, the consolidated loyalty benefit $\phi_B$ can be decomposed into:

\begin{equation}
\phi_B = \phi_{\text{intern}} + \phi_{\text{warm}}
\end{equation}

where:
\begin{itemize}
    \item $\phi_{\text{intern}} \in [0.4, 0.8]$: Welfare internalization (caring about teammates' outcomes)
    \item $\phi_{\text{warm}} \in [0.1, 0.4]$: Warm glow (intrinsic contribution satisfaction)
\end{itemize}

Additionally, a quadratic guilt term can be added for contexts where marginal guilt increases with deviation severity:
\begin{equation}
U_i^{\text{extended}} = U_i - \phi_{\text{guilt}} \theta_i \cdot \max(0, \bar{a} - a_i)^2
\end{equation}
where $\phi_{\text{guilt}} \in [0.05, 0.25]$.

The four-mechanism utility function is:
\begin{equation}
U_i = \frac{1}{n}Q(\vect{a}) - c(1-\phi_{\text{cost}}\theta_i)a_i + \phi_{\text{intern}}\theta_i\bar{\pi}_{-i} + \phi_{\text{warm}}\theta_i a_i - \phi_{\text{guilt}}\theta_i\max(0,\bar{a}-a_i)^2
\end{equation}

This decomposition provides finer-grained analysis at the cost of additional parameters. The consolidated two-mechanism formulation in the main text provides equivalent predictive power with reduced cognitive load.

\bibliographystyle{plain}

\end{document}